\documentclass[twocolumn,10pt]{IEEEtran}
\usepackage{ifpdf, flushend,subfigure}
\ifCLASSINFOpdf
  \usepackage[pdftex]{graphicx}
  \graphicspath{{../pdf/}{../jpeg/}}
  \DeclareGraphicsExtensions{.pdf,.jpeg,.png}
\else
  \usepackage[dvips]{graphicx}
  \graphicspath{{../eps/}}
  \DeclareGraphicsExtensions{.eps}
\fi
\usepackage{epstopdf}
\usepackage{multirow}
\usepackage{float}
\usepackage[cmex10]{amsmath}
\usepackage {amssymb}
\usepackage{graphicx}
\usepackage{array}
\usepackage{mdwmath}
\usepackage{mdwtab}
\usepackage{eqparbox}
\usepackage{nomencl}
\usepackage{cite}
\usepackage{algorithm}
\usepackage{algorithmic}
\usepackage{makeidx}
\usepackage{ifthen}
\usepackage{subfigure}
\usepackage{caption}
\usepackage{pdflscape}
\usepackage{hyperref}
\newcommand{\argmin}{\operatornamewithlimits{argmin}}
\newcommand{\argmax}{\operatornamewithlimits{argmax}}
\newcommand{\beq}{\begin{equation}}
\newcommand{\eeq}{\end{equation}}
\newcommand{\beqn}{\begin{eqnarray}}
\newcommand{\eeqn}{\end{eqnarray}}
\newcommand{\beqno}{\begin{eqnarray*}}
\newcommand{\eeqno}{\end{eqnarray*}}
\newcommand{\bma}{\begin{displaymath}}
\newcommand{\ema}{\end{displaymath}}
\newcommand{\bnu}{\begin{enumerate}}
\newcommand{\enu}{\end{enumerate}}
\newcommand{\bce}{\begin{center}}
\newcommand{\ece}{\end{center}}
\newcommand{\btb}{\begin{tabular}}
\newcommand{\etb}{\end{tabular}}
\newcommand*{\qedb}{\hfill\ensuremath{\square}}%

\newtheorem{theorem}{Theorem}[section]

\newtheorem{lemma}[theorem]{Lemma}

\newtheorem{proposition}[theorem]{Proposition}
\newtheorem{corollary}[theorem]{Corollary}
\newtheorem{definition}[theorem]{Definition}

\begin{document}

\title{A Market-Based Framework for Multi-Resource Allocation in Fog Computing}


%

\author{Duong~Tung~Nguyen,~\IEEEmembership{Student Member,~IEEE,}
        Long~Bao~Le,~\IEEEmembership{Senior Member,~IEEE,}
        and~Vijay~Bhargava,~\IEEEmembership{Life~Fellow,~IEEE}
}

\maketitle

\begin{abstract}

Fog computing  is transforming the network edge into an intelligent platform
by bringing storage, computing, control, and networking functions closer to end-users, things, and sensors. How to allocate multiple resource types (e.g., CPU, memory,  bandwidth) of  
capacity-limited heterogeneous fog nodes  to competing services with diverse requirements and preferences in a fair and efficient manner is a challenging task. 
To this end, we propose a novel market-based  resource allocation framework in which the services act as buyers and fog resources act as divisible goods in the market. 
The proposed framework aims to compute a market
equilibrium (ME)  solution at which every service obtains its favorite resource bundle under the budget constraint while the system achieves high resource utilization.
This work extends the General Equilibrium literature by considering a practical case of satiated utility functions. Also, we introduce the notions of non-wastefulness and frugality for equilibrium selection, and rigorously demonstrate that all the non-wasteful and frugal ME are  the optimal solutions to a convex program.
Furthermore, the proposed equilibrium is shown to possess salient fairness properties including envy-freeness, sharing-incentive, and proportionality. 
Another major contribution of this work is to develop
 a privacy-preserving distributed algorithm, which is of independent interest, 
for computing an ME while allowing market participants to obfuscate their private information.
Finally, extensive performance evaluation is conducted to  verify our theoretical analyses.

\end{abstract}



\printnomenclature

\section{Introduction}

Fog Computing (FC), also known as Edge Computing (EC),
 is an emerging 
paradigm that complements the 
cloud   to enable a wide range of Internet of Things (IoT) applications, reduce network traffic, and enhance  user experience.
By distributing storage, computing, control, and networking functions closer to end-users and data sources, 
FC enjoys many remarkable capabilities, including local data processing and analytics, distributed caching, localization, 
resource pooling and scaling,  enhanced privacy and security, and reliable connectivity \cite{mchi16,ymao17}.
Additionally, FC is the key to meeting the stringent requirements of new systems 
 and low-latency applications 
such as  embedded artificial intelligence, 5G networks, virtual/augmented reality (VR/AR), 
and tactile Internet.

Despite the tremendous potential, FC is still in its infancy stage and many  
challenges remain to be addressed.
In this paper, we focus on the fog resource allocation problem 
where a resource pool consisting of multiple Fog Nodes (FN) is shared among different services. 
Here, we consider an FN as any edge node  consisting of one or more computing units (e.g., edge clouds, micro data centers in campus buildings, enterprises, hospitals, malls, and telecom central offices, servers at base stations, and idle PCs in research labs) \cite{mchi16,ymao17,duong}.
Unlike the traditional cloud with virtually infinite capacity, 
 FNs have limited computational power. They also come with different sizes and configurations. 
Furthermore, in contrast to a small number of  cloud data centers (DC), there are numerous 
distributed FNs \cite{duong}.
Due to the heterogeneity of the FNs in terms of location, specifications, reliability, and reputation, the services may have 
diverse preferences towards them.
For example, a service may prefer an FN with powerful hardware and geographically close to it.

Therefore, a primary concern is how to efficiently allocate the limited fog resources 
 to  the competing services with diverse characteristics and preferences, 
considering the service priority and fairness.
To address this problem, we propose a new market-based  framework whose goal is to
harmonize the interests of different market participants so that every service is happy with its allotment while  the system maintains high resource utilization. 
The core idea is to assign different prices to resources of different FNs.
Specifically, under-demanded resources are priced low while  over-demanded resources have
high prices. In our model, each service has 
a certain budget for resource procurement, which represents the  service's priority level  and can  be interpreted as the market power of the service \cite{duong}. 
Given the resource prices, each service  buys an optimal resource bundle to maximize its utility under the budget constraint.
When the market clears, the resulting prices and allocation form a \textit{market equilibrium} (ME) \cite{eco,AGT}. 

The proposed model is motivated by many real-world scenarios   \cite{duong}. 
For instance, to divide fog resources among multiple network slices \cite{slicing},
a Telco can grant different budgets to the slices depending on their importance
and potential revenue generation (e.g., the total fee paid by the users of each slice).
Similarly, using virtual budgets, different labs/departments in a university 
can fairly share the fog resources located on their campus. 
Additionally, a service provider (e.g., Uber, Pokemon Go, a sensor network), who owns a set of FNs in a city, may allocate virtual budgets to different groups of users/sensors based on  their populations in different areas. Another example  inspired by the group buying concept is that several companies (i.e., services) may  agree  upfront  on their individual budgets, then buy/rent a  set of FNs together. Instead, these companies can pay  subscription fees (i.e., budgets) to use fog resources of a platform that manages a resource pool. This  is quite similar to the current postpaid mobile plan models.

For the settings described above, it is necessary to consider 
both system efficiency and fairness. Therefore, traditional solutions such as welfare maximization, maxmin fairness, and auction  may not be suitable \cite{duong}. For instance, a welfare maximization solution, which is also often the  design goal of many auction models \cite{nluo17}, typically allocates most of the resources to agents with high marginal gains while giving very little, even nothing, to agents with low marginal utilities. Thus, it may not be fair to some agents. On the other hands, a maxmin fairness allocation often gives too many resources to agents with low marginal utilities, hence, it may not be efficient.  
Different from these approaches, our proposed market-based solution is both fair and efficient. 
In particular, we formulate the fog resource allocation problem as a Fisher market where the services act as buyers and the fog resources are divisible goods.

Since there are multiple goods and the purchase decision of each buyer depends on all these goods (both prices and the buyer's preferences towards them), the proposed problem is  inherently  a General Equilibrium (GE) problem \cite{eco}.
 The existence of an ME in the GE model was established 
under some mild conditions (e.g., locally non-satiated utilities) in the seminal work of Arrow and Debreu \cite{karr54}. 
Unfortunately, their proof relies on fixed-point theorem and  does not  
give an algorithm to compute an equilibrium \cite{AGT}.
Hence, the algorithmic aspects of the GE theory have recently drawn a lot of interests from theoretical computer scientists \cite{ndev08,vvaz11,xche17,jgag15,bcod04}.
However, there is still no general technique for equilibrium computation \cite{xche17}. The strategic behavior of agents in the Fisher market game has also been studied extensively recently \cite{bads10,nche16,sbra14,sbra17}.

Conventionally, the dual variables associated with the resource capacity constraints 
are often interpreted as the resource prices \cite{boyd}. Thus, without budget consideration,  popular techniques such as network utility maximization (NUM) \cite{NUM}
can be employed to compute an ME. However, these techniques do not work for the Fisher market. 
The main challenge stems from the budget constraints and the market clearing condition that couple the allocation decision (i.e., primal variables) and the prices (i.e.,  dual variables). 
Fortunately, for a wide class of utility functions, the ME in a Fisher market can be found by solving an Eisenberg-Gale (EG) program \cite{EG,AGT}.

Unlike the previous works, we focus on a practical setting where the services have utility limits due to their limited demands  (e.g., the request rate of a service in a given area is not unlimited). This violates the implicit assumption in the GE literature that requires the utility function is non-satiated (i.e., for any bundle of goods, there is always another bundle of goods 
that is strictly improve the utility \cite{eco}). Also, because of the limited demand constraint, the services' utility functions do not satisfy the conditions of the EG program. Our work aims to understand the ME concept in this setting. 

While an ME solution  is highly desirable, multiple  equilibria may exist. Indeed, when we consider the  utility limit, the allocation at some equilibrium can be wasteful and non-Pareto efficient as discussed in \textit{\ref{finite}}. Hence, we impose additional criteria to select an ME with some good properties. Specifically, an allocation should be non-wasteful and frugal. An allocation is \textit{non-wasteful} if no resources desired by any service are left unallocated and no service receives resources that it has no use for \cite{ppou17,agut12}. 
The \textit{frugality} property implies that services want to maximize their utilities while spending less money. Note that the \textit{non-wastefulness} and \textit{frugality} concepts do not exist in the traditional Fisher market. The proposed framework allows us to compute a non-wasteful and frugal ME, which implies the allocation is efficient. 

We further connect the equilibrium to the fair division literature \cite{hmou04}.  
Specifically, the proposed ME is shown to have appealing fairness properties including envy-freeness, sharing-incentive, and proportionality \cite{ppou17,hmou04}. Indeed, these properties were seldom studied  in the ME literature. 
Note that, in general, not every ME satisfies these  properties.
In this work, we do not pursue strategy-proofness and assume that the services are price-takers. This assumption  is reasonable in an environment with a large number of agents. 
Furthermore, we develop a  privacy-preserving distributed algorithm for equilibrium computation, 
which prevents adversaries from learning private information (even statistical information) of the services, hence,  alleviates the necessity of the strategy-proofness requirement. 
To the best of our knowledge, this is one of the first parallel and distributed  algorithms implemented in a privacy-preserving manner.
This algorithm is quite general and may find applications in
other scenarios as well.

Although this work is closely related to our previous work \cite{duong}, there are several fundamental differences. First,  \cite{duong} considers the Fisher market with linear utilities and single resource type. Here, we study more complex utility functions and multiple resource types. Second, \cite{duong} aims to address the non-uniqueness of optimal bundles in designing distributed algorithms and equilibrium computation in the net profit maximization case. On the other hands, the purpose of this work is to investigate the ME concept in a market with utility limits. Additionally,  in \cite{duong}, the  utility of every  service is unique across all the market equilibria. However, it is not the case with our model. This work also introduces the new notions of non-wastefulness and frugality for equilibrium selection. Furthermore, we devise a privacy-preserving distributed algorithm for computing an ME. 
Our main contributions include:

\begin{itemize}

\item \textit{Modeling}. We formulate a new multi-resource allocation (MRA) problem in FC with budget-constrained agents. The problem is cast as a Fisher market and we advocate the GE. We are among the first explicitly tackle the MRA problem in the general setting with multiple nodes, each has multiple resource types, and multiple agents with finite demands \cite{ppou17}.

\item \textit{Centralized solution}. We theoretically extend the GE literature to consider the practical case where services have utility limits.
We  show that all non-wasteful and frugal market equilibria can be captured by a convex program.
Furthermore, the proposed equilibrium is proved to have remarkable fairness properties. 
 
\item \textit{Decentralized algorithms}. 
By blending the alternating direction method of multipliers (ADMM) method with auxiliary variables and an efficient privacy-preserving averaging technique, we derive a  parallel and distributed  algorithm that converges to a non-wasteful and frugal ME while  preserving the services' privacy.

\item \textit{Performance Evaluation.} Extensive simulations are conducted to confirm our theoretical results and demonstrate the efficacy of the proposed techniques. 

\end{itemize}

The rest of the paper is organized as follows. The system model and problem formulation are given in Section \ref{sysfor}. Section \ref{sol} presents  centralized solutions for computing an ME and analyzes the properties of the ME. The privacy-preserving distributed algorithm for computing the ME is introduced in Section \ref{PPDA}.
Finally, simulation results are shown in Section \ref{sim} followed by the discussion of related work in Section \ref{related}, and conclusions and future work in Section \ref{concl}.

\section{System Model and Problem Formulation}
\label{sysfor}
\subsection{System Model}
\label{sys}
\vspace{-0.2cm}

\begin{figure}[ht!]
	\centering
		\includegraphics[width=0.30\textwidth,height=0.21\textheight]{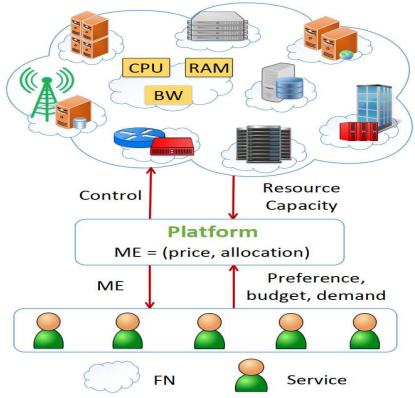}
	\caption{System model}
	\label{fig:model}
\end{figure} \vspace{-0.1cm}

With FC, besides local execution and remote processing at cloud DCs, requests and data from end-devices (e.g., smartphones, PCs, sensors) can be handled by a fog network.  Typically, a request first goes to a Point of Aggregation (e.g., switches/routers, base stations, access points), then it will be routed to an FN for processing. 
In practice, enterprises, factories, organizations (e.g., malls, schools,  hospitals),
 and other third parties (e.g., sensor network owners) can  outsource their services and computation to a fog network.  Also, content/application/service providers (e.g., Google, Netflix, Uber, AR companies) can proactively place their data and applications onto FNs to serve better their customers.

In this work, we 
 consider a system that consists of  multiple services and a set of resource-constrained heterogeneous FNs. The system size may correspond to a metropolitan area network.  Each service has a certain budget for resource procurement and wants to offload as many requests as possible to the fog network. The services are assumed to be price-takers, which is reasonable for a large market with multiple services.
Each FN provides multiple resource types such as CPU, memory,  and bandwidth (BW). 
 We assume that there is a platform lying
between the services and the FNs (i.e., the \textit{fog resource pool}). Based on the information
collected from the FNs (e.g.,  capacities) and the
services (e.g., budget, preference, demand), the platform computes
an ME solution that assigns a price to every fog resource and allocates an optimal resource bundle to every service. 
Each service can obtain information
about the prices and its allocated resources from the
platform. Also, the platform may control and manage the
resource pool to implement the allocation decision.
Fig.~\ref{fig:model} presents the system model.
 
 We  assume that the requests (e.g., Apple Siri, Google Voice Search,  weather/maps queries, AR, IoT data collection) are independent 
 and each request requires resources in a specific ratio  \cite{agho11,cjoe13,wwan15,ppou17,dpar15}. 
The amount of resources at an FN needed to handle a request of a service is called the \textit{base demand} of that service at the FN.
For example, if a request of service A needs 0.2 units of CPU and 0.1 units of RAM at FN B, its base demand vector at FN B is (CPU: 0.2, RAM: 0.1). 
The number of requests of a service can be served is proportional to the amount of resources allocated to it. For instance, if service A receives 0.4 units of CPU and 0.3 units of RAM from  FN B, it can process two requests (i.e., min$\{\frac{0.4}{0.2}, \frac{0.3}{0.1}\}$). 
Our model allows a service to specify different base demand vectors at different FNs 
to express its diverse preferences towards the fog resources. 

\subsection{Service Utility Model}
We denote the sets of  FNs and  services by  $\mathcal{M}$  and $\mathcal{N}$, respectively. The corresponding numbers of FNs and services are $M$ and $N$.
Let $\mathcal{R} = \big\{1, 2, \ldots, R \big\}$ be the set of $R$ different types  of resources. 
Define $i$, $j$, and $r$ as the service index, FN index, and resource type. 
The capacity of resource type $r$ at FN $j$ is $C_{j,r}$. 
 The \textbf{base demand} vector of service $i$ for resources of FN $j$ is $D_i^j = \big(a_{i,j,1}, \ldots, a_{i,j,R} \big)$ where $a_{i,j,r}$ is the amount of resource type $r$ at FN $j$ required to process one request of service $i$.
 For simplicity in presentation, we assume that  $a_{i,j,r} > 0, ~\forall i,j,r$. Note that all  results in this paper still hold if $a_{i,j,r} = 0, ~\text{for some}~ i,j,r$.
Due to different hardware specifications and locations of the FNs, the base demand vectors of a service at different FNs can be different. For example, a service may set  higher values for bandwidth and computing resources in its base demand vector at an FN far away from it to ensure low latency.

 Define $x_{i,j,r}$ as the amount of resource type $r$ at FN $j$ allocated to service $i$ and $x_{i,j} = \big(x_{i,j,1}, \ldots, x_{i,j,R} \big)$ as the vector of resources allocated to service $i$ from FN $j$. The \textbf{resource bundle} that service $i$ receives from the FNs is $x_i = \big(x_{i,1}, x_{i,2}, \ldots, x_{i,M} \big) \in \mathbb{R}^{M\text{x}R}$. 
Each bundle $x_i$ is a matrix where each row represents resources of an FN and each column represents a resource type.
As explained in the system model, given $x_{i,j}$, the number of requests $y_{i,j}$ of  service $i$ that can be processed by FN $j$ is \cite{agho11,cjoe13,wwan15,dpar15}

\beqn
y_{i,j}(x_{i,j}) = \min \Big\{\frac{x_{i,j,1}}{a_{i,j,1}},
\ldots,\frac{x_{i,j,R}}{a_{i,j,R}}\Big\} = \min_r \frac{x_{i,j,r}}{a_{i,j,r}}, ~ \forall i,~j. 
\eeqn 
Thus, given resource bundle $x_i$,
the total number of requests $y_i$ that the fog network can process for service $i$ is 
\beqn
y_i(x_i) = \sum_j y_{i,j}(x_{i,j})  =  \sum_j \min_r \frac{x_{i,j,r}}{a_{i,j,r}}, \quad \forall i.
\eeqn 

In practice, a service may not have infinite demand (i.e., number of requests over a certain period).
 Let $D_i$ be the maximum demand of service $i$. 
Then, the actual  number of requests $y_i(x_i)$ of service $i$ processed by the FNs is
\beqn
y_i(x_i)
= \min \Big\{ \sum_j \min_r \frac{x_{i,j,r}}{a_{i,j,r}}, ~D_i \Big\},~ \forall i.
\eeqn \vspace{-0.1in}

The utility function  of a service
 maps each resource bundle to a number quantifying 
the service's valuation for that bundle. Define the utility function of service $i$ as $u_i(x_i): \mathbb{R}^{M\text{x}R} \rightarrow \mathbb{R}$.
Let $e_i$ be the gain of service $i$ for each request processed by the fog network.  We have:
\beqn
\label{cap_u1}
u_i(x_i) = e_i~y_i(x_i) = \min \Big\{ e_i  \sum_j \min_r \frac{x_{i,j,r}}{a_{i,j,r}}, ~e_i  D_i \Big\}, ~\forall i.  
\eeqn 
 Without loss of generality, we assume $e_i$ = 1, $\forall i.$ As can be seen later, the proposed allocation scheme is scale-free and independent of $e_i$.
Then, we have:
\beqn
\label{cap_u}
u_i(x_i) &=&  \min \Big\{   \sum_j \min_r \frac{x_{i,j,r}}{a_{i,j,r}},   D_i \Big\},   \nonumber \\ 
&=& \min \Big\{ \sum_j u_{i,j}(x_{i,j}), ~u_i^{\sf max} \Big\},~\forall i,
\eeqn 
where $u_i^{\sf max} =   D_i$ and  $u_{i,j}(x_{i,j}) =  \min_r \frac{x_{i,j,r}}{a_{i,j,r}}$. We call $u_i^{\sf max}$ the \textit{utility limit} of service $i$.

Note that a service $i$ may want to be served by only a  subset of FNs $\mathcal{A}_i \in \mathcal{M}$ that satisfy certain criteria (e.g., in terms of hardware specifications, reliability, and communication distance).  For instance, a latency-sensitive service may be interested in only FNs that are geographically close to it. In this case, we have $u_i(x_i) = \min \Big\{ \sum_{j \in \mathcal{A}_i} u_{i,j}(x_{i,j}), ~u_i^{\sf max} \Big\},~\forall i$ and $x_{i,j,r} = 0,~\forall  r,~j \in \mathcal{M}/\mathcal{A}_i$.  For simplicity of notations, we do not consider this constraint and assume $\mathcal{A}_i = \mathcal{M}, ~\forall i$. Indeed, the subsequent analysis and solution approaches remain unchanged if we consider this constraint.

\subsection{Fog Computing Resource Allocation Problem}
\label{problem}
We cast the problem  as a Fisher market  where  the services act as buyers\footnote{The words "service" and "buyer" are used  interchangeably in this paper.} and the fog resources act as divisible goods.  The market aims to  price and allocate  the resources to the services effectively. Denote by $p_j = \big(p_{j,1}, p_{j,2}, \ldots, p_{j,R}\big)$  the price vector of resources of FN $j$ where  $p_{j,r}$ is the  price of of one unit of resource type $r$ at FN $j$. 
Let $p = \big(p_1, p_2, \ldots, p_M\big) \in \mathbb{R}^{M\text{x}R}$ capture the resource prices of all the FNs. Define $B_i$ as the budget of service $i$. 
The optimal (i.e., utility-maximizing) resource bundle of service $i$ at prices $p$ can be found by solving the following \textbf{service utility maximization} problem.
\beqn
\label{uobj1}
\underset{x_i}{\text{maximize}}  \quad  && u_i(x_i) \\
\label{ueq1}
\text{subject to } && \sum_j \sum_r x_{i,j,r} p_{j,r} \leq B_i  \\
\label{ueq2}
&& x_{i,j,r} \geq 0, \quad \forall j,~r,
\eeqn
where (\ref{ueq1}) captures the budget limit of service $i$. 
Note that each service computes its optimal resource bundle based simply on the resource prices and it does not care about the resource capacity constraints of the FNs. From the definition of the service utility function given in (\ref{cap_u}), we can rewrite the optimization problem (\ref{uobj1})-(\ref{ueq2}) of each service $i$ as follows. 
\beqn
\label{opt1}
\underset{x_i,~u_i}{\text{maximize}}  &&  u_i \\
\label{opt11}
\text{subject to } && u_i = \min \Big\{ \sum_j u_{i,j},~ u_i^{\sf max} \Big\} \\
\label{opt12}
&& u_{i,j} = \min_r \frac{x_{i,j,r}}{a_{i,j,r}}, \quad \forall j   \\
\label{opt13}
&& \sum_j \sum_r x_{i,j,r} p_{j,r} \leq B_i  \\
\label{opt14}
&& x_{i,j,r} \geq 0, \quad \forall j,~r.
\eeqn

\textbf{Problem Statement:} The proposed scheme aims to  find an ME solution ($p^*$, $X^*$) consisting of equilibrium prices $p^* = \big(p_1^*, p_2^*, \ldots, p_M^* \big)$ and equilibrium allocation $X^* = \big( x_1^*, x_2^*, \ldots, x_N^* \big)$. 
At the equilibrium, every service receives its optimal resource bundle. 
In other words, $x_i^*$ is a utility-maximizing resource bundle of service $i$ at the equilibrium prices $p^*$.
Additionally, we would like to achieve high utilization of the available fog resources.
The definition of an ME is formally given as follows \cite{AGT}.

\begin{definition}
\label{MEdef} 
($p^*$,$X^*$) is an ME if the two following conditions hold \cite{AGT}.
\end{definition}

\begin{itemize}

\item \textit{Service Satisfaction Condition}: 
Given a non-negative  price vector $p^*$, every service receives its favorite resource bundle $x_i^*$ that maximizes its utility under the  budget constraint. In other words, $x_i^*$ is an optimal solution to the problem (\ref{uobj1})-(\ref{ueq2}) 
with $p = p^*$, i.e., we have:
\beqn
\forall i: ~x_i^* \in \argmax_{x_i \geq 0;~ \sum_j \sum_r p_{j,r}^* x_{i,j,r} \leq B_i} u_i(x_i). 
\eeqn 

\item \textit{Market-Clearing Condition}:  Every fog resource is either fully allocated or has zero price,  i.e., we have:  $(\sum_{i} x_{i,j,r}^* - C_{j,r} )~p_{j,r}^* = 0,  ~\forall j,~r$. 
\end{itemize}

These two conditions ensure that the equilibrium allocation maximizes the happiness of every service while maintaining  high resource utilization.
The second condition presents \textit{Walras' law} \cite{AGT,eco} which states that all fog resources with positive prices ($p_{j,r}^* > 0$) are fully allocated ($\sum_{i} x_{i,j,r}^* = C_{j,r}$) and the remaining resources (i.e., not fully allocated) have zero prices.
Indeed, due to zero prices, these resources can be allocated arbitrarily to the services without violating the budget constraints. Also, this additional allocation does not strictly improve the utility of any service since $x_i^*$ is already a utility-maximizing resource bundle of service $i$ among all   its affordable bundles (i.e.,  following the service satisfaction condition). Hence, we can understand that the market clears.

 In our problem, the services are players competing for the limited fog resources, while the
platform tries to satisfy the market clearing condition. Prices are used to coordinate the market. 
Without loss of generality, we normalize the resource capacities of the FNs, i.e., we have: 
\beqn
C_{j,r} = 1, \quad \forall j,~r. \vspace{-0.1in}
\eeqn
The  prices and the base demand vectors of the services  can be scaled accordingly. 
 This normalization is just  to follow the standard market equilibrium literature \cite{AGT,ndev08}, and to simplify expressions and equations in the paper.

It is  worth emphasizing that  (\ref{uobj1})-(\ref{ueq2}) or equivalently (\ref{opt1})--(\ref{opt14}) is the optimization problem of individual services while the market/platform (not the services) needs to ensure the global constraints including the market-clearing condition and the resource capacity constraints as follows:
   \beqn
	\label{g1eq}
	&& \Big(\sum_{i} x_{i,j,r}^* - 1 \Big)~p_{j,r}^* = 0,  ~\forall j,~r \\ 
	\label{g2eq}
	&& \sum_{i} x_{i,j,r}^* \leq 1, \quad \forall j,~r.	 \vspace{-0.1in}
	\eeqn	
Our design goal is to find an ME ($p^*$, $X^*$) that simultaneously maximizes the utility of every service (i.e., $x_i^*$ is an optimal solution of (\ref{uobj1})-(\ref{ueq2}) with $p = p^*$,~$\forall i$) while satisfying the global constraints (\ref{g1eq})-(\ref{g2eq}).

\section{Centralized Solution Approach}
\label{sol}

This section presents centralized solutions
for computing market equilibria of the proposed 
problem. First, we briefly introduce the Fisher market  and the Eisenberg--Gale (EG) program for computing ME when the services' utility functions are \textit{concave} and \textit{homogeneous of degree one}\footnote{A function $f(x)$ is homogeneous of degree $d$ if $f(\alpha x) = \alpha^d f(x), ~ \text{for all} ~\alpha > 0$.} \cite{AGT,eco}. However, due to the finite demands of the services, their utilities are not homogeneous. 
We show that the optimal solution of the EG program for services with infinite demands is still an ME of our model for services with finite demands. However, this equilibrium 
allocation can be wasteful.  Then, we rigorously demonstrate that  \textit{all the non-wasteful and frugal market equilibria of our model can be captured by a convex program} which is a natural generalization of the EG program. Furthermore, we show compelling fairness features of the  ME solution obtained from this convex program. 

\subsection{The Eisenberg-Gale Program}

Consider a Fisher market with $N$ buyers and $K$ divisible goods with unit capacities. 
Let $i$ and $k$ be the buyer index and goods index. Buyer $i$ is characterized by budget $B_i$ and a utility function $U_i(x_i)$ where $x_i = (x_{i,1},\ldots,x_{i,K})$ is the vector of resources allocated to buyer $i$. If the buyers' utility functions are all concave and homogeneous of degree one, the optimal solution of the following EG program is an exact ME of this Fisher market \cite{EG,AGT}.
\beqn
\label{EGprogram1}
\vspace{-0.1in} 
\underset{X}{\text{maximize}} &&  \sum_i B_i \ln~ U_i(x_i) \\
\label{EQ12}
\text{subject to} && \sum_i x_{i,k} \leq  1, \quad \forall k \\
\label{EQ13}
&& x_{i,k} \geq 0, \quad \forall i,~k.
\eeqn
Specifically, let  ($p^*,X^*$) be an optimal solution of the EG program (\ref{EGprogram1})-(\ref{EQ13}) where $p^*$ is the dual variables associated with constraint (\ref{EQ12}). Then, for every buyer $i$, $x_i^*$ maximizes $U_i(x_i)$ under the budget constraint $\sum_k x_{i,k}~ p_k^* \leq B_i$ and $x_{i,k} \geq 0,~\forall k$. Furthermore, the market-clearing condition is satisfied, i.e., $\Big(\sum_i x_{i,k}^* - 1 \Big)~ p_{k}^* = 0,~\forall k.$

Many well-known functions, such as linear, Cobb-Douglas, and Leontief,
satisfy both concavity and homogeneity of degree one \cite{AGT}.
A function $u(x)$ is linear if $u(x) = \sum_k a_k x_k.$ A Cobb-Douglas function has the form $u(x) = \prod_k x_k^{a_k}$ where $\sum_k a_k = 1$. Finally, a Leontief function has the form $u(x) = \min_k (x_k/a_k)$.
Another popular class of homogeneous utility functions 
is the Constant Elasticity of Substitution (CES) function \cite{AGT,eco}.

\subsection{Centralized Solution}

We now return to our resource allocation problem. 
It can be observed that
the service utility function in  (\ref{cap_u}) is a combination of linear and Leontief functions (i.e., sum of Leontief functions $u_{i,j}(x_{i,j})$). 
Additionally, the utility limit makes this function more complex. 

\subsubsection{Services with Infinite Demands}
First, we consider the case in which services have unlimited demands. 
Then, from (\ref{cap_u}), the utility of service $i$ becomes
\beqn
\label{nocap_u}
u_i(x_i) = u_i^{\sf inf}(x_i) = \sum_j u_{i,j}(x_{i,j}) =   \sum_j \min_r \frac{x_{i,j,r}}{a_{i,j,r}}, \quad \forall i. \\ \nonumber
\eeqn 

\begin{proposition} 
\textit{If  a service has infinite demand, its utility function  is concave and homogeneous of degree one.
}\end{proposition}

\textit{Proof.} Please refer to \textit{Appendix \ref{prop1}}. 
\qedb 

Thus, we can directly apply  the EG program (\ref{EGprogram1})-(\ref{EQ13}), where goods $k$ corresponds to  resource type $r$ of FN $j$ and each service is a buyer, and have the following
 corollary.
\begin{corollary}
\label{coro1}
\textit{If all the services have infinite demands, 
the optimal solution of the following problem} 
\beqn
\label{EGinf}
\vspace{-0.1in} 
\underset{X}{\text{maximize}} &&  \sum_i B_i \ln \Big( \sum_j \min_r \frac{x_{i,j,r}}{a_{i,j,r}} \Big)\\
\label{EGinf1}
\text{subject to} && \sum_i x_{i,j,r} \leq  1, \quad \forall j,~r \\
\label{EGinf2}
&& x_{i,j,r} \geq 0, \quad \forall i,~j,~r.
\eeqn
\textit{is an exact ME of our fog resource allocation problem.}
\end{corollary}

\subsubsection{Services with Finite Demands}
\label{finite}
When the services have limited demands, which is often the case in practice, it is easy to see that the  utility function 
  in (\ref{cap_u}) is not homogeneous. 
For example, when a service reaches its utility limit, its utility remains unchanged when we double the amount of resources allocated to it. 
We show that  the optimal solution to 
(\ref{EGinf})-(\ref{EGinf2}) is still an ME of  the proposed resource allocation model where services have finite demands (i.e., utility limits). However, this solution may lead to a wasteful allocation because some services may receive too many resources at the equilibrium without improving their utilities.
An allocation is \textbf{non-wasteful} if: i) the remaining resources after the allocation cannot improve the utility of any service;  ii) no service receives more resources than its need \cite{ppou17,agut12}.

\begin{proposition}
\label{theo1}
\textit{The optimal solution to the problem  (\ref{EGinf})-(\ref{EGinf2})  (i.e., without considering utility limit) is an ME of the proposed  resource allocation problem for services with finite demands. 
However, this solution may lead to a wasteful and non-Pareto optimal allocation.} 
\end{proposition}

\textit{Proof.} 
Let ($p^*,X^*$) be an optimal solution to (\ref{EGinf})-(\ref{EGinf2})   and
 $p_{j,r}^*$ is the dual variable associated with the capacity constraint (\ref{EGinf1}). From 
 Corollary \ref{coro1}, ($p^*,X^*$) is an ME of our model for services with \textit{infinite} demands. We will show ($p^*,X^*$) is also an ME of our model for services with \textit{finite} demands.

According to Definition \ref{MEdef}, at  prices $p^*$, $x_i^*$ maximizes $ \sum_j \min_r \frac{x_{i,j,r}}{a_{i,j,r}}$, $\forall i$, under the service's budget constraint.
	For services with utility limits (i.e., finite demands), since   $x_i^*$ maximizes $\sum_j \min_r \frac{x_{i,j,r}}{a_{i,j,r}}$ at prices $p^*$, $x_i^*$ also maximizes $\min \big\{\sum_j \min_r \frac{x_{i,j,r}}{a_{i,j,r}}, u_i^{max} \big\}$ at $p^*$ while respecting the budget constraint. Additionally, ($p^*,X^*$) still satisfies the market clearing condition.	
	Therefore, ($p^*,X^*$) is an exact ME of the proposed problem for services with finite demands.

However, $X^*$ can be a wasteful allocation since some services with limited utilities may get redundant resources at the equilibrium (i.e., $\sum_j \min_r \frac{x_{i,j,r}^*}{a_{i,j,r}} > u_i^{max}$). These resources can be reallocated to other services that have not reached utility limit to improve their utilities. Thus, 
 $X^*$ may not be Pareto-optimal since some service can get strictly better utility without affecting the utility of other services. 
This issue is illustrated in 
the simulation (e.g., see Fig.~\ref{fig:EGGEG}).
\qedb

In addition to the result presented in Fig. 2 in the
simulation, interested readers can refer to Appendix F
and Appendix G for some toy examples of the wasteful
equilibrium allocation phenomenon.
Since multiple market equilibria can exist, we need to impose some criteria to select a good ME. 
First, we are interested in finding a non-wasteful ME, which implies the equilibrium allocation is efficient. Assume $X^*$ is a non-wasteful allocation and $(p^*,X^*)$ is a non-wasteful ME. 
The market clearing condition of an ME ensures that no resources desired by any service are left unallocated. Specifically, resources that are not fully allocated have zero prices. As explained in the paragraph following the \textit{Definition 2.1}, these resources can be given to any service without increasing the utility of the service. 
Thus, an ME always satisfies the first requirement of a non-wasteful allocation (defined before \textit{Proposition 3.3}). 
To satisfy the second requirement that no service receives resources that it has no use for, i.e., we need: 
\beqn
\label{NWcon}
 \sum_j \min_r \frac{x_{i,j,r}^*}{a_{i,j,r}} \leq u_i^{max},~\forall i.
\eeqn 
Additionally, resources allocated to a service from an FN must be proportional to the service's base demand vector (i.e., $x^*_{i,j,r} = \beta ~a_{i,j,r}$ for all $i,~j,~r$ and some constant $\beta$).

The second criterion is \textbf{frugality}. It is natural to assume that the services want to maximize their utilities while spending as least money as possible. 
Hence, a service would prefer an FN with the cheapest price for its base demand vector.
 Let $p$ be the resource prices. The price for the base demand vector $D_i^j$ of service $i$ at FN $j$ is 
\beqn
\label{bprice}
q_{i,j} = \sum_r p_{j,r} a_{i,j,r}, \quad \forall i,j. 
\eeqn
Define $q_i^{\sf min} = \min_j q_{i,j},~\forall i$. The frugality property states that service $i$ will buy resources only from the set of FNs $j$ such that $q_{i,j} = q_i^{\sf min}$. We also define $\alpha_i = \frac{1}{q_i^{\sf min}}$ as maximum bang-per-buck (MBB) of service $i$, which generalizes the notion of MBB for linear utility in \cite{ndev08}. 

Before presenting our main result, it is necessary to  understand the buying strategy (i.e., action) of the services at  a non-wasteful and frugal ME ($p^*, X^*)$. Consider service $i$. To maximize its utility as well as to satisfy the frugality condition,  service $i$ will buy resources from the cheapest FNs only (i.e., the FNs offer it MBB). In other words, service $i$ buys resources only from the set of FNs $j'$ such that $q_{i,j'} = q_i^{\sf min} = \min_j \sum_r p_{j,r}^* a_{i,j,r}$. Furthermore, service $i$ either spends full budget to buy resources from the cheapest FNs or spends money until reaching its utility limit. Intuitively, this buying strategy maximizes the utility of the service and satisfies the frugality condition.

Our main finding is that the following optimization problem,  which is a natural extension of the EG program, allows us to find a non-wasteful and frugal ME.
\beqn
\label{EGmain}
\underset{X,u}{\text{maximize}} &&\sum_i B_i \ln \sum_j u_{i,j} \\
\label{EGmain1}
\text{subject to} &&u_{i,j} a_{i,j,r} = x_{i,j,r}, \quad \forall i,~j,~r \\
\label{EGmain2}
&&\sum_j u_{i,j} \leq u_i^{\sf max}, \quad \forall i \\
\label{EGmain3}
&&\sum_i x_{i,j,r} \leq 1, \quad \forall j,~r \\
\label{EGmain4}
&&x_{i,j,r} \geq 0, \quad \forall i,~j,~r.
\eeqn

While the idea of adding utility limit constraints 
is natural, the following theorem and its proof are not trivial since the existing results on ME in the Fisher market (and  the GE theory in general)  rely heavily on the \textit{critical assumptions} of non-satiated utility functions  \cite{EG,AGT}, which is not the case in our model.
Furthermore,  the \textit{non-wastefulness} and \textit{frugality} notions do not exist in the traditional Fisher model.  
\textbf{Theorem \ref{main}} includes two main parts. The first part states that the optimal solutions to the problem (\ref{EGmain})-(\ref{EGmain4}) are non-wasteful and frugal market equilibria of our problem. \textit{More importantly}, the second part states that the set of all  the non-wasteful and frugal market equilibria are exactly the set of optimal solutions to the problem (\ref{EGmain})-(\ref{EGmain4}).
 
Based on the analysis of the optimal buying strategy of the services, to prove the first part, we need to show: 
i) the optimal solutions to (\ref{EGmain})-(\ref{EGmain4}) satisfy the market-clearing condition in \textit{Definition \ref{MEdef}};~ii) the services buy resources from the cheapest FNs only (i.e., FNs offer them MBB);~iii) every service either reaches its utility limit or spends full budget. Note that constraints (\ref{EGmain1})-(\ref{EGmain2})  
imply that any optimal solution $X^*$ to this problem is a non-wasteful allocation.
To prove the second part of the theorem, we need to show that from the conditions of an ME, non-wastefulness, and frugality, we can reconstruct the problem   (\ref{EGmain})-(\ref{EGmain4}).

\begin{theorem}
\label{main}
\textit{The optimal solutions to the problem (\ref{EGmain})-(\ref{EGmain4}) are exactly non-wasteful and frugal market equilibria. 
At the equilibra, every service either spends all budget or attains its utility limit. The ME always exists and the utilities are unique across all such equilibria. Finally, all non-wasteful and frugal market equilibria are captured by the problem (\ref{EGmain})-(\ref{EGmain4}).}
\end{theorem}

\textit{Proof.} 
Please refer to Appendix \ref{theorem}.

To understand the importance and impact of non-wastefulness and frugality as well as utility limit, we have several remarks as follows.

\textbf{Remark 1}: we again emphasize that from Proposition 3.3, it seems to be natural to add the utility limit constraint to achieve non-wasteful ME. Hence, one may think that ``every non-wasteful ME is an optimal solution to the problem (\ref{EGmain})-(\ref{EGmain4}).'' However, this statement may not hold. A non-wasteful ME may not necessarily be a solution to (\ref{EGmain})-(\ref{EGmain4}). In other words, this convex program may not necessarily capture the set of all non-wasteful market equilibria. Indeed, you can observe that the proof of Theorem 3.4 relies heavily on the frugality property.

\textbf{Remark 2}: For a system with one FN  only (i.e., buyers with Leontief utility functions), we may not need the frugality criteria since a buyer has only one node to buy resources (i.e., goods) from. The non-wasteful properties may be sufficient since it enforces every buyer to either fully exhaust his budget or reach his utility limit. However, when there are multiple FNs (i.e., the hybrid linear-Leontief utility function in our model), at the equilibrium prices (or any price vector), a buyer who can reach his utility limit may have multiple options to buy resources. Note that in the Fisher market, money has no intrinsic value to the buyers. Hence, he may not need to buy resources from the cheapest FNs to obtain his utility limit. Therefore, frugality is important for our result. Note that if a buyer does not fully spend his budget, the surplus budget can be accumulated/redistributed into later time slots to give more priority to him in the later slots. The detailed policy and implementation are flexible and not the focus of this work. Our work focuses on one time slot only when the system parameters (e.g., budget, limit, preferences, capacities) are given.

\textbf{Remark 3}: Without utility limit, obviously, every buyer will spend full budget to buy more resources to improve his utility. Consequently, every ME is non-wasteful. However, it is not true when we consider utility limit. As shown in the theorem above, some buyers do not fully exhaust their budgets at the equilibrium.

\textbf{Remark 4}: As shown in Proposition 3.1., without utility limit, every ME is an optimal solution to the EG program \cite{AGT}. Furthermore, the utilities of each buyer at such equilibria are unique \cite{AGT}. However, when we consider utility limit, there can exist multiple market equilibria. The utilities of each buyer at the equilibria may not be unique (e.g., wasteful ME in Proposition 3.3 and non-wasteful ME in Theorem 3.4). It is also worth noting that there can be other market equilibria which are not the optimal solutions to either the problem (\ref{EGinf})-(\ref{EGinf2})  or the problem (\ref{EGmain})-(\ref{EGmain4}).

Based on the proof above, we can further show that the market equilibria obtained from (\ref{EGmain})-(\ref{EGmain4}) are Pareto-optimal. 
An allocation is \textit{Pareto-optimal} if there is no other allocation that would make some service better off without making some other service worse off \cite{eco}, which means there is no strictly ``better'' allocation. 
It is worth noting that the famous \textit{first fundamental theorem of welfare economics}  \cite{karr54,eco} states that under some mild conditions, an ME is always Pareto-optimal. However, its proof relies on the crucial assumption of locally non-satiated utility functions  (e.g., see Proposition 16.C.1
in \cite{eco}). Without considering the utility limit, the service
utilities become locally non-satiated, hence, every ME is
Pareto-efficient. However, due to the utility limit, the utility
functions in our problem are locally satiated. Hence, not
every equilibrium allocation in our model is Pareto-optimal
(e.g., the equilibria mentioned in Proposition 3.3. Therefore, our result on Pareto-optimality is interesting since the  utility functions in our problem are locally satiated due to the utility limit.

\begin{lemma}
\textit{The market equilibria obtained from the problem (\ref{EGmain})-(\ref{EGmain4}) are Pareto-optimal. 
}\end{lemma}
\textit{Proof.} We show this by contradiction. Let $(p^*, X^*)$ be an optimal solution to (\ref{EGmain})-(\ref{EGmain4}). Assume $X'$ is a strictly better allocation than $X^*$. Hence, $u_i(x'_i) \geq u_i(x_i^*), ~\forall i$ and strict inequality holds for some $i$.  Let $h$ be the index of the service that has strictly better utility at $X'$. Obviously, if service $h$ reaches its utility limit at $X^*$, it cannot strictly increase its utility. Hence, at $X^*$, service $h$ has not reached its utility limit. Thus, from the results of \textit{Theorem} \ref{main}, service $h$ spends full budget at $X^*$. Furthermore, we proved that every service buys resources  only from cheapest FNs. Hence, at prices $p^*$, service $h$ cannot improve its utility and $X^*$ is Pareto-optimal.
 \qedb

In summary,  a non-wasteful and frugal ME $(p^*, X^*)$  is an optimal solution to 
(\ref{EGmain})-(\ref{EGmain4}). Since this is a convex optimization problem, efficient techniques such as the interior-point method can be used to solve it effectively \cite{boyd,conv1}. The computational aspect of this convex program is further
discussed in Appendix E.

\subsection{Fairness Properties of the ME}

In this section, we show that the proposed ME, which is an optimal solution to (\ref{EGmain})-(\ref{EGmain4}),  possesses  appealing fairness properties that encourage the services to participate in the proposed  scheme. An allocation is \textit{envy-free} if no service envies with the allocation of any other service (i.e., if every service weakly prefers its allocation to the allocation of any other service). In other words, every service is happy with its allocation and does not want to swap its allocation with that of another service. When the budgets are equal, an envy-free allocation $X$ implies $u_i(x_i) \geq u_i(x_{i'})$, $\forall i,~ i' \in \mathcal{N}$ \cite{hmou04}. It is well-known that every competitive equilibrium
from equal income (CEEI) is envy-free \cite{hvar74}. For the
Fisher market, equal income means the buyers have equal
budgets. If the budgets are not equal, it is easy to see that a
market equilibrium can be not envy-free (e.g., two agents
with the same utility function and one good, the agent
with lower budget may envy with the allocation of the
agent with higher budget). It is because different budgets
results in different sets of affordable resource bundles of
different agents at the equilibrium prices.
 Since the services may have different budgets, we need to extend 
 the classical definition of envy-freeness. 
An allocation $X$ is envy-free if $ u_i(x_i) \geq u_i ( \frac{B_{i}}{B_i'} x_{i'}) $,
$\forall i,~ i' \in \mathcal{N}$. 

\textit{Sharing-incentive} is another popular fairness criterion. An allocation satisfies this property if it gives every service a better utility than its utility in the \textit{proportional sharing} scheme. Proportional sharing is an intuitive way to share resources fairly (in terms of resource quantity), which allocates every resource to the services proportional to their budgets. 
Let $\hat{x}$ be the allocation in which every service receives resources from the FNs proportional to its budget, i.e., $\hat{x}_{i,j,r} = \frac{B_i}{\sum_i' B_{i'}}$, $\forall i,~ j$. Indeed, $\hat{x}$ can be 
interpreted as a resource-fair allocation. 
The \textit{sharing-incentive} property implies $u_i(x_i) \geq u_i(\hat{x}_i),~\forall i.$ 
Additionally, in the fog federation setting where each service $i$ contributes a portion $\hat{x}_{i,j,r}$
of every resource to FN $j$ in a resource pool that consists of all the FNs,
sharing-incentive ensures that every service prefers the equilibrium allocation to its initial resources contributed to the pool. This property can be interpreted as resource-fairness.

Finally, \textit{proportionality} is a well-known fairness notion in Economics \cite{hmou04}. 
 Let $C \in \mathcal{R}^{MxR}$  be the set of all the fog resources. Obviously, $u_i(C)$ is the maximum utility that service $i$ can achieve by being allocated all the available resources (i.e., $C_{j,r} = 1,~\forall j,r$). It is natural for service $i$ to expect to obtain a utility of at least $\frac{B_i}{\sum_{i'} B_{i'}} u_i (C)$. 
The \textit{proportionality} property guarantees that the utility of  every service at the equilibrium is at least proportional to its budget. In other words, an allocation $X$ satisfies \textit{proportionality} if  $u_i(x_i) \geq \frac{B_i}{\sum_{i'} B_{i'}} u_i (C),~\forall i.$
Thus, this property can be interpreted as the utility-fairness  
(i.e., every service feels fair in terms of the achieved utility). 

\begin{theorem}
\textit{The proposed ME is envy-free, and satisfies the sharing-incentive and proportionality properties.} 
\end{theorem}

\textit{Proof.} Please refer to \textit{Appendix \ref{appfair}}.
\qedb 

\section{Decentralized Solution}
\label{PPDA}

In this section, we develop a privacy-preserving distributed algorithm for computing a non-wasteful and frugal ME that is an optimal solution to  the problem
(\ref{EGmain})-(\ref{EGmain4}). The developed 
algorithm converges to the ME while preserving the services' privacy. Specifically, no service needs to reveal any private information to other parties including the other services and the platform, which would significantly mitigate the strategic behavior of the market participants.
For example, if a service can learn private information of some other services, it may be able to guess the utility functions of these services and submit strategic bids to the platform to gain benefits. Our scheme prevents this issue. The proposed scheme 
is also robust to common attacks such as collusion and eavesdropping. 

We assume that all the services are semi-honest, often called \textit{honest-but-curious}, which is a common security model in security, privacy, and cryptography research \cite{pinkas,ayao82}. In the honest-but-curious setting, all parties follow the protocol (i.e., the distributed algorithm) honestly, but they are still curious about the information of other parties.
To achieve the design goal of the privacy-preserving distributed algorithm, we first convert  (\ref{EGmain})-(\ref{EGmain4}) to a standard ADMM form such that the distributed algorithm can be conducted in parallel by the services. This algorithm requires only the average of the optimal demand vectors of the services at each iteration. Thus, our task is to derive an effective procedure for calculating the exact average vector in a privacy-preserving manner.

It is worth noting that if there is only one FN (i.e., a small DC), 
the ME can be found in a distributed manner using the dual decomposition method \cite{boyd}.  Since the objective function in the problem  (\ref{EGmain})-(\ref{EGmain4}) is not strictly concave in $u_i$, we leverage the ADMM technique, which works for non-strictly convex functions.
Due to space limitation, here we do not study the single FN case. 

First, we rewrite the problem  (\ref{EGmain})-(\ref{EGmain4}) as follows: 

\beqn
\label{EGadmm}
\underset{X}{\text{maximize}} &&\sum_i B_i \ln \sum_j \frac{x_{i,j,1}}{a_{i,j,1}}  \\ \nonumber
\label{EGadmm1}
\text{subject to} && \sum_j   \frac{x_{i,j,1}}{a_{i,j,1}} \leq u_i^{\sf max}, \quad \forall i \\ \nonumber
\label{EGadmm2}
&&\sum_i x_{i,j,r} \leq 1,~\forall j,~r;~~x_{i,j,r} \geq 0,~ \forall i,~j,~r.
\eeqn

The traditional ADMM approach \cite{admm} cannot be directly applied to tackle this problem since it can only solve optimization problems without inequality constraints. More importantly, the problem (\ref{EGadmm}) has  multi-block  variables (i.e., $x_i$'s)  while the traditional ADMM approach can only solve problems with two blocks of variables.
An intuitive solution is to extend the traditional ADMM to multi-block variables and update $x_i$ sequentially. However, such \textit{Gauss-Seidel} ADMM may not converge for $N \geq 3$ \cite{cche13}.  Furthermore, in the traditional ADMM, the blocks are updated one after another, hence, it is not amenable for parallelization.
Another solution is to convert the multi-block problem into an equivalent two-block problem via variable splitting \cite{admm}. However, this method substantially increases the number of variables and constraints in the problem, especially when $N$ is large. In \cite{wden17}, the authors propose the  \textit{Jacobi-Proximal ADMM} approach, which is suitable for parallel and distributed optimization, by adding additional proximal terms to the variable-update step. Our approach does not utilize proximal terms.
Finally, the existing approaches require the agents to share their private information with each other in every iteration. Thus, they are not suitable for our design goal.

To make use of ADMM, we first clone all variables $x_i$'s using auxiliary variables  $z_i = (z_{i,1},\ldots,z_{i,M})$ and obtain the following  problem that is equivalent to the problem  (\ref{EGadmm}).

\beqn
\label{admm}
\underset{x,~z}{\text{maximize}}&& \sum_i B_i \ln \sum_j \frac{x_{i,j,1}}{a_{i,j,1}} \\ \nonumber
\label{admmeq1}
\text{subject to}  && x_{i,j,r} - z_{i,j,r} = 0, \quad \forall i,~j,~r \\ \nonumber
\label{admmeq2}
&&\sum_j   \frac{x_{i,j,1}}{a_{i,j,1}} \leq u_i^{\sf max}, \quad \forall i \\ \nonumber
\label{admmeq4}
&&\sum_i z_{i,j,r} \leq 1, \quad \forall j,~r \\ \nonumber
\label{admmeq3}
&&x_{i,j,r} \geq 0,\quad \forall i,~j,~r.
\eeqn 

To convert this problem into ADMM form, we define:
\beqn
&&f_i(x_i) := v_i(x_i) + h_i(x_i), ~\forall i \nonumber \\ \nonumber
&&v_i(x_i) := - B_i \ln ~ \displaystyle  \sum_j \frac{x_{i,j,1}}{a_{i,j,1}}, ~ \forall i \\ \nonumber
&& h_i(x_i) := 
	\begin{cases} 
      0 & \text{if ~ $x_i \in \mathcal{X}_i$} \\ \nonumber
      +\infty & \text{otherwise}
   \end{cases}\\
&& g (\displaystyle \sum_i z_i ) :=
\begin{cases} 
      0 & \text{if ~ $z \in \mathcal{Z}$} \\ \nonumber
      +\infty & \text{otherwise}
   \end{cases} \\ \nonumber
%
&& \mathcal{X}_i := \Big\{ x_i ~ \big|~  \sum_j   \frac{x_{i,j,1}}{a_{i,j,1}} \leq u_i^{\sf max}, ~~\forall i  ; \\ \nonumber
&& \quad \quad \quad x_{i,j,r} \geq 0, \forall i,~j,~r \Big\}  \\ \nonumber
&& \mathcal{Z} :=  \Big\{ z ~ \big|~ \sum_i z_{i,j,r} \leq 1, ~ \forall j,~r  \Big\},
\eeqn 
where $h_i(.)$ and $g(.)$ are indicator functions. Then, the problem (\ref{admm}) can be written in ADMM form as follows:
\beqn
\label{admmsharing}
\underset{x,z}{\text{minimize}} && \sum_i f_i(x_i) + g(\displaystyle \sum_i z_i  ) \\ \nonumber
\label{admmeq5}
\text{subject to}  && x_{i,j,r} - z_{i,j,r} = 0, \quad \forall i,~j,~r,
\eeqn
with variables $x_i$, $z_i \in \mathbb{R}^{M\text{x}R}, ~\forall i.$ 
To simplify the notation, we consider each resource type $r$ at FN $j$ as an item $k$ and we have a total of K = M \text{x} R items.  
Then, $x_i$, $z_i \in \mathbb{R}^K, ~\forall i$, and the problem (\ref{admmsharing}) can be  expressed as below.
\vspace{-0.04in}
\beqn
\label{admmsharing1}
\underset{x,~z}{\text{minimize}} && \sum_i f_i(x_i) + g(\displaystyle \sum_i z_i  ) \\  
\label{admmeq6}
\text{subject to}  && x_i - z_i = 0, \quad \forall i.
\eeqn

We have eliminated the coupling constraints between variables $x_i$'s.  Furthermore,  the variables now can be divided into two blocks (i.e., one contains all $x_i$'s, and the other contains all $z_i$'s).
Hence, we are ready to use the  ADMM approach to construct a parallel and distributed algorithm for computing the ME. 
First, define $\overline{x}, ~\overline{z} \in \mathbb{R}^K$ as follows
\beqn
\overline{x} = \frac{1}{N}\sum_i x_i; \quad \overline{z} = \frac{1}{N}\sum_i z_i.
\eeqn
Also, let  $p_i = \big(p_{i,1}, p_{i,2}, \ldots, p_{i,K} \big) \in \mathbb{R}^K, ~\forall i,$ be the dual variables associated with constraint (\ref{admmeq6}). Indeed, we can show that $p_i = p,~\forall i$ (see Appendix \ref{admm_apen}).
 The ADMM-based decentralized implementation is summarized in \textbf{Algorithm 1}. 
Due to the space limitation, we do not present the construction of this algorithm. Please refer to  \textit{Appendix \ref{admm_apen}}
for the detailed derivation.

In this algorithm, $r^{\sf primal,t+1}$ and $r^{\sf dual,t+1}$ are the primal and dual residuals at iteration $t+1$, respectively. As defined in \cite{admm}, we have $ ||r^{\sf primal,t+1}||_2 = \sqrt{N}~||\overline{z}^{t+1} - \overline{x}^{t+1}  ||_2  $ and $||r^{\sf dual,t+1}||_2 = \rho~ ||\overline{z}^{t+1} - \overline{z}^t ||_2$. 
At each iteration $t$, given $\overline{z}$ and price signal $p$, each service $i$ solves a convex optimization problem to find its optimal resource bundle $x_i^{t+1}$ and sends it to the platform.
After collecting the optimal demands of all the services, the platform updates $\overline{z}^{t+1}$ by solving (\ref{z-up}). Based on the updated values of $\overline{x}$ and $\overline{z}$, the resource prices can be updated by the platform or the FNs. 
The  algorithm terminates when the primal and dual residuals are sufficiently small \cite{admm} or the number of iterations becomes sufficiently large. Otherwise, the new values of $\overline{x}, \overline{z},$ and $p$ are sent to every service. The steps described above repeat until the algorithm converges. The convergence properties of \textit{Algorithm 1} follows directly the standard ADMM method \cite{admm}.

\textit{Remark:} It is worth noting that the \textit{z-update} and \textit{dual-update} steps can be carried out by every service and the services only need to exchange messages to update
 $\overline{x}$. By this way, the algorithm can be implemented in a fully distributed manner by the services (i.e., without the platform).

\begin{algorithm}[H]
\footnotesize
\caption{\textsc{Distributed Implementation}}
\label{dist}
\begin{algorithmic}[1]

\STATE Initialization: set  $\{x_i^1\}$, $\{z_k^1\}$, $\{p_k^1\}$, $\rho$, $\gamma_1$, $\gamma_2$, t = 0.  

\REPEAT 
\STATE 	 At iteration $t:= t +  1$, do:
\STATE 	\textbf{x-update:} each service $i$ solves the following sub-problem to compute $x_i^{t+1}$ \\

 $\underset{x_i} \min ~v_i(x_i) + \frac{\rho}{2}~\Big|\Big| x_i - x_i^t +  \overline{x}^t -\overline{z}^t + \frac{1}{\rho}~ p^t \Big|\Big|_2^2, \text{s.t.}, x_i \in \mathcal{X}_i.$ \\

\STATE  \textbf{z-update:} to compute $\overline{z}^{t+1}$, the platform solves

\beqn
\label{z-up}
\underset{\overline{z}} \min ~  \Big|\Big| \overline{z} - \overline{x}^{t+1} - \frac{1}{\rho} ~p^t \Big|\Big|_2 ~~ \text{s.t.}, N~ \overline{z} \in \mathcal{Z}.
\eeqn

\STATE \textbf{Dual update:} the platform or FNs updates $p^{t+1}$ \\ 

 $p_k^{t+1} := p_k^t + \rho~ \overline{x}_k^{t+1} -  \rho ~\overline{z}_k^{t+1},~\forall k.$ \\ 
 
\UNTIL {$ \big( ||r^{\sf primal,t+1}||_2 \leq \gamma_1$ and  $ ||r^{\sf dual,t+1}||_2 \leq \gamma_2$ \big) or $t$ is too large.}

\STATE Output: equilibrium allocation $X^*$ and equilibrium prices $p^*$.

\end{algorithmic}
\end{algorithm}

Now we consider the privacy-preserving issue. It can be observed that in \textit{Algorithm 1}, the utility function of a service is only known to that service. However, $\overline{x}^{t+1}$ is required for the \textit{z-update} step. A conventional approach to obtain $\overline{x}^{t+1}$ is to ask every the services to  report $x_i^{t+1}$ to the platform and $\overline{x}^{t+1} = \frac{1}{N} \sum_i x_i^{t+1}$. However, this approach reveals $x_i^{t+1}$, which is undesirable. 
For instance, by learning a sequence of $x_i$, which is the optimal resource bundle of service $i$ given the price signal $p$, an attacker (e.g., the platform or another service) may be able to guess the utility function of service $i$. A privacy-preserving protocol should not disclose any private information of the services. If we can find a method to obtain $\overline{x}^{t+1}$ without asking the exact value of $x_i^{t+1}$ from every service $i$, we can turn \textit{Algorithm 1} into a privacy-preserving distributed algorithm.
  This procedure is executed right before the \textit{z-update} step and does not affect the properties (e.g., convergence rate) of the ADMM-based 
	distributed algorithm.

Conventional cryptographic techniques for average computation are quite sophisticated \cite{ayao82,cgen09,pinkas}.
On the other hands, differential privacy approaches do not produce the exact average value \cite{cdwo14}. Recently, \cite{ymo17} proposes a novel privacy-preserving average consensus method. This method runs in an iterative manner to find the average value, thus,
may not be suitable for our iterative optimization problem. 
Here, we introduce a simple and efficient technique for privacy-preserving average computation.
Specifically, after the \textit{x-update} step, each service $i$ holds a vector $x_i = (x_{i,1},\ldots,x_{i,K})$. The iteration index 
is removed for simplicity. We need to compute the average vector $\overline{x} = (\overline{x}_{1},\ldots,\overline{x}_{K})$  while 
keeping each vector $x_i$ known to service $i$ only. 
The proposed \textit{privacy-preserving average computation} algorithm works as follows. 
First, each service $i$ selects a random set $\mathcal{N}_i$ of $n_i$ other services. Then, it generates a set of $n_i + 1$ random numbers which includes $y_i^0$ and $y_i^l$, for all $l \in \mathcal{N}_i$, such that sum of these numbers is zero, i.e., $y_i^0 + \sum_{l \in \mathcal{N}_i} y_i^l = 0, \forall i$. Also, $y_i^l = 0$, $\forall l \notin \mathcal{N}_i$.

 Each service $i$ keeps $y_i^0$ and sends $y_i^l$ to the corresponding service $l$ in $\mathcal{N}_i$. To enhance security, the message containing $y_i^l$ between service $i$ and $l$ can be encrypted to prevent attackers from eavesdropping. In the next step, each service $i$ waits to receive all the messages from the other services 
who selected $i$. 
If the message sent from service $l$ to service $i$ is encrypted, service $i$ has to decrypt the message using the shared key between them to extract $y_l^i$. Then, service $i$ computes  $z_i = x_i + y_i^0 + \sum_{l \in \mathcal{N}} y_l^i$ and sends the resulting vector $z_i$ to the platform. Obviously, service $i$ can also encrypt $z_i$ before sending it the platform. It can be observed that
\beqn
\sum_i z_i &=& \sum_i \Big(x_i + y_i^0 + \sum_l y_l^i \Big)  \\ \nonumber
&=& \sum_i x_i +   \sum_i  \Big( y_i^0 + \sum_{l \in \mathcal{N}_i}  y_i^l + \sum_{l \notin \mathcal{N}_i}   y_i^l \Big)= \sum_i x_i.
\eeqn

Thus, the platform can take the average of all the vectors $z_i$'s that it receives, which is precisely $\overline{x}$. 
The steps above are summarized in \textbf{Algorithm 2}, which allows us to find the exact average vector $\overline{x}$ without disclosing $x_i$. It is worth noting that the generated noises $y_i^l$, $\forall i,~l$, cannot be reused. In other words, we need to rerun the privacy-preserving average computation algorithm at every iteration in \textit{Algorithm 1} to prevent an attacker to learn $x_i^{t+1} - x_i^{t}$, which is private information of service $i$. In particular, if the attack can overhear $z_i$ information, she can infer $z_i^{t+1} - z_i^{t} = (x_i^{t+1} + y_i^0 + \sum_l y_l^i) - (x_i^t + y_i^0 + \sum_l y_l^i) = x_i^{t+1} - x_i^{t}.$ 
Finally, for the fully distributed version of \textit{Algorithm 1} (i.e, without the platform), \textit{Algorithm 2} can be slightly modified by selecting a service to act as the platform to compute $\overline{x}$. Then, this service will broadcast the $\overline{x}$ to the other services.

 \textit{Algorithm 2} is resistant to eavesdropping and collusion attacks.  
Specifically, the private data of a service (e.g., service $i$) might be exposed in two situations: i) attackers can eavesdrop on the communication channels between this service and some other services; ii) a group of services (may include the platform) collude to infer $x_i$. Indeed, the collusion case is equivalent to the case where the communication channels among service $i$ and the collusive agents are eavesdropped.  
Let $\mathcal{I}_i$ be the set of services that sends noise signals $y_l^i$, $\forall ~l \in \mathcal{I}_i$, to service $i$. Define $\mathcal{L}_i = \mathcal{I}_i \cup \mathcal{N}_i, ~\forall i$, as the set of services that communicate with service $i$.

\begin{algorithm}[H]
\footnotesize
\caption{\textsc{Privacy-Preserving Averaging}}
\label{average}
\begin{algorithmic}[1]

\STATE Input: N services. Each service $i$ keeps a vector $x_i = (x_{i,1},\ldots,x_{i,K})$. \\

\STATE \textbf{Step 1}: each service $i$ picks randomly a set $\mathcal{N}_i$ of $n_i$ other services. \\
 
\STATE \textbf{Step 2}: each service $i$ locally generates $n_i+1$ random numbers including $y_i^0$ and $y_i^l$, $\forall l \in \mathcal{N}_i$, such that 
$y_i^0 + \sum_{l \in \mathcal{N}_i} y_i^l = 0$. For every service $l \notin \mathcal{N}_i$, $y_i^l = 0.$ Also, service $i$ keeps $y_i^0$ for herself.
 \\

\STATE \textbf{Step 3:} each service $i$ sends $y_i^l$ to service $l \in \mathcal{N}_i$ and does not send anything to any service  $l \notin \mathcal{N}_i$. 

\STATE \textbf{Step 4:} after receiving all $y_l^i$ from other services, service $i$ computes $z_i = x_i + y_i^0 + \sum_{l \in \mathcal{N}} y_l^i$ and sends $z_i$ to the platform.

\STATE \textbf{Step 5:} the platform computes $\overline{x} = \frac{1}{N} \sum_i z_i$.

\STATE Output: The average vector $\overline{x}$

\end{algorithmic}
\end{algorithm}

Observe that $z_i$ is the only message of service $i$ that contains vector $x_i$. 
However, even if an attacker can overhear 
and decrypt this message to obtain $z_i$, it is still difficult for her to infer $x_i$. To recover $x_i$, the attacker needs to know $y_i^0$ and $y_l^i$,~$\forall l \in \mathcal{I}_i$. Note that $y_i^0$ depends on $y_i^l,~\forall l \in \mathcal{N}_i$.   
Thus, $x_i$ is revealed only if all the communication channels between service $i$ and the platform as well as the services in $\mathcal{L}_i$ are compromised. Define $P_i$ as the probability that the communication channel between service $i$ and another service (or the platform) is cracked. Let $Q_i(m)$ be the probability that the number of services in set $\mathcal{L}_i \setminus  \mathcal{N}_i$ is $m$. 
Then, given $P_i$, the probability that $x_i$ is disclosed can be approximated by $P_i^{n_i+1} \sum_{m=0}^{N} Q_i(m) P_i^m$ if the platform is not corrupt, and by $P_i^{n_i} \sum_{m=0}^{N} Q_i(m) P_i^m$ if the platform is corrupt. 

In summary, by embedding  \textit{Algorithm 2} 
into  \textit{Algorithm 1}, we achieve a fully parallel and distributed privacy-preserving algorithm. The additional step of running the privacy-preserving averaging protocol (i.e., Algorithm 2) is independent and does not affect the convergence property of \textit{Algorithm 1}, which follows directly the standard ADMM algorithm. Obviously, the added privacy-preserving feature comes at a cost, which is mainly the communication cost of exchanging messages among the services in \textit{Algorithm 2}. This cost decreases as the size of set $\mathcal{N}_i$ decreases. For simplicity, assume $n_i = |\mathcal{N}_i| = b,~\forall i.$ Then, the communication cost of \textit{Algorithm 2} reduces significantly as $b$ decreases. On the other hands, the proposed algorithm becomes more robust as $b$ increases. Thus, there is a tradeoff between the robustness of the privacy-preserving feature and the communication cost.
Depending on the specific system and design goal, the platform can decide a suitable value of $b$ to balance between the privacy and communication overhead.

\section{Performance Evaluation}
\label{sim}

\subsection{Simulation Setting}

We generate data for a system that consists of 100 FNs and 40 services (e.g., in a metropolitan area network).
Each FN is chosen randomly from the set of M4 and M5 Amazon EC2 instances\footnote{https://aws.amazon.com/ec2/instance-types/}. 
There are three types of resources (e.g., CPU, RAM, and bandwidth) 
associated with each FN. For example, an M4 Amazon EC2 machine of type \textit{m4.4xlarge} has 16 vCPUs, 64 GiB of memory, and 2000 Mbps of bandwidth.  It is assumed that each service has the same base demand vector at different FNs. The base demand vectors of different services can be different. For each service, the amount of resource type 1 (vCPU), type 2 (RAM), and type 3 (bandwidth) in its base demand vector are generated randomly in the ranges of [0.1, 0.5] vCPUs, [0.4, 2.0] GiB, and [10, 50] Mbps, respectively.

The maximum demand of every service is set to be 600 (i.e., maximum number of requests over a certain time period). Hence, all the services have the same utility limit of 600 (i.e., $u_i^{\sf max} = u^{\sf max} = 600, \forall i$). 
From the generated data, we can normalize the resource capacity of the FNs and 
compute  $a_{i,j,r}$ accordingly, $\forall i, j, r$. 
By normalization, every resource type at an FN has a capacity of one unit (i.e., $C_{j,r} = 1,~\forall j,r$). Then, note that one memory unit of  FN 1 may correspond to 64 GiB (e.g.,  FN 1 is an \textit{m4.4xlarge} instance) while one memory unit of FN 2 corresponds to 160 GiB (e.g.,  FN 2 is an \textit{m4.10xlarge} instance). Thus, one  unit of the same resource type at different FNs may have different values to a service.

For the sake of clarity in the figures and analysis, 
in the \textbf{base case}, we consider a small system with 40 FNs and 8 services (i.e., M = 40, N = 8), which are selected randomly from the original set of 100 FNs and 40 services. Furthermore, it is assumed that all the services have equal budget  in the base case (i.e., $B_i = 1, ~\forall i$). This default setting is used in most of the simulations unless mentioned otherwise. It is worth noting that the simulation served to illustrate our theoretical results only.  
To reproduce our test cases and results, interested readers can refer to our code using Matlab and CVX/MOSEK \cite{code}. 
The simulation was implemented on a Macbook Air 2017
1.8GHz dual-core Intel Core i5 and 8GB of RAM.
We utilize the same seed (i.e., same dataset) to run the  simulations. Our code can be used
 to  create arbitrary datasets/settings. We have run the simulations using different datasets and observed similar trends and conclusions for all the figures.

\subsection{Numerical Results}

We consider the  following five allocation schemes.

\begin{itemize}

\item \textit{Generalized  Eisenberg-Gale scheme} (\textbf{GEG}): This is the proposed scheme where  the allocation is an optimal solution to the generalized EG program (\ref{EGmain})-(\ref{EGmain4}).

\item \textit{Eisenberg-Gale scheme} (\textbf{EG}): In this scheme, the resource allocation 
is an optimal solution to  (\ref{EGinf})-(\ref{EGinf2}).

\item \textit{Proportional sharing scheme} (\textbf{PROP}): 
Every buyer is allocated a portion of every resource proportional to her budget (i.e., buyer $i$ receives $\frac{B_i}{\sum_i B_i}$ of every resource).

\item \textit{Social welfare maximization scheme} (\textbf{SWM}) : In this scheme, the platform determines an optimal allocation that maximizes the total utility of all the buyers (i.e., $\sum_i u_i(x_i)$) subject to the resource capacity constraints of the FNs. The budget constraint is not considered.

\item \textit{Maxmin fairness scheme} (\textbf{MM}): The system tries to maximize the utility of the buyer with the lowest utility (i.e., max $\min_i u_i(x_i)$) under the FNs' capacity constraints. This scheme also ignores the budget constraint.

\end{itemize}

Note that GEG and EG are the two allocation schemes  studied in this paper. The  others  are considered as benchmark schemes. In all the five schemes, the resulting  utilities of the buyers are truncated by their corresponding utility limits.
From \textit{Proposition \ref{theo1}} and \textit{Theorem \ref{main}}, the solutions in both EG and GEG schemes are market equilibria in  our resource allocation problem for buyers with utility limits. Obviously, if all the buyers have infinite demands (i.e., $u^{\sf max} = \infty$), the solutions produced by EG and GEG are the same. 

\begin{figure}[ht!]
	\centering
		\includegraphics[width=0.33\textwidth,height=0.10\textheight]{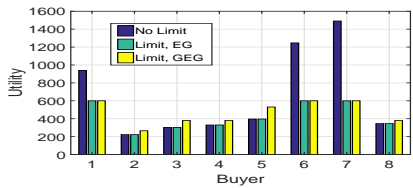} 
			\caption{Comparison between EG and GEG schemes}
	\label{fig:EGGEG}
\end{figure} 

Fig.~\ref{fig:EGGEG} compares the ME solutions in the EG and GEG schemes. 
 The bars labeled \textit{``No Limit''} present the buyers' utilities at the market equilibrium  in our resource allocation problem where buyers have no utility limits. 
Indeed,  the buyers' utilities in the EG scheme are their utilities in the \textit{``No Limit''} setting truncated by the utility limits.
We can observe that 
 when the utility limit is considered,  the optimal utilities in the GEG scheme tend to dominate those in the EG scheme. This is because some buyers in the EG scheme receive too many resources while they already reach their utility limits (e.g., buyers 1, 6, and 7), 
 which reduces the resources available to other buyers.  
 Since the GEG scheme takes this issue into account by redistributing the wasteful resources to other buyers, it outperforms the EG scheme. 
In the EG scheme, if the redundant resources of buyers 1, 6, and 7 are reallocated to the other buyers, it can strictly improve the utilities of some buyers. Thus, the ME produced by the EG scheme is not Pareto-efficient. 
These results align with our statements in \textit{Proposition \ref{theo1}}.

\begin{figure}[h!]
	\centering
		\subfigure[$u^{\sf max} = \infty$]{
		  \includegraphics[width=0.24\textwidth,height=0.10\textheight]{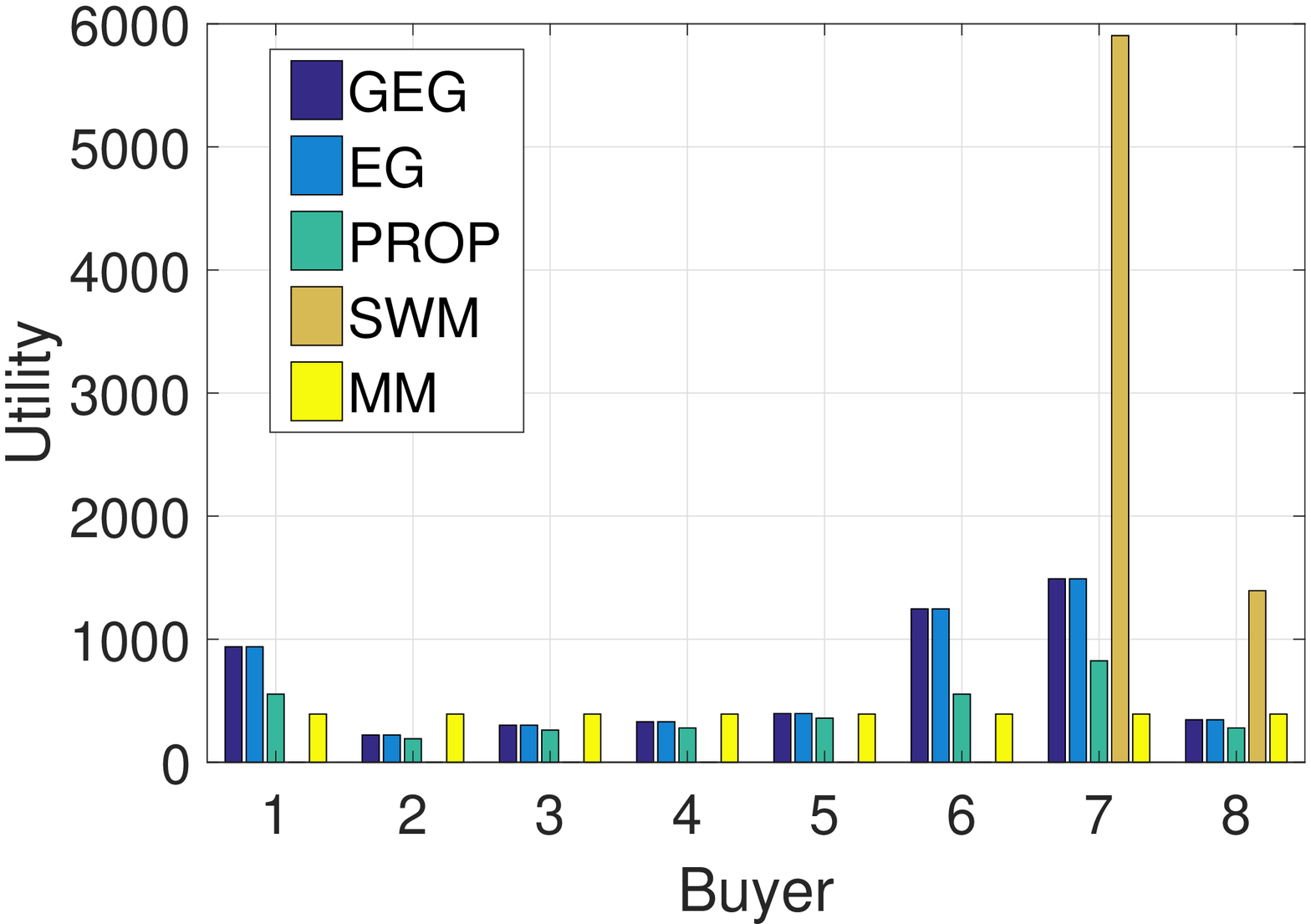}
	    \label{fig:5nocap}
	}   \hspace*{-2.1em} 
		 \subfigure[$u^{\sf max} = 600$]{
	     \includegraphics[width=0.24\textwidth,height=0.10\textheight]{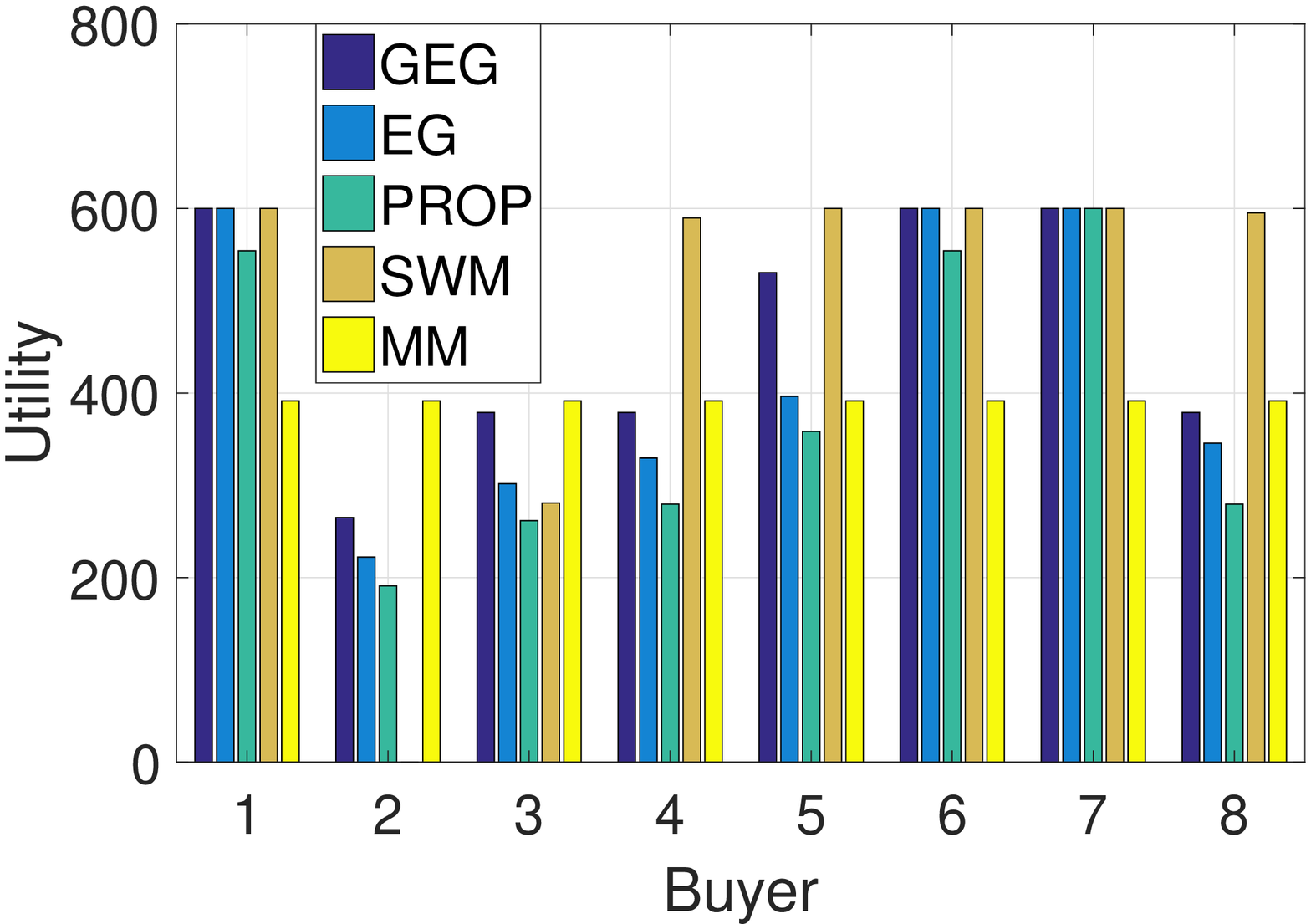}
	     \label{fig:5cap}
	}  \vspace{-0.2cm}
	\caption{Individual utility comparison}
\end{figure}

\begin{figure}[h!]
		\subfigure[$u^{\sf max} = \infty$]{
		  \includegraphics[width=0.24\textwidth,height=0.10\textheight]{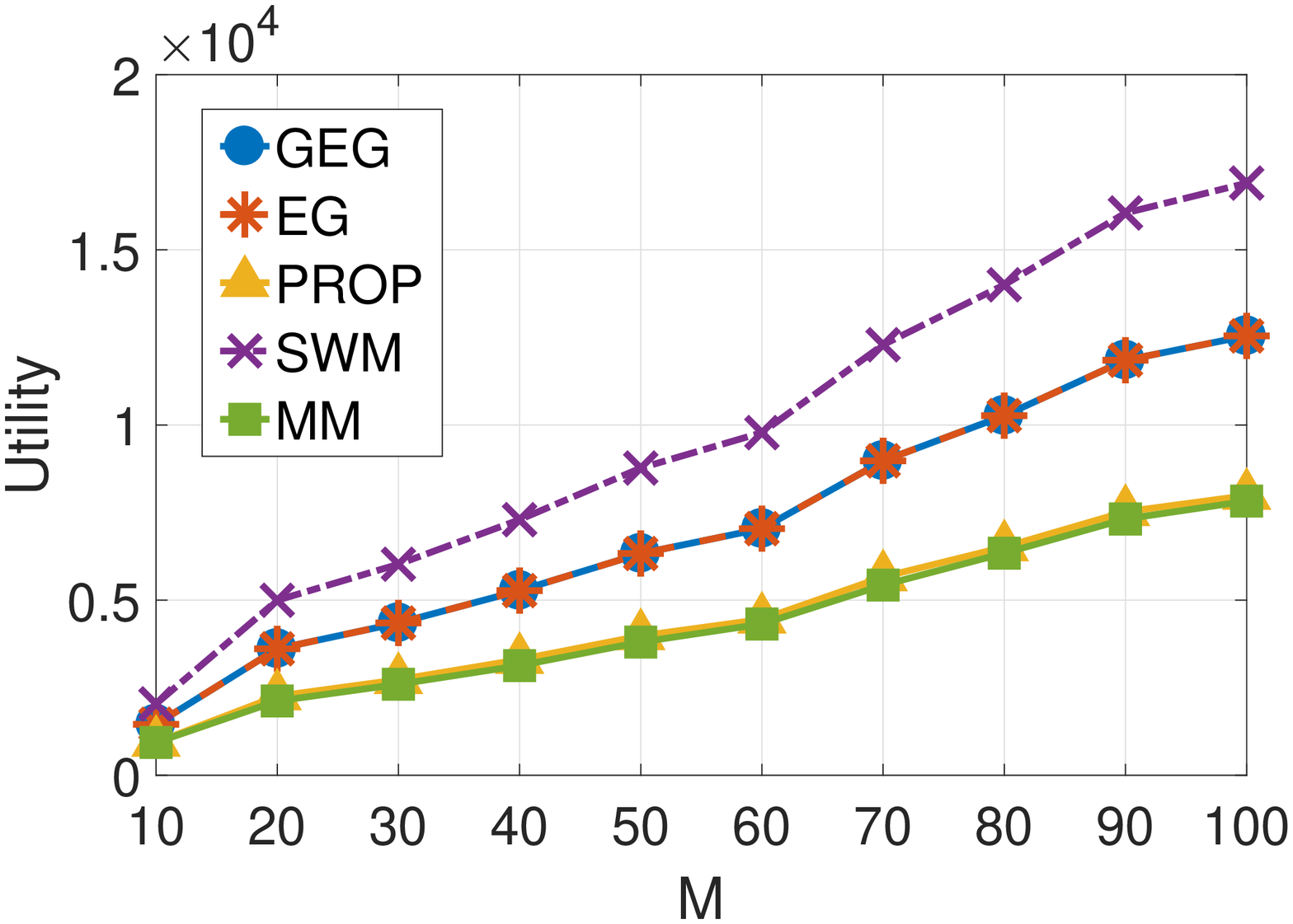}
	    \label{fig:nocaptu}
	}   \hspace*{-2.1em} 
		 \subfigure[$u^{\sf max} = 600$]{
	     \includegraphics[width=0.24\textwidth,height=0.10\textheight]{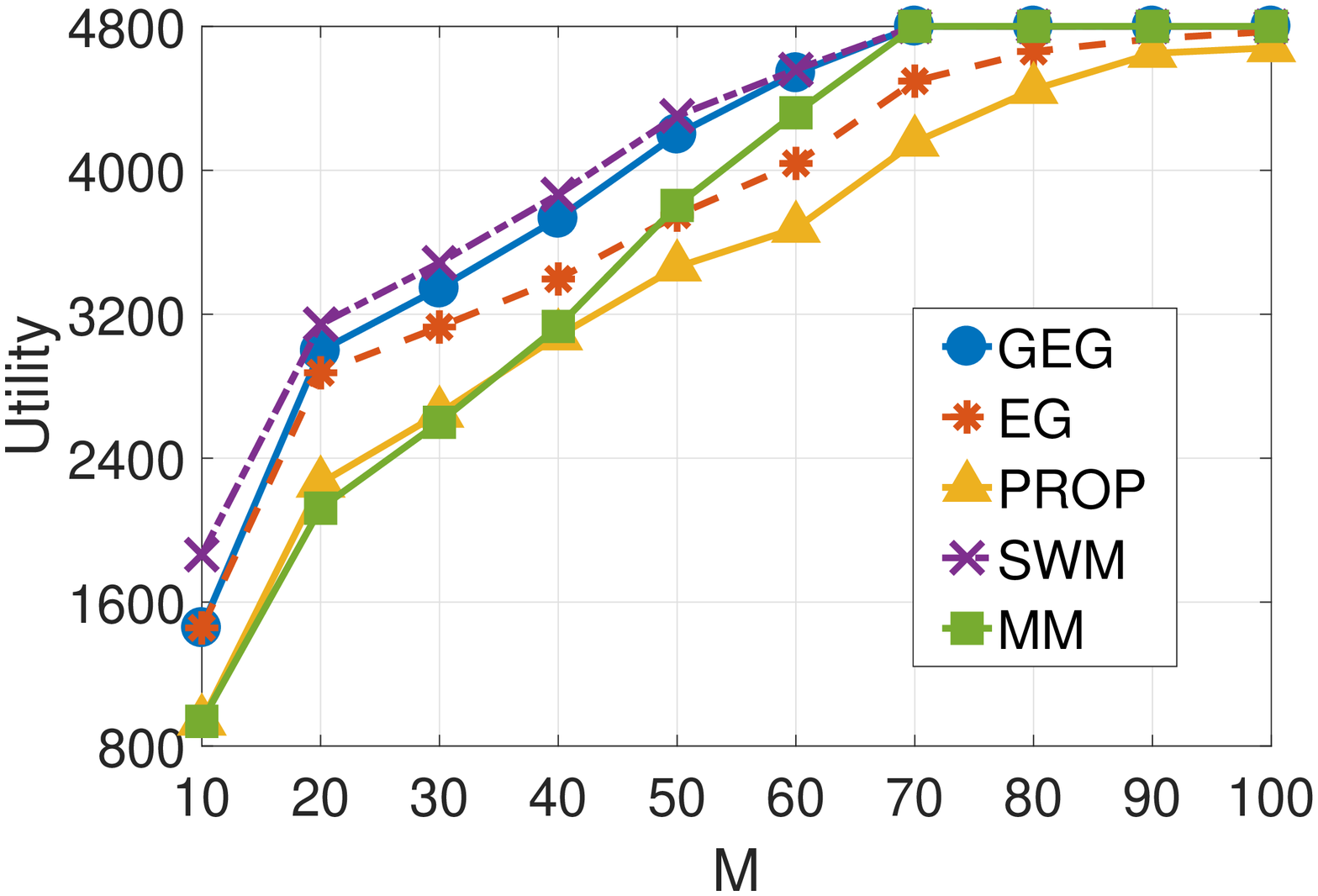}
	     \label{fig:captu}
	}  \vspace{-0.2cm}
	\caption{Comparison of the total utility (N = 8)}
\end{figure}

\begin{figure}[ht]
		\subfigure[ $u^{\sf max} = \infty$]{
		  \includegraphics[width=0.24\textwidth,height=0.10\textheight]{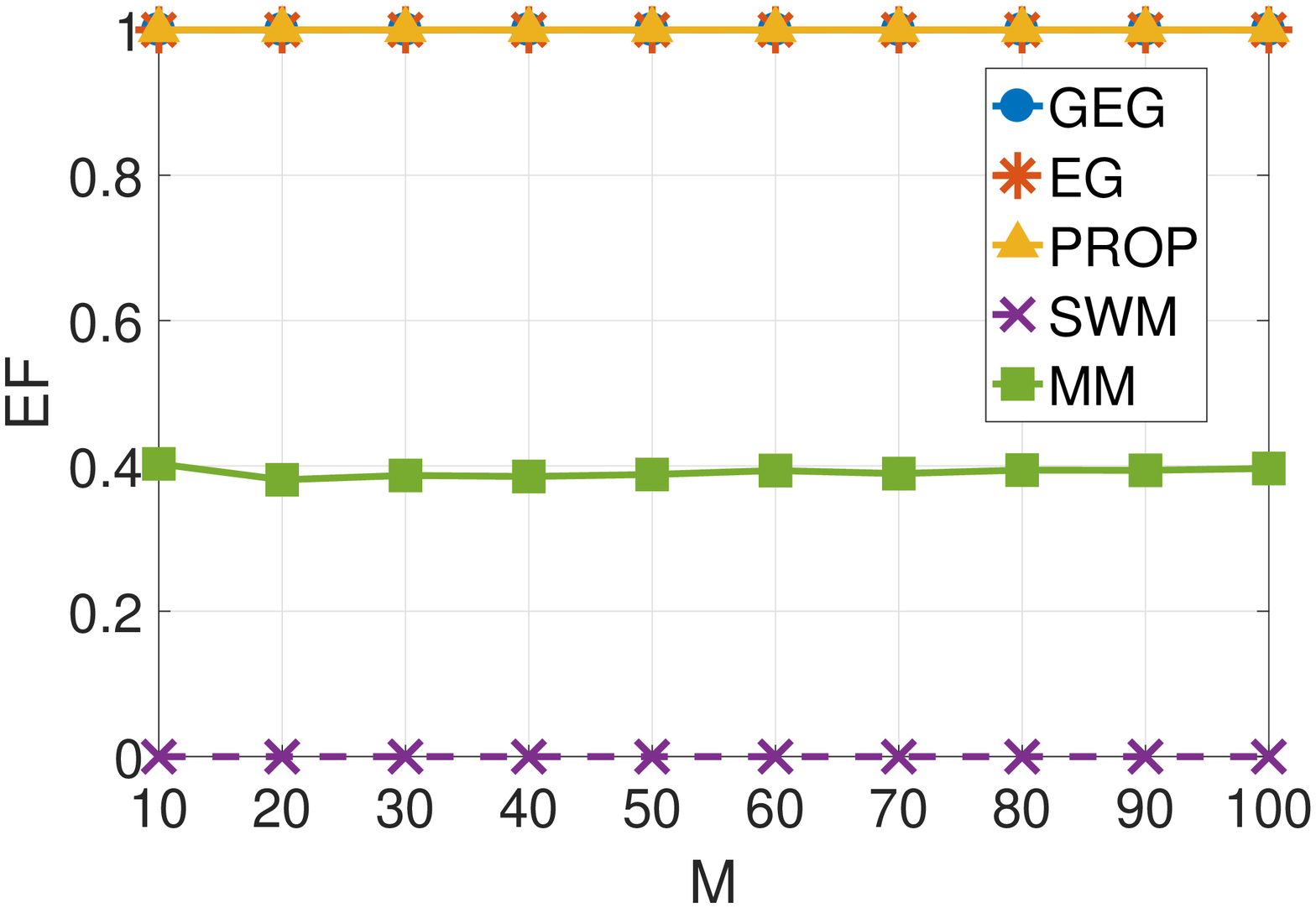}	    
	    \label{fig:uncapEF}
	}   \hspace*{-2.1em}  
		 \subfigure[ $u^{\sf max} = 600$]{
	     \includegraphics[width=0.24\textwidth,height=0.10\textheight]{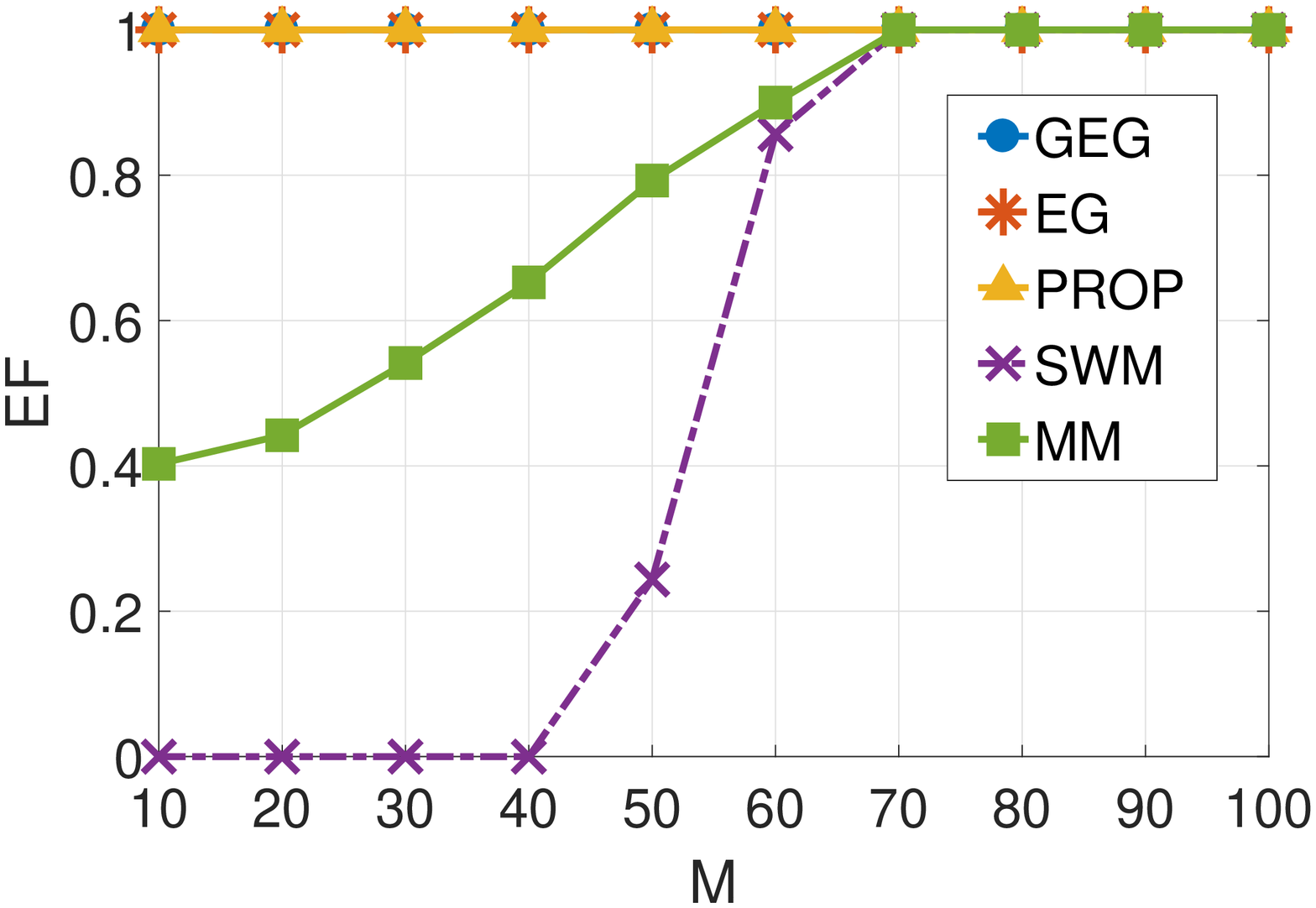}
	     \label{fig:capEF}
	} \vspace{-0.2cm}

	\caption{Envy-freeness comparison (N = 8)}
\end{figure}

\begin{figure}[ht]
		\subfigure[$u^{\sf max} = \infty$]{
		  \includegraphics[width=0.24\textwidth,height=0.10\textheight]{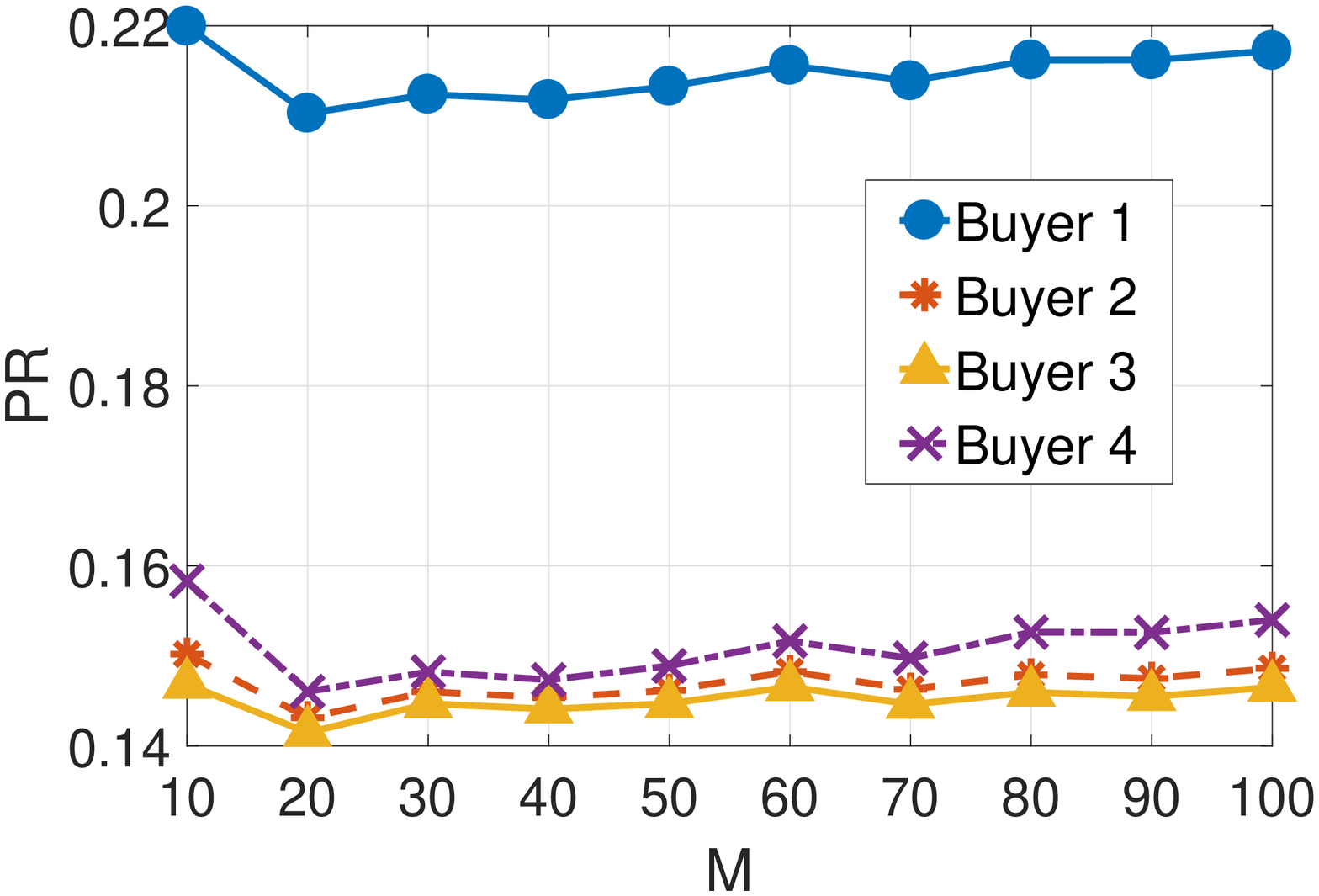}
	    \label{fig:uncapProp}
	}   \hspace*{-1.8em} 
		 \subfigure[$u^{\sf max} = 600$]{
	     \includegraphics[width=0.24\textwidth,height=0.10\textheight]{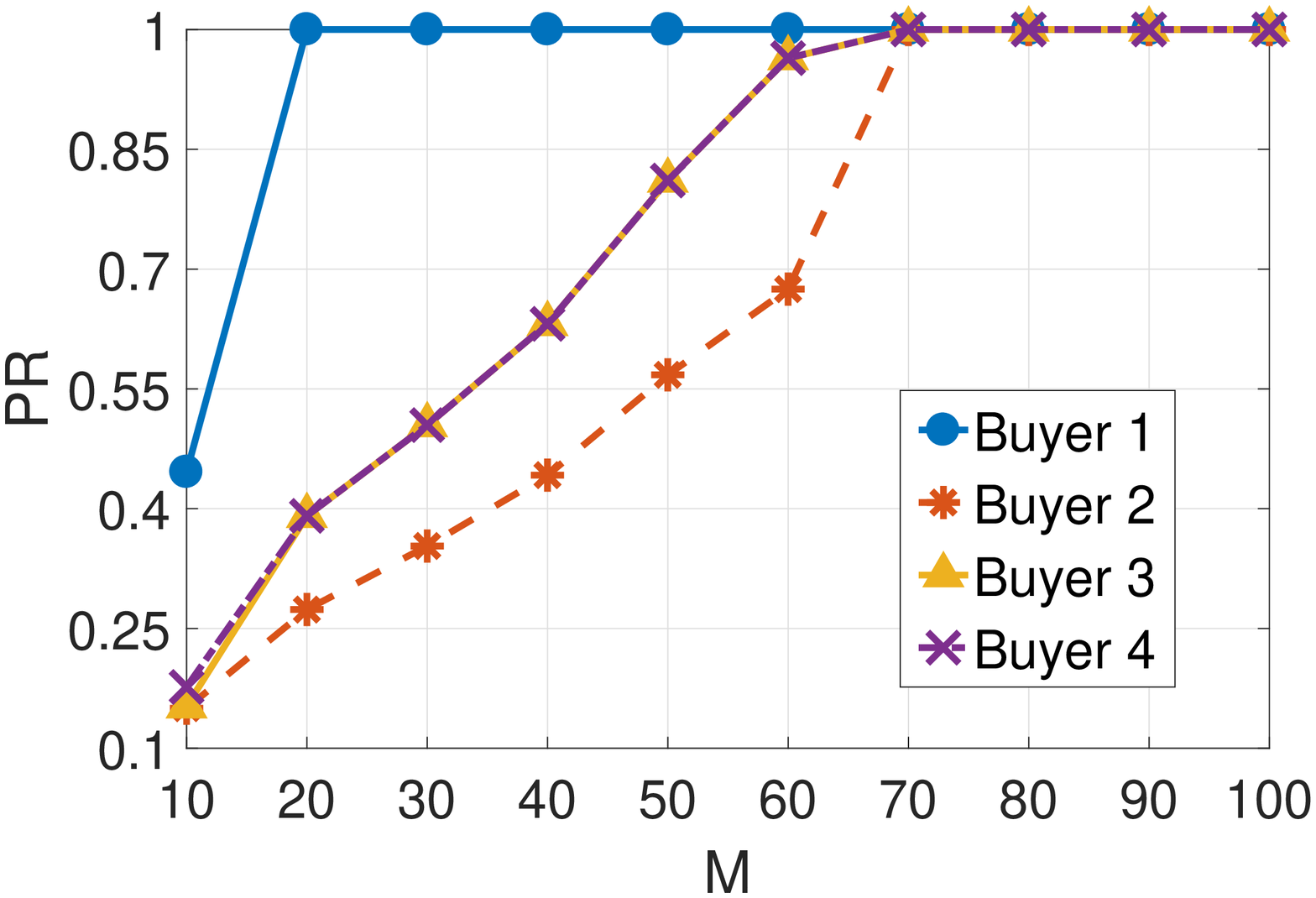}
	     \label{fig:capProp}
	}  \vspace{-0.2cm}
	\caption{Proportionality property (N = 8)}
\end{figure}

Figs.~\ref{fig:5nocap}--\ref{fig:capProp} present the performance comparison among the five  schemes with and without the utility limit. We can see that GEG balances well the tradeoffs between the system efficiency and fairness.
Specifically, Figs.~\ref{fig:5nocap}--\ref{fig:5cap} show that the utility of every buyer in GEG is always greater or equal to that in the PROP scheme. Additionally, the total utility of all the buyers in GEG is significantly higher than that in PROP as illustrated in Figs.~\ref{fig:nocaptu}--\ref{fig:captu}. These observations \textit{imply the sharing-incentive property of the proposed GEG scheme}. 

Although the SWM scheme gives high utilities for some buyers, it may result in unfair allocations. For example, some buyers have very low or even zero utilities as shown in Figs.~\ref{fig:5nocap}--\ref{fig:5cap}. Finally, the MM scheme inherently generates a fair allocation in terms of utility but it may lead to low system efficiency. Also, MM may not be fair in terms of resource quantity allocated to the buyers since some buyers receive too many resources to compensate for their low marginal utilities. In Fig.~\ref{fig:captu}, the total utility of the buyers becomes saturated as number of FNs increases.
It is because when resources are abundant, all the buyers can reach their utility limits.

Figs.~\ref{fig:uncapEF}--\ref{fig:capEF} show the envy-freeness indices of the five schemes.
The envy-free index (EF) of an allocation $X$ is \cite{duong}
\beqn
EF(X) = \min_{i,i'} \frac{u_i(x_i)}{u_i\big( \frac{B_i}{B_{i'}} x_{i'}\big)},~\forall i,i' \in \mathcal{N}. \nonumber
\eeqn
An allocation $X$ is envy-free if $EF(X)  = 1$. The higher EF is, the better an allocation is in terms of envy-freeness. Obviously, PROP is envy-free by definition. We can observe that both EG and GEG are also envy-free, which \textit{confirms the envy-freeness property} of the proposed GEG scheme as theoretically proved. These figures also reveal that both SWM and MM are not envy-free. Noticeably, SWM may produce a very unfair allocation (e.g.,  in Fig.~\ref{fig:capEF}, EF = 0 as $M \leq 40$) when the number of FNs is small (i.e., when the  fog resource is scarce). Similarly, given the same amount of fog resources, the EF index of  SWM   decreases as the number of buyers increases. This result is not shown here due to space limitation. 
Thus, our proposed \textit{ME solution significantly outperforms SW and MM schemes in terms of envy-free fairness.}

Figs.~\ref{fig:uncapProp}--\ref{fig:capProp} illustrate the proportionality fairness property of the proposed GEG scheme. In particular, \textit{PR} of a buyer is defined as the ratio between her actual utility and her maximum possible utility by receiving all the resources (i.e., $PR_i = u_i(x_i)/u_i(C)$). When there are eight buyers (i.e., N = 8) with the same budget, an allocation satisfies the proportionality property if the \textit{PR} of every buyer is greater or equal to $B_i/\sum_{i'} B_{i'} = 1/8$. Hence, these figures support our claim that the \textit{equilibrium allocation produced by GEG satisfies the proportionality property. }

\begin{figure}[ht!]
	\centering
		\includegraphics[width=0.33\textwidth,height=0.10\textheight]{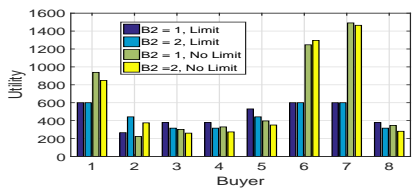} 
			\caption{Impact of budget on the equilibrium utilities}
	\label{fig:budget}
\end{figure}

In the following, we study the proposed GEG scheme only.
The impact of  budget on the utilities of the buyers is presented in Fig.~\ref{fig:budget} where we vary the budget $B_2$ of buyer 2  
and fix the budgets of other buyers to be one (i.e., $B_1 = B_3 = B_4 = B_5 = B_6 = B_7 = B_8 = 1$). 
It can be seen that with and without considering the utility limit, when we double the budget of buyer 2, her utility increases significantly while the utilities of the other buyers tend to decrease. Thus, the proposed scheme is \textit{effective in capturing the service priority} 
in making allocation decision.

\begin{figure}[h!]
		\subfigure[N = 8]{
		  \includegraphics[width=0.245\textwidth,height=0.10\textheight]{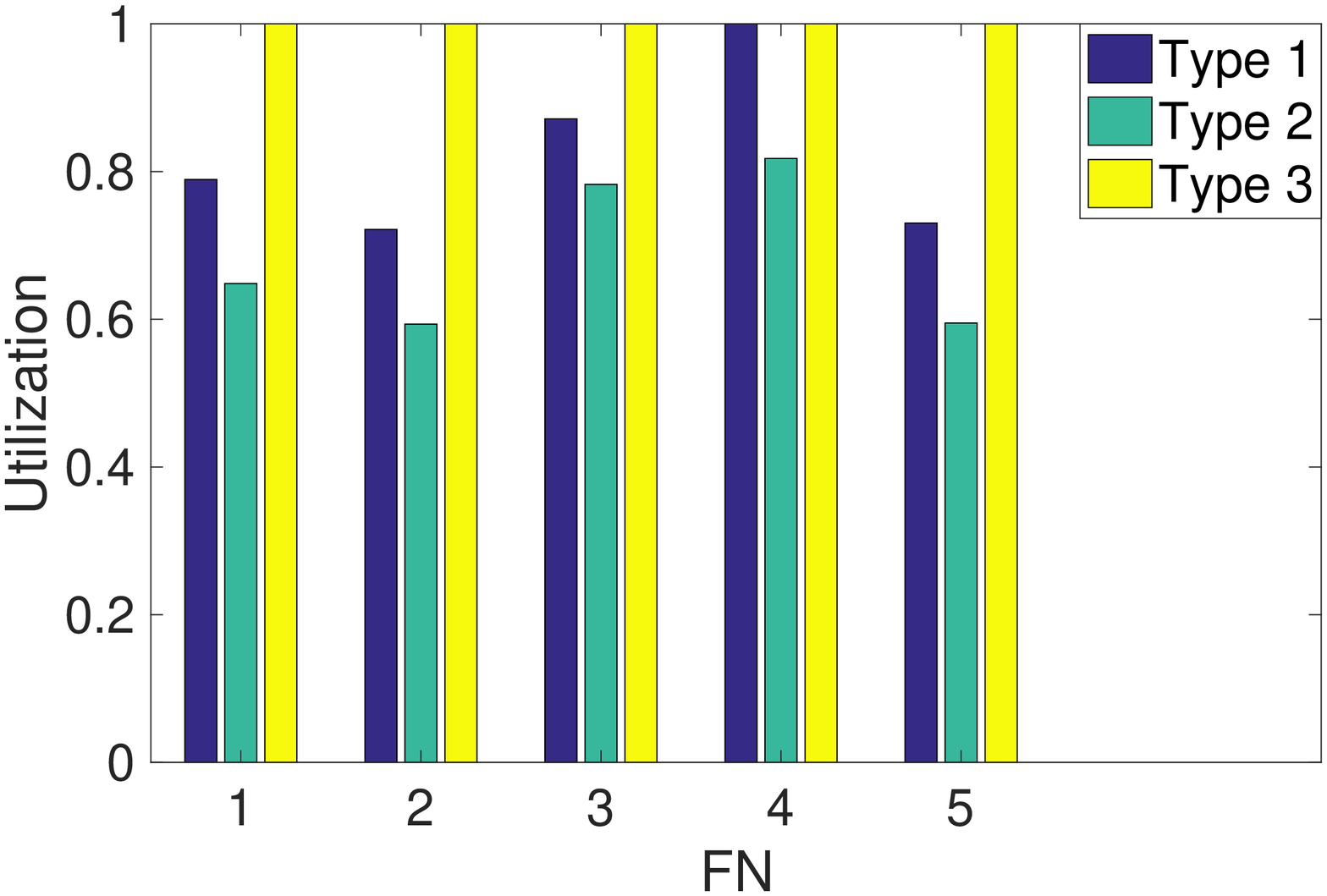}
	    \label{fig:ru1}
	}   \hspace*{-2.1em} 
		 \subfigure[N = 40]{
	     \includegraphics[width=0.245\textwidth,height=0.10\textheight]{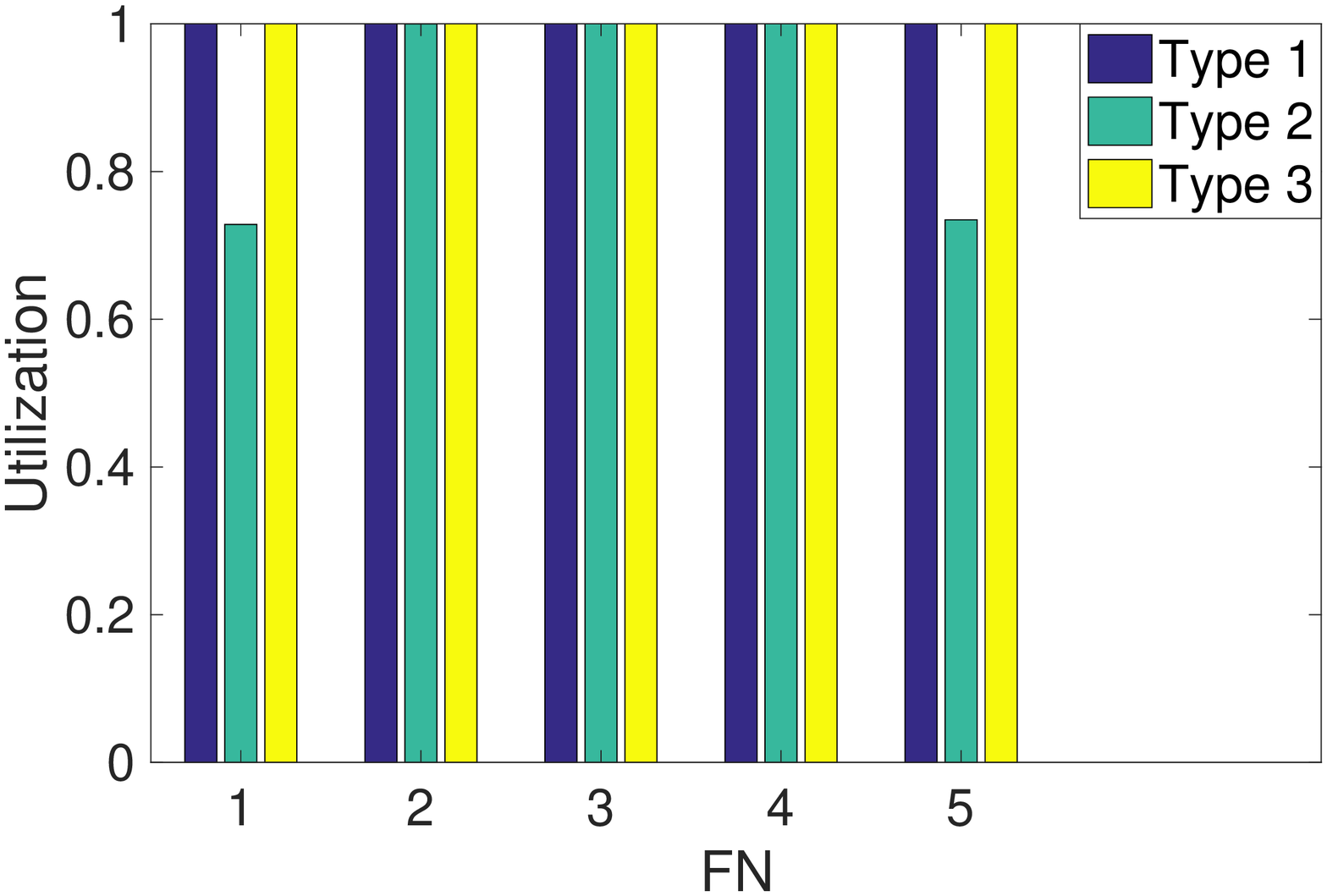}
	     \label{fig:ru2}
	}  \vspace{-0.2cm}
	\caption{Resource utilization (M = 40)}
\end{figure}

Figs.~\ref{fig:ru1}--\ref{fig:ru2} depict the utilization of different resource types
at several FNs. As we can see, the proposed scheme produces an ME with \textit{high resource utilization}. Additionally, at least one resource type at every FN is fully utilized at the equilibrium. Thus, the remaining resources cannot improve the utility of any buyer and the allocation is efficient and  non-wasteful. Note that non-fully utilized resources have zero prices at the equilibrium. 
Also, as the number of buyers N increases, the resource utilization tends to increase.

\begin{figure}[ht]
		\subfigure[Residual, M = 40, N = 8]{
		  \includegraphics[width=0.24\textwidth,height=0.10\textheight]{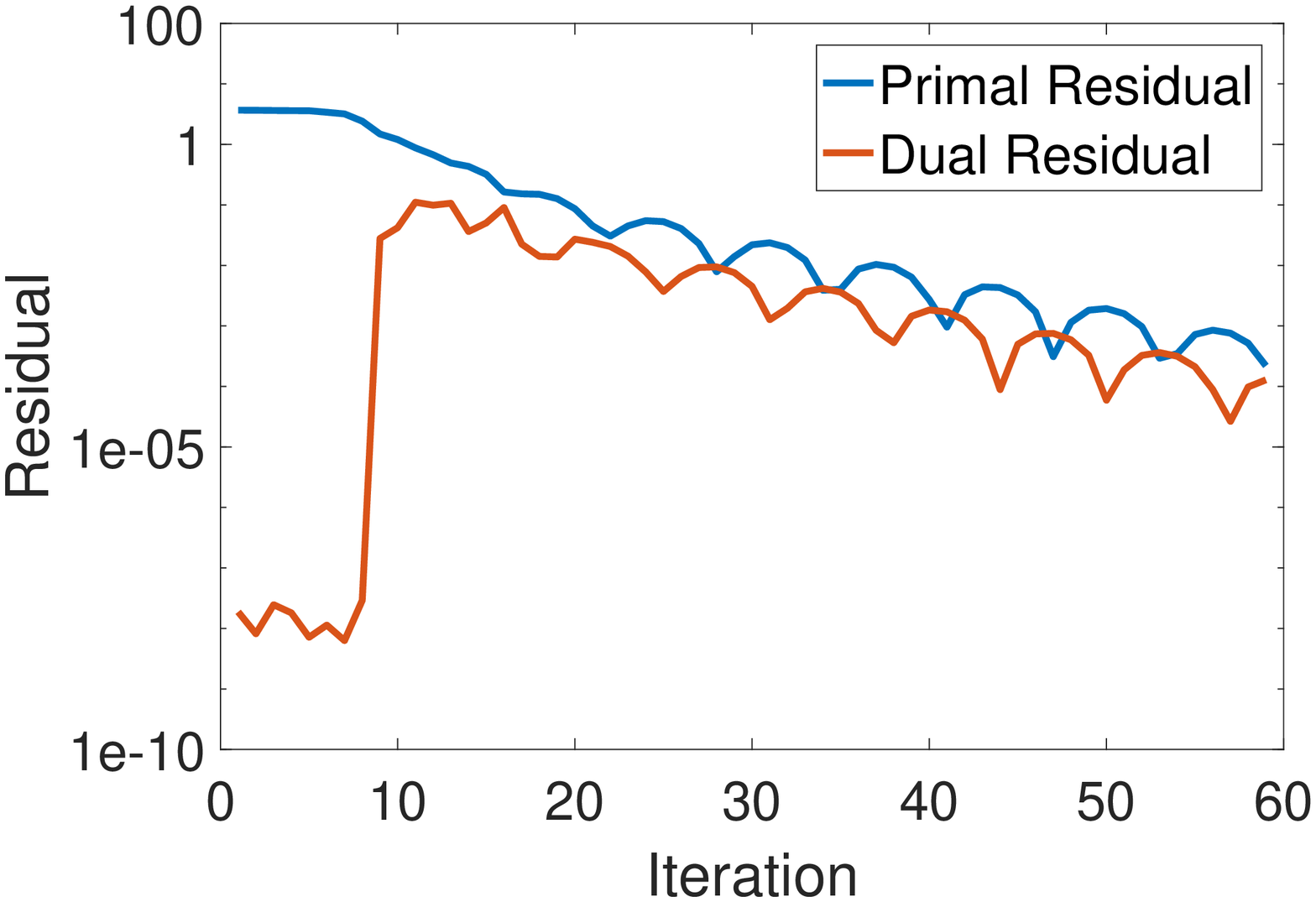}
	    \label{fig:capADMMr}
	}   \hspace*{-2.1em} 
		 \subfigure[Utility, M = 40, N = 8]{
	     \includegraphics[width=0.24\textwidth,height=0.10\textheight]{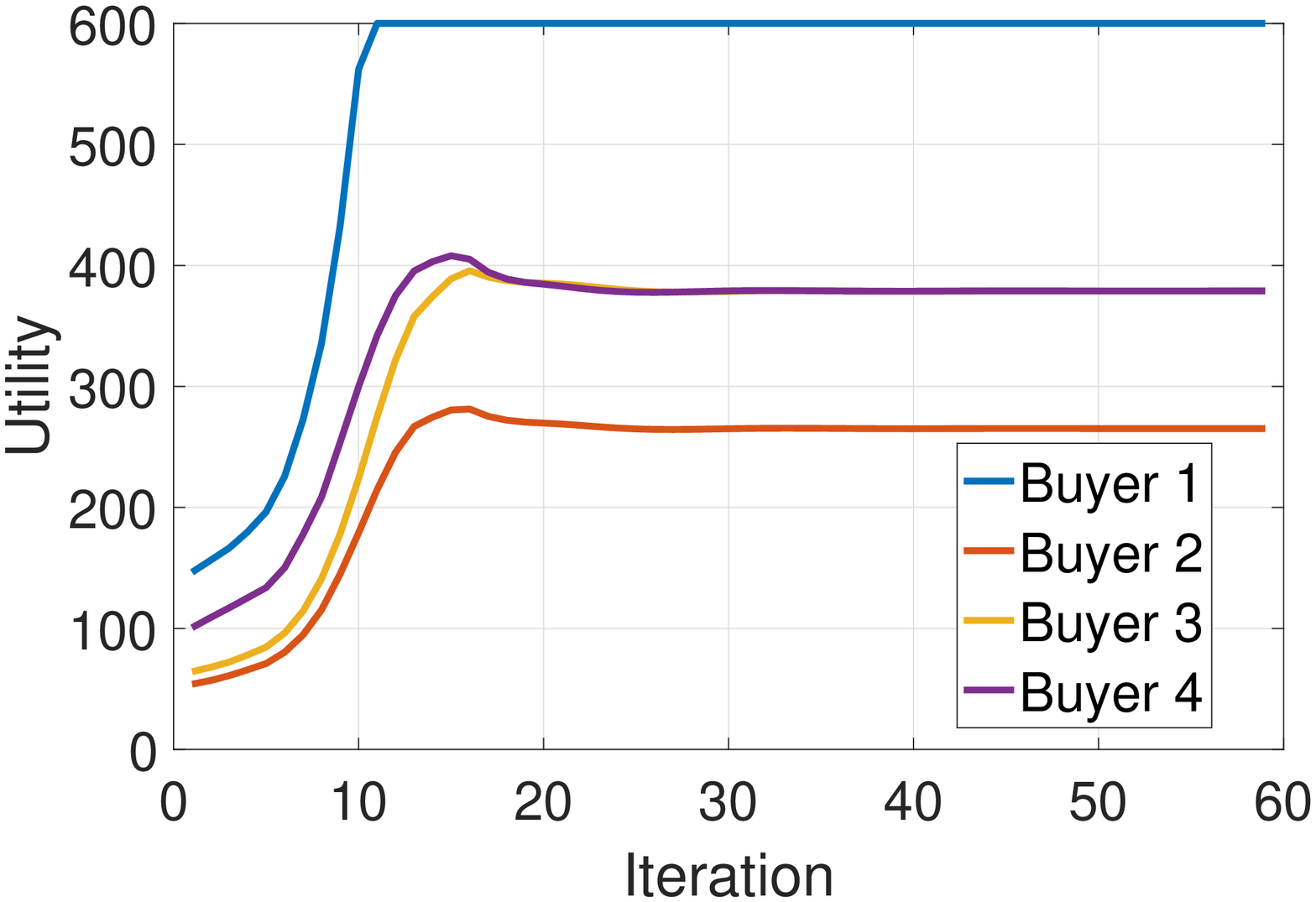}
	     \label{fig:capADMMu}
	} 
			\subfigure[Residual, M = 100, N = 20]{
		  \includegraphics[width=0.24\textwidth,height=0.10\textheight]{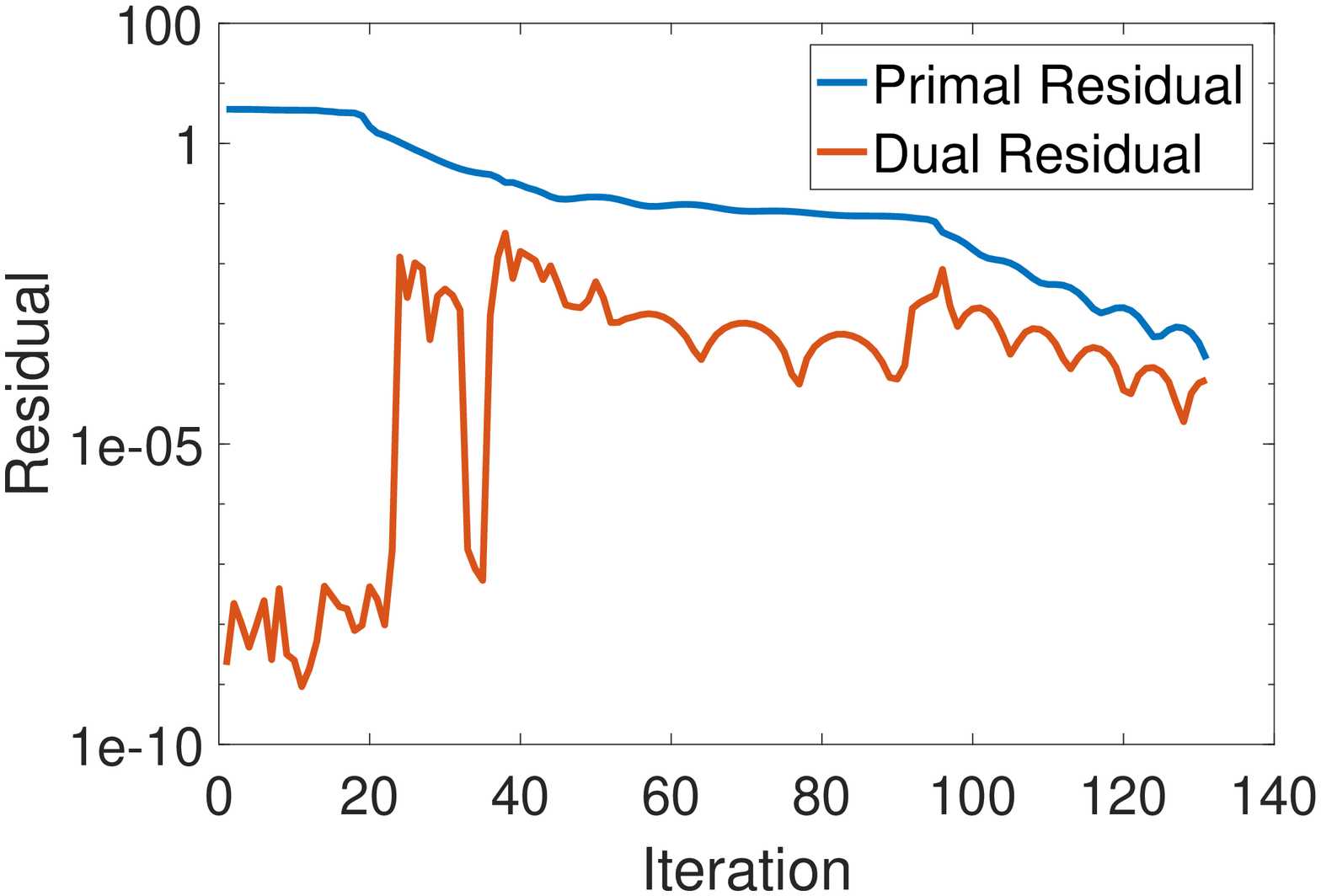}
	    \label{fig:capADMMr1}
	}   \hspace*{-2.1em} 
		 \subfigure[Utility, M = 100, N = 20]{
	     \includegraphics[width=0.24\textwidth,height=0.10\textheight]{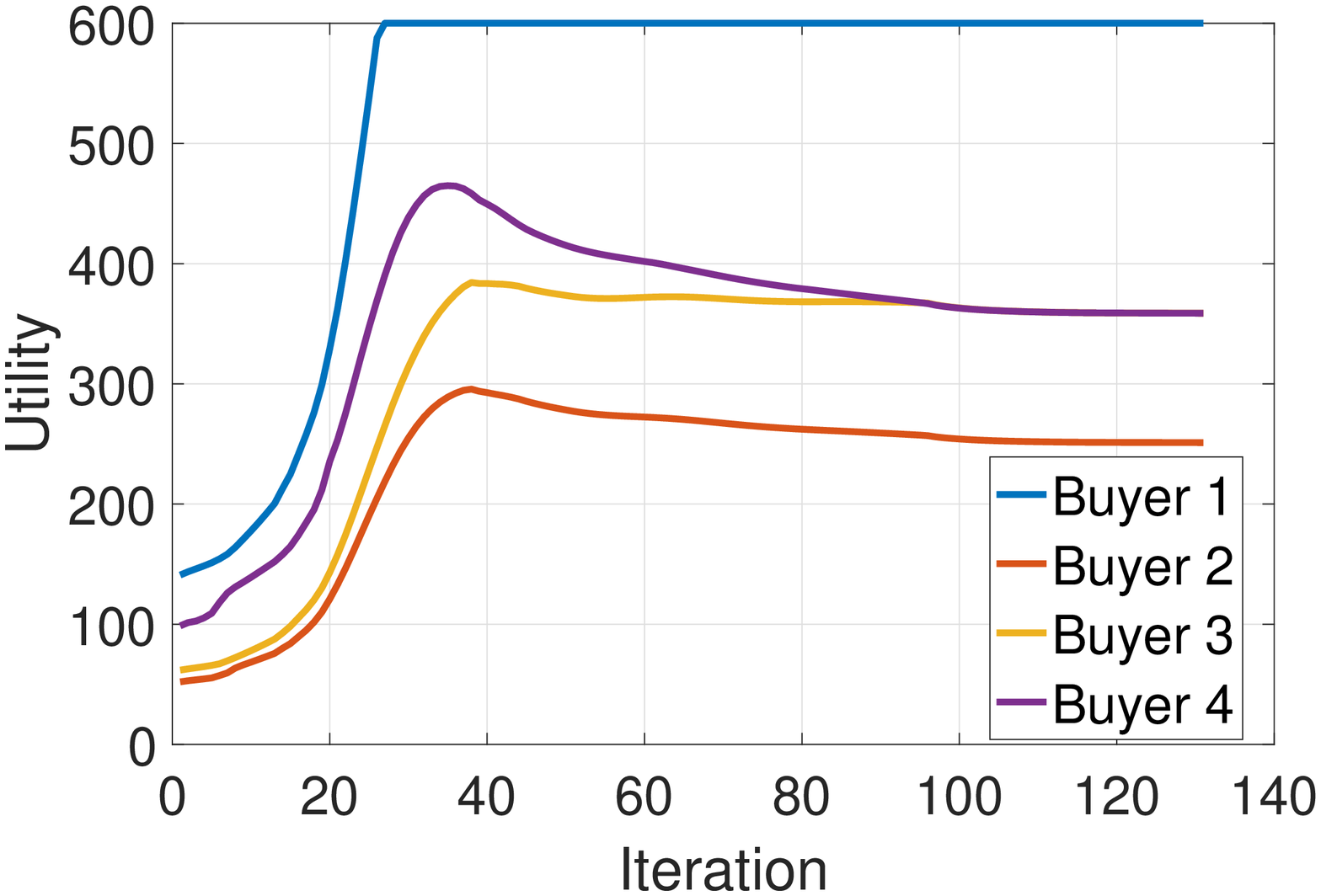}
	     \label{fig:capADMMu1}
	}  \vspace{-0.2cm}
		\caption{Convergence of the distributed algorithm}
\end{figure}

Finally, the convergence  of \textit{Algorithm 1} is illustrated in Figs.~\ref{fig:capADMMr}-\ref{fig:capADMMu1} for different system sizes. Here, we set the initial values of $x_{i,k}$ and $\overline{z}_k$ to be $\frac{1}{N}$ as in the proportional sharing scheme, and $p_k^1 = 1,~\forall k.$ As we can see, the algorithm converges to the optimal solution within a reasonable number of iterations, which implies the effectiveness and practicality of the proposed decentralized implementation. 
For instance, for a system with 100 FNs and 20 services, 
it takes about 130 iterations for convergence (less than 1s for each iteration).

\section{Related Work}
\label{related}

FC and related concepts like EC and cloudlet	have drawn a lot of attention recently.
In \cite{aces17}, A. Ceselli {\em et al.} present a comprehensive mobile edge-cloud  framework
considering various design factors including cloudlet placement, user mobility, and service-level agreement. In \cite{dzen16}, the authors 
jointly optimize task image placement and task scheduling in a fog network with dedicated storage and computing servers to minimize the task completion time. 
By exploiting the Priced Timed Petri Nets concept, 
\cite{lni17} proposes a dynamic resource allocation strategy that assists users to  
select suitable fog resources from a group of pre-allocated resources.
In \cite{rden16}, the authors formulate a workload allocation problem 
in a hybrid fog-cloud system, which aims to minimize the 
energy cost under latency constraints. The online primal-dual approach is employed in \cite{duong1} to examine the edge resource crowdsouring problem.
Another major line of research considers the joint allocation of communication and computational resources for task offloading in  wireless networks \cite{ymao17}.
Different from the existing literature, we study FC from the market design and algorithmic game theory perspectives \cite{AGT}, with a specific focus on \textit{pricing design and resource allocation in a multi-FN multi-buyer environment}.

Indeed, cloud resource allocation and pricing  have been studied extensively in the literature \cite{nluo17}. In \cite{hxu13}, 
a dynamic pricing scheme is introduced  to maximize the cloud provider's revenue. 
The resource and profit sharing problem among providers in a cloud federation is investigated in \cite{lmas152}. 
In \cite{dard17}, the interaction between cloud providers and multiple services is modeled as a generalized Nash game.  Reference \cite{vcar18} formulates  the single-cloud multi-service resource provisioning and pricing problem as a Stackelberg  game that aims to maximize the provider's revenue while minimizing the services' costs. 
Additionally, various auction models have been proposed to study cloud resource allocation \cite{jli17,mnej15,wshi162} 
with the goal of maximizing the social welfare or the cloud provider's profit. 
Typically, only winners in an auction receive  resources. 
Also, most of the existing auction models do not consider flexible demands \cite{jli17}. For instance, bidders are often assumed to be single-minded, who are only interested in a specific bundle and have zero value for other bundles.
Unlike the existing works on cloud economics and resource allocation, 
we examine a market with multiple nodes and \textit{budget-constrained} buyers. 
This model captures practical aspects such as requests can be served at different nodes and the buyers' demands can be defined flexibly. 
More importantly, the \textit{new ME solution concept}, which optimizes both the buyers' utilities and the resource utilization, is the salient feature of our work.

This work is also closely related to the MRA literature,  which has 
received considerable attention recently. The state-of-the-art research on MRA for data centers is surveyed in \cite{ppou17}. Noticeably, \textit{Dominant Resource Fairness} (DRF) \cite{agho11} is the most prominent approach to MRA fairness in data centers. In a nutshell, DRF is a generalization of max-min fairness for multiple resources, which aims to maximize the minimum dominant resource share across all the users. 
For the simplistic setting where all resources are pooled together in a single location and with some strong assumptions \cite{agho11}, DRF offers many compelling properties. 
This approach has been extended in the follow-up works \cite{cjoe13,wwan15,agut12,dpar15} to address more realistic assumptions such as user demands are finite and resources are distributed over multiple nodes. Both of these aspects are captured in our work.
To the best of our knowledge, this is the \textit{first comprehensive work that employs the Fisher market model and competitive equilibrium 
 to address the MRA} problem in a general setting with \textit{multiple nodes}, each of which has multiple resource types, and multiple users with \textit{finite demands}. While similar market-based approaches have been examined in \cite{agut12,agho13}, 
they consider simpler settings (e.g., a single resource pool, a single resource type, and/or infinite demands).

Finally, although numerous distributed optimization techniques \cite{boyd,admm, NUM,cche13,wden17} are available in the literature, most of them require agents to exchange and reveal their estimates explicitly to neighboring agents in every iteration to reach consensus on the final optimal solution. We are among the \textit{first  to propose a privacy-preserving parallel and distributed algorithm}, which converges to the optimal solution 
without asking the agents to reveal any private data. 
 Unlike differential privacy based methods \cite{cdwo14} that add carefully-designed noises to cover sensitive information and
are subject to a trade-off between privacy and accuracy, our proposed scheme enables privacy preservation without sacrificing
accuracy. 

Another approach to enable data privacy is to employ cryptographic techniques such as Garble Circuit (GC) \cite{ayao82} and Fully Homomorphic Encryption (HE) \cite{cgen09}, which are often used in secure multi-party computation \cite{pinkas}. However, HE is computationally expensive due to public key operations and GC has expensive communication costs.
Hence, these cryptographic approaches are not suitable for distributed optimization which typically needs many iterations to converge.
Indeed, \cite{czha18} is the only work that we are aware of studying privacy-preserving decentralized optimization. Different from their proposed  algorithm that based on ADMM and  partially homomorphic cryptography, \textit{our construction of the privacy-preserving} part is \textit{non-cryptographic} and simpler.

\section{Conclusion}
\label{concl}
In this work, we introduced a new solution approach for  allocating multiple resource types of different FNs to competing services with diverse priorities and preferences. The proposed solution 
 produces a non-wasteful and frugal ME that makes every service happy with the allocation decision while maximizing the resource utilization efficiency.
Also, we showed that the equilibrium has appealing fairness properties including envy-freeness,  sharing-incentive, and proportionality, which encourages the services to engage in the proposed scheme. Furthermore, a privacy-preserving distributed algorithm was developed to compute the ME while obfuscating the private information of the services, which significantly limits the strategic capabilities of the market participants. 
 The proposed framework has potential to be applied in other settings such as allocating virtual machines in data centers to cloud users and  allocating virtual network functions and middleboxes to different network slices \cite{slicing} (e.g., consider each network function at a network node as a resource type and each slice as a buyer).

\bibliographystyle{IEEEtran}

\begin{thebibliography}{19}


\bibitem{mchi16}
M. Chiang and T. Zhang, ``Fog and IoT: an overview of research opportunities,'' {\em IEEE Internet Things J.},  vol. 3, no. 6, pp. 854--864, Dec. 2016.


\bibitem{ymao17}
Y. Mao, C. You, J. Zhang, K. Huang, and K.B. Letaief, ``A survey on mobile edge computing: the communication perspective,'' {\em  IEEE Commun. Surv. Tut.}, vol. 19, no. 4, pp. 2322--2358, Fourthquarter 2017.

\bibitem{duong}
D.T. Nguyen, L.B. Le, and V. Bhargava, ``Price-based resource allocation for edge computing: a market equilibrium approach'', {\em IEEE Trans. Cloud Comput.}, to appear.



\bibitem{eco}
A. Mas-Colell, M.D. Whinston, and J. R. Green, ``Microeconomic Theory'', 1st ed. New York: Oxford Univ. Press, 1995.


\bibitem{AGT}
 N. Nisan, T. Roughgarden, E. Tardos, and V. Vazirani, ``Algorithmic Game Theory'', Cambridge, U.K.: Cambridge Univ. Press, 2007.

\bibitem{slicing}
S. Vassilaras et al., ``The algorithmic aspects of network slicing,'' {\em IEEE Commun. Mag.}, vol. 55, no. 8, pp. 112--119, 2017.


\bibitem{nluo17}
N.C. Luong, P. Wang, D. Niyato, Y. Wen, and Z. Han, ``Resource management in cloud networking using economic analysis and pricing models: a survey,'' {\em IEEE Commun. Surv. Tut.}, vol. 19, no. 2, pp. 954--1001, Secondquarter 2017.



\bibitem{karr54}
 K.J. Arrow and G. Debreu, ``Existence of equilibrium for a  competitive economy,'' {\em Econometrica}, vol. 22, no. 3, pp. 265--290, 1954.
 
 
 
\bibitem{ndev08}
N.R. Devanur, C.H. Papadimitriou, A. Saberi, and V.V. Vazirani, ``Market equilibrium via a primal--dual algorithm for a convex program,'' J. ACM vol. 55, no. 5, article 22, Nov. 2008.

\bibitem{vvaz11}
V.V. Vazirani and M. Yannakakis, ``Market equilibrium under separable, piecewise-linear, concave utilities,'' {\em J. ACM}, vol. 58, no. 3, article. 10, May 2011.


\bibitem{xche17}
X. Chen, D. Paparas, and M. Yannakakis, ``The complexity of non-monotone markets'', {\em J. ACM}, vol. 64, no. 3, Article 20, Jun. 2017.

\bibitem{jgag15}
J. Garg, R. Mehta, M. Sohoni, and V.V. Vazirani, ``A complementary pivot algorithm for market equilibrium under separable, piecewise-linear concave utilities'', {\em SIAM J. Comput.}, vol. 44, no. 6, pp. 1820--1847, 2015.

\bibitem{boyd}
S. Boyd and L. Vandenberghe, ``Convex Optimization'', Cambridge, U.K.: Cambridge Univ. Press, 2004.


\bibitem{NUM}
D.P. Palomar and M. Chiang, ``A tutorial on decomposition methods for network utility maximization,'' {\em IEEE J. Sel. Areas Commun.}, vol. 24, no. 8, pp. 1439--1451, Aug. 2006.

\bibitem{EG}
E. Eisenberg and D. Gale, ``Consensus of subjective probabilities: The pari-mutuel method,'' {\em Annals of Mathematical Statistics}, vol. 30, pp. 165--168, 1959.


\bibitem{agut12}
A. Gutman and N. Nisan, ``Fair allocation without trade'', in {\em Proc. AAMAS}, pp. 719--728, Valencia, Spain, Jun. 2012.


\bibitem{ppou17}
P. Poullie, T. Bocek, and B. Stiller, ``A survey of the state-of-the-art in fair multi-resource allocations for data centers,'' {\em IEEE Trans. Netw. Service Manag.}, vol. 15, no. 1, pp. 169--183, Mar. 2018.



\bibitem{hmou04}
 H. Moulin, ``Fair division and collective welfare,'' {\em MIT Press}, 2004.

\bibitem{agho11}
A. Ghodsi, M. Zaharia, B. Hindman, A. Konwinski, S. Shenker, and I. Stoica, ``Dominant resource fairness: fair allocation of multiple resource types,'' in {\em Proc. USENIX NSDI},  PP. 323--336, Boston, MA, 2011.


\bibitem{cjoe13}
C. Joe-Wong, S. Sen, T. Lan, and M. Chiang, ``Multiresource allocation: fairness–efficiency tradeoffs in a unifying framework,'' {\em IEEE/ACM Trans. Netw.}, vol. 21, no. 6, pp. 1785--1798, Dec. 2013.



\bibitem{wwan15}
W. Wang, B. Liang, and B. Li, ``Multi-resource fair allocation in heterogeneous cloud computing systems,'' {\em IEEE Trans. Parallel Distrib. Syst.}, vol. 26, no. 10, pp. 2822-2835, Oct. 1 2015.

\bibitem{dpar15}
D.C. Parkes, A.D. Procaccia, and N. Shah, ``Beyond dominant resource fairness: extensions, limitations, and indivisibilities,'' {\em ACM Trans. Econ. Comput.}, vol. 3, no. 1, Art. 3, Mar. 2015. 


\bibitem{ayao82}
A.C. Yao, ``Protocols for secure computations,'' in {\em Proc. FOCS}, pp. 160--164, 1982.


\bibitem{pinkas}
Y. Lindell and B. Pinkas, ``Secure multiparty computation for privacy-preserving data mining,'' {\em J. Privacy Confidentiality}, vol. 1, no. 1, pp. 59--98, 2009.



\bibitem{admm}
S. Boyd, N. Parikh, E. Chu, B. Peleato, and J. Eckstein, ``Distributed optimization and statistical learning via the alternating direction method of multipliers,'' {\em Foundations and Trends in Machine Learning}, 3(1):1–122, 2010.

\bibitem{cche13}
C. Chen, B.S. He, Y. Ye, and X. Yuan, ``The direct extension of admm for multi-block convex minimization problems is not necessarily convergent,'' {\em Mathematical Programming}, 2016.

\bibitem{wden17}
W. Deng, M. Lai, Z. Peng, and W. Yin,`` Parallel multi-block ADMM with o(1/k) convergence,'' {\em  J. Scientific Computing}, 2017.





\bibitem{cgen09}
C. Gentry, ``Fully homomorphic encryption using ideal lattices, {\em STOC}, vol. 9, pp. 169--178, 2009.


\bibitem{cdwo14}
C. Dwork and A. Roth, ``The algorithmic foundations of differential privacy,'' {\em Found. Trends Theor. Comput. Sci.}, vol. 9, no. 3--4, pp. 211--407, 2014.



\bibitem{ymo17}
Y. Mo and R.M. Murray, ``Privacy preserving average consensus,'' {\em IEEE Trans. Automatic Control}, vol. 62, no. 2, pp. 753--765, Feb. 2017.



\bibitem{aces17}
A. Ceselli, M. Premoli and S. Secci, ``Mobile edge cloud network design optimization,'' {\em IEEE/ACM Trans. Netw.}, vol. 25, no. 3, pp. 1818--1831, Jun. 2017.



\bibitem{dzen16}
D. Zeng, L. Gu, S. Guo, Z. Cheng, and S. Yu, ``Joint optimization of task scheduling and image placement in fog computing supported software-defined embedded system,'' {\em IEEE Trans. Comput.}, vol. 65, no. 12, pp. 3702--3712, Dec. 2016.


\bibitem{lni17}
L. Ni, J. Zhang, C. Jiang, C. Yan, and K. Yu, ``Resource allocation strategy in fog computing based on priced timed petri nets,''  {\em IEEE Internet  Things J.}, vol. 4, no. 5, pp. 1216--1228, Oct. 2017.

\bibitem{rden16}
R. Deng, R. Lu, C. Lai, T. H. Luan, and H. Liang, ``Optimal workload allocation in fog-cloud computing toward balanced delay and power consumption,'' {\em IEEE Internet Things J.}, vol. 3, no. 6, pp. 1171--1181, Dec. 2016.


\bibitem{duong1}
D.T. Nguyen, L.B. Le, and V. Bhargava, ``Edge computing resource procurement: an online optimization approach,'' in {\em Proc. IEEE WF-IoT}, pp. 807--812, Singapore, 2018.













\bibitem{hxu13}
H. Xu and B. Li, ``Dynamic cloud pricing for revenue maximization,'' {\em IEEE Trans. Cloud Comput.}, vol. 1, no. 2, pp. 158--171, Jul.--Dec. 2013.

 
\bibitem{lmas152}
L. Mashayekhy, M. M. Nejad, and D. Grosu, ``Cloud federations in the sky: formation game and mechanism,'' {\em  IEEE Trans. Cloud Comput.},  vol. 3, no. 1, pp. 14--27, Jan.-Mar. 2015.

\bibitem{dard17}
D. Ardagna, M. Ciavotta, and M. Passacantando, ``Generalized Nash equilibria for the service provisioning problem in multi-cloud systems,'' {\em IEEE Trans. Serv. Comput.}, vol. 10, no. 3, pp. 381--395, May-Jun. 2017.


\bibitem{vcar18}
V. Cardellini, V. Di Valerio, and F. Lo Presti, ``Game-theoretic resource pricing and provisioning strategies in cloud systems,'' {\em IEEE Trans. Serv. Comput.},  to appear.

\bibitem{mnej15}
M.M. Nejad, L. Mashayekhy, and D. Grosu, ``Truthful greedy mechanisms for dynamic virtual machine provisioning and allocation in clouds,'' {\em IEEE Trans. Parallel Distrib. Syst.}, vol. 26, no. 2, pp. 594--603, Feb. 2015.

\bibitem{jli17}
J. Li, Y. Zhu, J. Yu, C. Long, G. Xue, and S. Qian, ``Online auction for IaaS clouds: towards elastic user demands and weighted heterogeneous VMs,'' in {\em Proc. IEEE INFOCOM}, Atlanta, GA, 2017.


\bibitem{wshi162}
W. Shi, L. Zhang, C. Wu, Z. Li, and F.C.M. Lau, ``An online auction framework for dynamic resource provisioning in cloud computing,'' {\em IEEE/ACM Trans. Netw.}, vol. 24, no. 4, pp. 2060--2073, Aug. 2016.




\bibitem{agho13}
A. Ghodsi, M. Zaharia, S. Shenker, and I. Stoica, ``Choosy: max-min fair sharing for datacenter jobs with constraints,'' in {\em Proc. EuroSys},pp. 365--378, Prague, Czech Republic, Apr. 2013.





\bibitem{czha18}
C. Zhang and Y. Wang, ``Privacy-preserving decentralized optimization based on ADMM'', {\em arXiv:1707.04338}, 2017. 



\bibitem{code}
Code. Available [Online]: \url{https://github.com/duongtungnguyen/MEFOG}



\bibitem{bcod04} 
B. Codenotti and K. Varadarajan, ``Efficient computation of equilibrium prices for markets with Leontief utilities'', in {\em Proc. ICALP}, 2004.


\bibitem{bads10}
B. Adsul, C.S. Babu, J. Garg, R. Mehta, and M. Sohoni, ``Nash equilibria in fisher market,'' in {\em Proc. SAGT}, pp. 30--41, 2010.

\bibitem{sbra14}
S. Branzei, Y. Chen, X. Deng, A.F. Ratsikas, S.K.S. Frederiksen, and J. Zhang, ``The fisher market game: equilibrium and welfare,'' in {\em Proc. AAAI}, pp. 587--593, Quebec City, Quebec, Canada, Jul. 2014. 

\bibitem{nche16}
N. Chen, X. Deng, B. Tang, and H. Zhang, ``Incentives for strategic behavior in Fisher market games'', in {\em Proc. AAAI}, pp. 453--459, Phoenix, Arizona, USA, Feb. 2016.

\bibitem{sbra17}
S. Branzei, V. Gkatzelis, and R. Mehta, ``Nash social welfare approximation for strategic agents,'' in {\em Proc. ACM EC}, pp. 611--628, Cambridge, Massachusetts, USA, June 2017.



\bibitem{conv1}
Convex Optimization and Machine Learning 10-725, CMU, Fall 2018. \url{http://www.stat.cmu.edu/~ryantibs/convexopt/}

\bibitem{hvar74}
H.R. Varian, ``Equity, envy, and efficiency,'' {\em J. Economic Theory}, vol. 9, pp. 63--91, 1974.

\bibitem{nche11}
N. Chen, X. Deng, and J. Zhang, ``How profitable are
strategic behaviors in a market?'' in {\em Proc. the 19th
European Symposium on Algorithms (ESA)}, pp. 106--118, 2011.

\bibitem{nche12}
N. Chen, X. Deng, H. Zhang, and J. Zhang, ``Incentive ratios of fisher markets'', in {\em Proc. the 39th International Colloquium on Automata, Languages and Programming (ICALP)}, pp. 464--475, 2012.


\end{thebibliography}

\newpage

\appendix
\subsection{Proof of Proposition 3.1}
\label{prop1}
The utility function $u_i(x_i)$ of service $i$ with unlimited demand is given in  (\ref{nocap_u}). It is easy to verify the concavity of $u_i(x_i)$ by checking the definition of a concave function \cite{boyd}. Furthermore, for every $\alpha > 0$, we have
\beqn
u_i(\alpha x_i) =    \sum_j \min_r \frac{\alpha x_{i,j,r}}{a_{i,j,r}} = \alpha \sum_j \min_r \frac{ x_{i,j,r}}{a_{i,j,r}} = \alpha u_i(x_i),\nonumber \vspace{-0.3in}
\eeqn
which confirms $u_i(x_i)$ is homogeneous of degree one.

\subsection{Proof of Theorem 3.4}
\label{theorem}
Notice that the problem (\ref{EGmain})-(\ref{EGmain4})  always has an interior feasible solution by simply setting $x_{i,j,r}$ positive and sufficiently small, $\forall i, j, r$, so that all the constraints (\ref{EGmain2})-(\ref{EGmain4}) are satisfied
with strict inequalities. Hence, Slaters condition holds and the Karush--Kuhn--Tucker (KKT) conditions are necessary
and sufficient for optimality \cite{boyd}.
Define $\lambda_{i,j,r}, ~\mu_i,~ p_{j,r}$, and $\gamma_{i,j,r}$ as the dual variables associated with (\ref{EGmain1}), (\ref{EGmain2}), (\ref{EGmain3}), and (\ref{EGmain4}), respectively.
Consider the problem (\ref{EGmain})-(\ref{EGmain4}). The Lagragian is 
\beqn
&&L\big( \mathcal{X}, u, \lambda, \mu, p, \gamma \big) = \sum_i B_i \ln \sum_j u_{i,j}    \nonumber \\ \nonumber
&&+ \sum_i \mu_i \big(u_i^{\sf max} - \sum_j u_{i,j} \big) + \sum_{j,r} p_{j,r} \big(1 - \sum_i x_{i,j,r} \big)    \\  
&&+ \sum_{i,j,r} \lambda_{i,j,r} \big(x_{i,j,r} - u_{i,j} a_{i,j,r} \big) + \sum_{i,j,r} x_{i,j,r} \gamma_{i,j,r}.
\eeqn
The KKT conditions of the problem (\ref{EGmain})-(\ref{EGmain4}) include
\beqn
\label{kkt1}
\frac{\partial L}{\partial u_{i,j}} =  \frac{B_i}{\sum_j u_{i,j}} - \sum_r \lambda_{i,j,r} a_{i,j,r} - \mu_i = 0, ~\forall i,~j \\
\label{kkt2}
\frac{\partial L}{\partial x_{i,j,r}} = \lambda_{i,j,r} + \gamma_{i,j,r} - p_{j,r} = 0, ~\forall i,~j,~r \\ 
\label{kkt3}
u_{i,j} a_{i,j,r} = x_{i,j,r},~ \forall i,j,r;~\mu_i \big(u_i^{\sf max} - \sum_j u_{i,j} \big) = 0,~\forall i \\
\label{kkt4}
p_{j,r}\big(1 - \sum_i x_{i,j,r} \big) = 0,~\forall j,r; ~ x_{i,j,r} \gamma_{i,j,r} = 0,~\forall i,j,r \\
\label{kkt5}
\mu_i \geq 0,~ \forall i; p_{j,r} \geq 0,~\forall j,r; \gamma_{i,j,r} \geq 0,~\forall i,j,r.
\eeqn
and primal feasibility conditions.\\
Since $u_i = \sum_j u_{i,j}$, from (\ref{kkt1}), we have
\beqn
\label{apkkt6}
\forall i,~j: \frac{B_i}{u_i} - \sum_r \lambda_{i,j,r} a_{i,j,r} - \mu_i = 0.
\eeqn
From (\ref{kkt2}), we have: 
\beqn
\label{akkt1}
\lambda_{i,j,r} = p_{j,r} - \gamma_{i,j,r}, ~\forall i,~j,~r.
\eeqn
Since $\gamma_{i,j,r} \geq 0,~\forall i,j,r$, we have
\beqn
\label{apkkt7}
\forall i,j,r: \lambda_{i,j,r} \leq p_{j,r}.
\eeqn
Also, from the second equality in (\ref{kkt4}), if $x_{i,j,r}  > 0 $, then $\gamma_{i,j,r} = 0$. Combined with (\ref{akkt1}), we have
\beqn
\label{apkkt8}
\forall i,j,r: \text{if} ~x_{i,j,r} > 0  \Rightarrow~ \lambda_{i,j,r} = p_{j,r} 
\eeqn
The first equality in (\ref{kkt4}) implies
\beqn
\label{apkkt9}
\forall j,r: \text{if}~ p_{j,r} > 0 \Rightarrow ~ \sum_i x_{i,j,r} = 1 \\
\label{kkt10}
\forall j,r: \text{if}~ \sum_i x_{i,j,r} < 1 \Rightarrow ~p_{j,r} = 0 
\eeqn
Finally, from the second equality in (\ref{kkt3}), we have
\beqn
\label{apkkt11}
\forall i: \text{if}~ \mu_i > 0 \Rightarrow~ u_i = \sum_j u_{i,j} = u_i^{\sf max}.
\eeqn
In summary, from the KKT conditions, we can infer
\beqn
\label{kkt6}
\forall i,~j: \frac{B_i}{u_i} - \sum_r \lambda_{i,j,r} a_{i,j,r} - \mu_i = 0 \\
\label{kkt7}
\forall i,j,r: \lambda_{i,j,r} \leq p_{j,r} \\
\label{kkt8}
\forall i,j,r: \text{if} ~x_{i,j,r} > 0  \Rightarrow \lambda_{i,j,r} = p_{j,r} \\
\label{kkt9}
\forall j,r: p_{j,r} \Big( \sum_i x_{i,j,r} -  1 \Big) = 0 \\
\label{kkt11}
\forall i: \text{if}~ \mu_i > 0 \Rightarrow u_i = \sum_j u_{i,j} = u_i^{\sf max}.
\eeqn

Indeed, \textit{(\ref{kkt9})  is exactly the market clearing condition}. Also, from (\ref{kkt9}), 
if $ p_{j,r} > 0$, then $\sum_i x_{i,j,r} = 1,~\forall j,r$, which means all 
resources with positive prices are fully allocated. Additionally,
if $ \sum_i x_{i,j,r} < 1$, then $p_{j,r} = 0, ~\forall j, r$, which means all non-fully allocated resources have zero prices.
 From (\ref{kkt6})-(\ref{kkt8}) and the definition of $q_{i,j}$ in (\ref{bprice}), we have: 
\beqn
\label{kkt61}
\forall i,~j : q_{i,j} = \sum_r p_{j,r} a_{i,j,r} \geq \sum_r \lambda_{i,j,r} a_{i,j,r} = \frac{B_i}{u_i} - \mu_i \\
\label{kkt81}
\forall i,~j: \text{if} ~x_{i,j,r} > 0 \Rightarrow  q_{i,j} = \sum_r p_{j,r} a_{i,j,r}  = \frac{B_i}{u_i} - \mu_i.
\eeqn
The inequality in (\ref{kkt61}) is due to (\ref{kkt7}). The last equality in (\ref{kkt61}) is from (\ref{kkt6}).
From  (\ref{kkt8}), if service $i$ buys resource type $r$ at FN $j$ (i.e., $x_{i,j,r} > 0$), then 
$\lambda_{i,j,r} = p_{j,r},~\forall i,j,r.$ Thus, from (\ref{kkt8}) and (\ref{kkt61}), we can obtain (\ref{kkt81}).
Denote $q_i^{\sf min} = \frac{B_i}{u_i} - \mu_i,~\forall i$. From (\ref{kkt61}) and (\ref{kkt81}), we have $q_{i,j} \geq q_i^{\sf min},~\forall i,j$, and if $x_{i,j,r} > 0$, then $q_{i,j} = q_i^{\sf min}$. Hence, the \textit{services buy resources only from the cheapest FNs}. 
Furthermore, from (\ref{kkt6}), we have:
\beqn
\label{kkt12}
B_i  - \mu_i ~ u_i &=& \Big( \sum_r \lambda_{i,j,r} ~a_{i,j,r} \Big)~ u_i  \nonumber \\
 &=& \sum_r \lambda_{i,j,r} ~a_{i,j,r} ~ \sum_j u_{i,j}  \nonumber \\  
&=& \sum_j  \sum_r \lambda_{i,j,r} \Big( a_{i,j,r} ~u_{i,j} \Big) \nonumber \\
&=&  \sum_j \sum_r \lambda_{i,j,r} ~x_{i,j,r} \nonumber  \\ 
&=& \sum_j \sum_r p_{j,r} ~x_{i,j,r}.
\eeqn

The fourth equality in (\ref{kkt12}) is from the feasibility constraint (\ref{EGmain1}). 
The last equality in (\ref{kkt12}) is due to (\ref{kkt8}) and $x_{i,j,r} \geq 0, ~\forall i,~j,~r.$
Thus, from (\ref{kkt12}), we have:
\beqn
\label{kkt121}
 \mu_i ~ u_i = B_i -  \sum_j \sum_r p_{j,r} ~x_{i,j,r}, \quad \forall i.
\eeqn
Since $ \sum_j \sum_r p_{j,r} x_{i,j,r}$ is the total money spent by service $i$ for procuring fog resources, $\mu_i u_i$ can be inferred as the budget surplus of service $i$ after purchasing the  resources. Consequently, if $\mu_i = 0$, we have $B_i = \sum_j \sum_r p_{j,r} x_{i,j,r}.$ In other words, if $\mu_i = 0$, service $i$ spends all of its budget to buy fog resources.
Also, from (\ref{kkt11}), if $\mu_i > 0$, then $u_i = u_i^{\sf max}$. 
Hence, \textit{a service either spends all money at the equilibrium or reaches its utility limit}.
Furthermore, we showed that the \textit{services buy resources from the cheapest FNs only}. Additionally, from (\ref{kkt121}), the services do not overspend since $\mu_i$, $u_i \geq 0, ~\forall i$. 
Therefore, the optimal solution $X$ to the 	problem (\ref{EGmain})-(\ref{EGmain4}) maximizes the utility of every service under the budget constraint. 

We have shown that all the conditions of a non-wasteful and frugal ME can be inferred from the KKT conditions of  (\ref{EGmain})-(\ref{EGmain4}). As a result, the optimal solutions to  (\ref{EGmain})-(\ref{EGmain4}) are indeed non-wasteful and frugal market equilibria.

 \textit{Now, we show that the utilities are unique across all such equilibria}. 
Obviously, the problem (\ref{EGmain})-(\ref{EGmain4}) is equivalent to
\beqn
\label{obj2}
\underset{X,~u}{\text{maximize}} &&\quad \sum_i B_i \ln u_i \\ \nonumber
\text{subject to} &&\quad u_i = \sum_j u_{i,j},~\forall i; ~~ (\ref{EGmain1})-(\ref{EGmain4}).
\eeqn
Since this problem has a strictly concave objective function, 
it has a unique optimal utility vector $u^*$. 

 \textit{Finally, we show that any non-wasteful and frugal ME is an optimal solution to the problem  (\ref{EGmain})-(\ref{EGmain4}).}  
Let $(p,X)$ be a non-wasteful and frugal ME in our resource allocation problem with the service utility function given in (\ref{cap_u}). 
By definition, $x_i$ is an optimal solution of the service maximization problem (\ref{opt1})-(\ref{opt14}) at the price vector $p$. 
We will show that $X$ is an optimal solution to the problem (\ref{EGmain})-(\ref{EGmain4}). Specifically, we prove that $X$ is feasible (i.e., satisfying (\ref{EGmain1})-(\ref{EGmain4})) and $(p, X)$ satisfies the KKT conditions (\ref{kkt6})-(\ref{kkt11}). 

Since the $X$ is non-wasteful, 
it does not over-allocate any fog resource, i.e., $\sum_i x_{i,j,r} \leq 1,~\forall j,r$. Also, from (\ref{NWcon}),  we have: $\sum_j \min_r \frac{x_{i,j,r}}{a_{i,j,r}} \leq u_i^{max},~\forall i$. 
Because a non-wasteful allocation allocates resources of an FN to a service proportional to the service's base demand vector, we have: $\frac{x_{i,j,r}}{a_{i,j,r}} = \frac{x_{i,j,r'}}{a_{i,j,r'}},~\forall r,~r'$. Hence, we can define $u_{i,j} = \min_r \frac{x_{i,j,r}}{a_{i,j,r}} =  \frac{x_{i,j,r}}{a_{i,j,r}},~\forall i,j,r$. Then,  $\sum_j u_{i,j} \leq u_i^{max},~\forall i$.
Thus, $X$ is a feasible solution to the problem (\ref{EGmain})-(\ref{EGmain4}).

Since the total money $\sum_j \sum_r p_{j,r} x_{i,j,r}$ spent by service $i$ is constrained by its budget (i.e., constraint (\ref{opt13})), we have:
\beqn
B_i &\geq& \sum_j \sum_r p_{j,r} x_{i,j,r} ~=~ \sum_j \sum_r p_{j,r} \Big( a_{i,j,r} \frac{x_{i,j,r}}{a_{i,j,r}} \Big)  \\
&=& \sum_j  \sum_r p_{j,r} a_{i,j,r}  u_{i,j}  ~=~ \sum_j  u_{i,j} \Big( \sum_r p_{j,r} a_{i,j,r}  \Big).            \nonumber
\eeqn
Note that $u_{i,j} = \frac{x_{i,j,r}}{a_{i,j,r}},~\forall i,j,r$. Also, the frugality condition imposes: $x_{i,j} > 0$  only if FN $j$ satisfies $\sum_r p_{j,r} a_{i,j,r} = q_i^{\sf min} $. If $\sum_r p_{j,r} a_{i,j,r} > q_i^{\sf min} $, then $x_{i,j} = 0$ and $u_{i,j} = 0$. Hence:
\beqn
B_i \geq \sum_j  u_{i,j} \Big( \sum_r p_{j,r} a_{i,j,r}  \Big)   =  \sum_j u_{i,j} q_i^{\sf min} = u_i q_i^{\sf min}.
\eeqn

Let $\mu_i = \frac{B_i}{u_i} -  q_i^{\sf min}$. Then, $\mu_i \geq 0,~\forall i.$
Also, we have $q_{i,j} = \sum_r p_{j,r} x_{i,j,r} \geq q_i^{\sf min} =  \frac{B_i}{u_i} - \mu_i,~\forall i,~j.$ Again, by the frugality property, if $x_{i,j,r} > 0$, then $q_{i,j} = \sum_r p_{j,r} x_{i,j,r} = q_i^{\sf min} =  \frac{B_i}{u_i} - \mu_i$. Hence, (\ref{kkt61}) and (\ref{kkt81}) hold. Consequently, by properly setting $\lambda_{i,j,r}$, we obtain  (\ref{kkt6})-(\ref{kkt8}).

By \textit{Definition \ref{MEdef}}, prices $p$ are  non-negative. Also, if $p_{j, r} > 0 $, resource type $r$ at FN $j$ is fully allocated (i.e., $\sum_{i} x_{i,j,r} = 1$). Thus, $(\sum_{i} x_{i,j,r} - 1)~p_{j,r} = 0, ~\forall i, j, r$, which is exactly  (\ref{kkt9}). The final step is to show (\ref{kkt11}).

Assume $\mu_i > 0.$ Then, $\frac{B_i}{u_i} -  q_i^{\sf min} = \mu_i > 0$. Hence
\beqn
 B_i &>& q_i^{\sf min} u_i ~=~ q_i^{\sf min} \sum_j u_{i,j} ~=~ \sum_j q_{i,j} u_{i,j} \\ \nonumber 
&=& \sum_j \Big(\sum_r p_{i,j,r} a_{i,j,r} \Big) u_{i,j} ~=~ \sum_j \sum_r p_{i,j} x_{i,j,r},
\eeqn
where the last equality is because $u_{i,j} = \frac{x_{i,j,r}}{a_{i,j,r}},~\forall i,j,r$. The second equality is from the frugality property of $X$, which imposes: $u_{i,j} = 0$ if $q_{i,j} > q_i^{\sf min}$; and $u_{i,j} > 0$ only if $q_{i,j} = q_i^{\sf min}$. 
Thus, if $\mu_i > 0$, the total money spent by service $i$ is \textit{strictly} less than its budget $B_i$. 
Define $\phi_i = \frac{B_i}{ \sum_j \sum_r q_{i,j} x_{i,j,r}} > 1$. 
We will prove (\ref{kkt11}) by contradiction. Assume $u_i = \sum_j u_{i,j} < u_i^{\sf max}$. Since bundle ($\phi_i x_i$) is affordable to service $i$, 
the service  can strictly improve its utility by purchasing  bundle ($\phi_i x_i$) because, obviously, $u_i(\phi_i x_i) > u_i(x_i)$. Thus, $X$ does not satisfy the service satisfaction condition, which contradicts to the assumption that $(p,X)$ is an ME. Therefore, if $\mu_i > 0$, then $u_i = \sum_j u_{i,j} = u_i^{\sf max}, ~\forall i$, which is exactly (\ref{kkt11}). 

\subsection{Proof of Theorem 3.6}
\label{appfair}
- \textit{Envy-freeness}: 
To prove that the allocation $X^*$ is envy-free, we need to show that $u_i(x_i^*) \geq u_i(\frac{B_i}{B_{i'}} x_{i'}^*), ~\forall i,~i'$.

Since $x_{i'}^*$ is the optimal resource bundle of service $i'$ at the equilibrium, service $i'$ can afford to buy bundle $x_{i'}^*$ at prices $p^*$. Hence, we have: 
 $\sum_j \sum_r p_{j,r}^* x_{i',j,r}^* \leq  B_{i'}$. Thus:
\beqn
 \sum_j \sum_r p_{j,r}^* \Big( \frac{B_i}{B_{i'}} x_{i',j,r}^* \Big) \leq  B_i, \quad \forall i.
\eeqn 

Therefore, bundle $\Big( \frac{B_i}{B_{i'}} x_{i'}^* \Big) $ is affordable to service $i$ at the equilibrium prices $p^*$.
However, $x_i^*$ is the favorite bundle of service $i$ at the equilibrium. 
Hence, we have:  $u_i(x_i^*) \geq u_i(\frac{B_i}{B_{i'}} x_{i'}^*), ~\forall i,~i'$. 

- \textit{Sharing-incentive:} At the equilibrium, no service spends more than its budget, i.e.,
 $\sum_j \sum_r p_{j,r}^* x_{i,j,r}^* \leq  B_i, ~\forall i$. Hence,
\beqn
 \sum_i \sum_j \sum_r p_{j,r}^* x_{i,j,r}^* \leq  \sum_i B_i. \nonumber
\eeqn 
In other words, we have $ \sum_j \sum_r p_{j,r}^* \sum_i x_{i,j,r}^* \leq  \sum_i B_i$.
As shown before, if $\sum_i x_{i,j,r}^* < 1$, then $p_{j,r}^* = 0$. Also, if $p_{j,r}^* > 0$, then $\sum_i x_{i,j,r}^* = 1$. Therefore, we have 
\beqn
\sum_j \sum_r p_{j,r}^* =  \sum_j \sum_r p_{j,r}^* \sum_i x_{i,j,r}^* \leq  \sum_i B_i. \nonumber
\eeqn 
Consequently, the resource bundle $\hat{x}_i$ costs service $i$: 
\beqn
&& \sum_j \sum_r \hat{x}_{i,j,r} p_{j,r}^* = \sum_j \sum_r \frac{B_i}{\sum_i B_{i}} p_{j,r}^* \nonumber \\
&&= \frac{B_i}{\sum_i B_{i}} \sum_j \sum_r  p_{j,r}^* \leq  \frac{B_i}{\sum_i B_{i}} \sum_i B_i = B_i, ~\forall i. \nonumber
\eeqn
Therefore, service $i$ can afford to buy bundle $\hat{x}_i$ at prices $p^*$. 
However, given the set of affordable (i.e., feasible) resource bundles to service $i$, her favorite (i.e., utility-maximizing) one is $x_i^*$.
Thus, $u_i(x_i^*) \geq u_i(\hat{x}_i), ~\forall i.$\\ \\
- \textit{Proportionality:} 
By definition
\beqn
\frac{B_i}{\sum_{i'} B_{i'}} u_i(\mathcal{C}) = \frac{B_i}{\sum_{i'} B_{i'}} \min \Big\{  \sum_j \min_r \frac{1}{a_{i,j,r}}, u_i^{\sf max}  \Big\} \nonumber \\ \nonumber
=  \min \Big\{  \sum_j \min_r \frac{  \frac{B_i}{\sum_{i'} B_{i'}}   }{a_{i,j,r}}, \frac{B_i}{\sum_{i'} B_{i'}}u_i^{\sf max}  \Big\}. \\ \nonumber
\eeqn
From the sharing-incentive property, we have
\beqn
u_i(x_i^*) &\geq& u_i\big(\frac{B_i}{\sum_{i'}B_{i'}} \mathcal{C}\big) \nonumber \\ \nonumber 
&=& \min \Big\{  \sum_j \min_r \frac{  \frac{B_i}{\sum_{i'} B_{i'}} }{a_{i,j,r}}, u_i^{\sf max}  \Big\} \\ \nonumber
&\geq&  \min \Big\{  \sum_j \min_r \frac{  \frac{B_i}{\sum_{i'} B_{i'}}   }{a_{i,j,r}}, \frac{B_i}{\sum_{i'} B_{i'}}u_i^{\sf max}  \Big\}. \\ \nonumber
\eeqn

Hence, $u_i(x_i^*) \geq \frac{B_i}{\sum_{i'} B_{i'}} u_i(\mathcal{C})$. Consequently, the equilibrium allocation satisfies the proportionality property.

\subsection{Derivation of Algorithm 1}
\label{admm_apen}
The augmented Lagrangian of the problem (\ref{admmsharing1})-(\ref{admmeq6}) is
\beqn
L(x,z,p) = \sum_i f_i(x_i) + g(\displaystyle \sum_i z_i  ) + \\ \nonumber
 \sum_i \Big( p_i^T (x_i - z_i ) + (\rho/2)||x_i - z_i ||_2^2 \Big). \nonumber
\eeqn 
The ADMM updates are
\beqn
&&x_i^{t+1} := \argmin_{x_i} \Big(f_i(x_i) +  (p_i^t)^T x_i +   (\rho/2)||x_i - z_i^t ||_2^2    \Big) \nonumber \\ \nonumber
&&z_i^{t+1} := \argmin_{z} \Big(  g(\displaystyle \sum_i z_i  ) - \sum_i (p_i^t)^T z_i   \\ 
&&\quad \quad \quad \quad \quad \quad+ \sum_i  (\rho/2)||z_i - x_i^{t+1} ||_2^2  \Big)  \\  \nonumber
\label{admm1}
&&p_i^{t+1} := p_i^t + \rho(x_i^{t+1} - z_i^{t+1}),
\eeqn
where $t$ is the iteration index and $z = \big(z_1, z_2, \ldots, z_N \big)$. Define $w_i = p_i/\rho$, which is called scaled dual variable, we have the following scaled form of ADMM 
\beqn
&&x_i^{t+1} := \argmin_{x_i} \Big(f_i(x_i) +  (\rho/2)||x_i - z_i^t + w_i^t||_2^2    \Big) \nonumber \\ 
&&z_i^{t+1} := \argmin_{z} \Big(  g(\displaystyle \sum_i z_i  ) \nonumber \\ 
&& \quad \quad \quad \quad \quad +  (\rho/2) \sum_i ||z_i - x_i^{t+1} - w_i^t ||_2^2 \Big) \nonumber \\ 
\label{admm2}
&&w_i^{t+1} := w_i^t + x_i^{t+1} - z_i^{t+1}. 
\eeqn
It is easy to see that the update steps of $x_i$ and $w_i$ (or $p_i$) can be carried out independently in parallel for
each user $i$. 

We can further reduce the number of variables and also simplify the update step of variable $z$ as follows.
Denote $\overline{z} = (1/N)\sum_i z_i$. We have $\overline{z} \in \mathbb{R}^K$. The z-update step in (\ref{admm2}) can be expressed as
\beqn
\label{zave}
\underset{z_1, z_2, \ldots, z_N}{\text{minimize}} &&   g(N \overline{z} ) +  (\rho/2) \sum_i ||z_i - x_i^{t+1} - w_i^t ||_2^2 \nonumber  \\ 
\label{zaveeq}
\text{subject to}  && \overline{z} = (1/N)\sum_i z_i.
\eeqn
For a given $\overline{z}$, the optimal solution of problem (\ref{zaveeq}) is
\beqn
z_i = x_i^{t+1} + w_i^t + \overline{z} - (\overline{x}^{t+1} + \overline{w}^t),~\forall i.
\eeqn
Hence, problem (\ref{zave}) can be rewritten as
\beqn
\label{zave2}
\underset{\overline{z}}{\text{minimize}} &&   g(N \overline{z} ) +  (\rho/2) \sum_i ||\overline{z} - \overline{x}^{t+1} -\overline{w}^t ||_2^2 
\eeqn
From (\ref{admm2}), we also have
\beqn
w_i^{t+1} = \overline{w}^t + \overline{x}^{t+1} - \overline{z}^{t+1},
\eeqn
which implies that the scaled dual variables $w_i$ are all equal. Denote $w = w_i, ~\forall i.$ 
The scaled form of ADDM becomes
\beqn
&&x_i^{t+1} := \argmin_{x_i} \Big(f_i(x_i)  \nonumber  \\ \nonumber 
&&\quad \quad \quad \quad \quad \quad +  (\rho/2)||x_i - x_i^t + \overline{x}^t -\overline{z}^t + w^t||_2^2    \Big) \nonumber \\ 
&&\overline{z}^{t+1} := \argmin_{\overline{z}} \Big(  g(\displaystyle N \overline{z}) +  (N \rho/2) ||\overline{z} - \overline{x}^{t+1} - w^t ||_2^2 \Big) \nonumber \\ 
\label{admm3}
&&w^{t+1} := w^t + \overline{x}^{t+1} - \overline{z}^{t+1}. 
\eeqn

From the definition of $f_i(x_i)$, we can write
\beqn
&&x_i^{t+1} := \argmin_{x_i} \Big(v_i(x_i)  +  (\rho/2)||x_i - x_i^t + \overline{x}^t -\overline{z}^t + w^t||_2^2    \Big) \nonumber \\ \nonumber
&& \text{subject to} \quad x_i \in \mathcal{X}_i.
\eeqn

By replacing $w$ with $\frac{1}{\rho} p$, we obtain \textbf{Algorithm 1}. 

\subsection{Computational Complexity of the Centralized Solution}

In the centralized solution, we need to solve the convex program (\ref{EGmain})-(\ref{EGmain4}).  
Indeed, convex optimization problems can be solved effectively by efficient techniques such as Interior Point Methods \cite{boyd}. For example, the optimality gap of the solution computed by the barrier method, which is an Interior Point Method,  after $k$ centering steps is smaller or equal to 
$m/(\mu^k t^{(0)})$ where $m$ is the number of inequality constraints and  $\mu$ and $t^{(0)}$ are constants (see \cite{conv1} for more details about convergence analysis, especially page 18 in the barrier method lecture).  In other words, to reach a desired accuracy level of $\epsilon$, it requires 
$log(m/(t^{(0)} \epsilon))/log \mu$ 
centering steps. Note that the Newton's method (see  \cite{boyd} or the Newton's method lecture in \cite{conv1})  is used in each centering step.

\begin{figure}[ht!]
	\centering
		\includegraphics[width=0.48\textwidth,height=0.19\textheight]{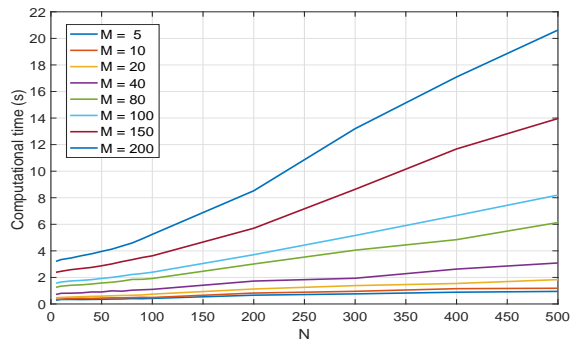} 
			\caption{Computational time with varying system size}
	\label{fig:convext}
\end{figure}

Note that there exist many efficient convex optimization solvers. In this work, we used CVX  with MOSEK solver option \cite{code} to solve the formulated convex optimization problem.
 The computational time under different system sizes (all with three resource types) is reported in Fig.~\ref{fig:convext}. For each system size, we run 10 simulations with different generated data and take the average computational time. As we can observe, the optimal solution can be computed quite efficiently. For example, for a system with 200 buyers (N = 200) and 100 FNs (M = 100), it takes less than 4 seconds to obtain the optimal solution. For moderate system sizes, the optimal solution can be found in less than 1 second.
 
\subsection{ Example with Two Buyers and One Single-resource FN}

\begin{figure}[h!]
	\centering
		\includegraphics[width=0.35\textwidth,height=0.25\textheight]{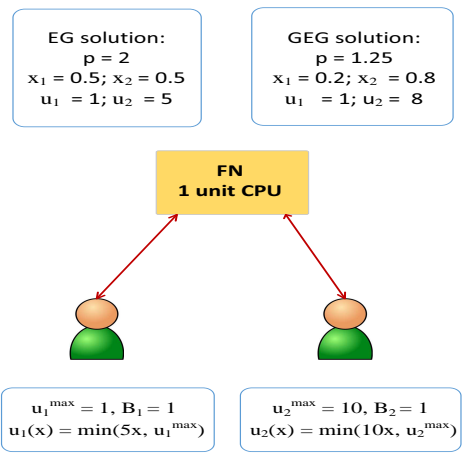} 
			\caption{Toy example (one single-resource FN, two services)}
	\label{fig:toy1}
\end{figure}

In addition to the example presented in the paper (see Fig.~\ref{fig:EGGEG})), here we present a simple example  with two buyers and a single-resource FN (e.g., consider computing resource only) to illustrate the wasteful equilibrium allocation phenomenon. 
The example is summarized in Fig.~\ref{fig:toy1}.
In particular, the budgets of the buyers are  $B_1 = B_2 = 1.$ The utility limits of the buyers are $u_1^{max} = 1$ and $u_2^{max} = 10$. Additionally, $a_1 = 0.2$ and $a_2 = 0.1$ (i.e., one request of buyer/service 1 needs 0.2 unit of CPU). Hence, we have the utility functions: 
\beqn
u_1(x_1) = min\big(\frac{x_1}{a_1}, 1\big) = min \big( 5 \times x_1, 1\big)  \nonumber \\ \nonumber
 u_2(x_2) = min\big(\frac{x_2}{a_2}, 10 \big) = min \big( 10 \times x_2, 10\big)
\eeqn

The optimal solution to the problem (22)--(24) (i.e., the EG scheme mentioned in the simulation) gives us the ME: $p = 2$, $x_1 = x_2 = 0.5$, and $u_1 = 1, u_2 = 5$. At the equilibrium, both buyers exhaust their budget since $p \times x_1 = 1 = B_1$ and $p\times x_2 = 1 = B_2$.  Obviously, ($p, x$) where p = 2 and $x = (x_1, x_2$) = (0.5, 0.5) is an ME to our model since both buyers maximize their utility at the equilibrium price, the fog resource is fully allocated to the buyers, and equilibrium price is positive (i.e., $p = 2$). 
It is easy to observe that buyer 1 receives more resources than he needs (i.e., $5 \times x_1 = 2.5 > 1 = u_1^{max}$). Thus, if we can reallocate the redundant resource of buyer 1 to buyer 2, it will not affect the utility of buyer 1 while improving the utility of buyer 2. 

For instance, the GEG scheme as mentioned in the simulation (i.e., the optimal solution to the problem (27)--(31)) results in a new ME as follows: $p = 1.25, x = (x_1,~x_2$) = (0.2, 0.8). This is obviously an ME since at the positive price $p = 1.25$, the fog resource is fully allocated (i.e., market clearing), and both buyers maximize their respective utilities at the this price. At this new ME, the utilities of the buyers become: $u_1 = 1$ and $u_2 = 8$, which are weakly better than the utilities in the previous ME computed by the EG scheme. There is no wasteful resources since $5 \times x_1 = 1 = u_1^{max}$ and $u_2 = 8 < u_2^{max}$. It is also worth mentioning that buyer 1 has a surplus budget of $1 - 0.2 \times 1.25 = 0.75$ at this new ME.

Note that there exist other ME in this example, which are not the optimal solutions to both the problem (22)--(24) (i.e., the EG scheme) and (27)--(31) (i.e., the GEG scheme). For instance, $p = 1.6, x = (x_1, x_2) = (0.375,~0.625)$. Clearly, at price $p = 1.6, x_2 = 0.625$ maximizes  the utility of buyer 2 subject to the budget constraint of this buyer. Also, at this price, $x_1$ maximizes the utility of buyer 1 and satisfies his budget constraint. At the positive price p = 1.6, the fog resource is fully allocated to the buyers (i.e., $x_1 + x_2 = 1$). Therefore, 
$(p,x) = (1.6, (0.375,~0.625))$ is another ME in this example. At this equilibrium, we have $u_1 = 1$ and $u_2 = 6.75$. Buyer 2 fully spends his budget while buyer 1 has a budget surplus of $1 - 1.6 \times 0.325 = 0.48$ at this  equilibrium.

\subsection{Example with Two Buyers and Two Single-resource FNs}
\label{toy2}
In the following, we present another toy example to show the wasteful equilibrium allocation issue in Proposition \ref{theo1}. Consider a market with two services and two single-resource FNs. Obviously, the utilities of the services become linear functions in the single-resource setting. Assume $u_1(x_1) = min \big\{8 \times x_{1,1} + 2 \times x_{1,2}, 1\big\}$,  $u_2(x_2) = 5 \times x_{2,1}+ 2 \times x_{2,2}$, $B_1 = 3$, and $B_2 = 1$. The optimal solution to the convex program (\ref{EGinf})-(\ref{EGinf2}) gives all resource of FN1 to service 1 and all resource of FN2 to service 2 (i.e., $x_{1,1} = 1$ and $x_{2,2} = 1$) and the prices are $p_1 = 3$ and $p_2 = 1$. It is easy to verify that this price vector and allocation form an ME. At the equilibrium, 
$u_1 = 1$ and $u_2 = 2$. Since $8 x_{1,1} + 2 x_{1,2} = 8 > 1$, service 1 receives more resources than it needs.
Consider another ME where p = ($p_1,~p_2) = (0.7843,~0.3137)$, $x_1 = (x_{1,1},~x_{1,2}) = (0.125,~0)$, $x_2 = (x_{2,1},~x_{2,2}) = (0.875,~1)$ (i.e., we reallocate some redundant resources of service 1 in the previous ME to service 2). At the new ME, $u_1 = 1$ and $u_2 = 6.375$.
Obviously, this new equilibrium allocation dominates the previous equilibrium allocation.

\subsection{Discussion of Frugal Market Equilibria}
Frugality is a natural and desirable property for any market mechanism.  Clearly, to obtain the same utility, a service prefers to spend as least money as possible. For example, consider  two single-resource FNs with prices $p_1 = 10$ and $p_2 = 1$. Assume that service A has the same utility for each FN (e.g., $u_1(x_1) = 2 \times x_{1,1}+ 2 \times x_{1,2}$). Obviously, given the  prices, service A will naturally buy resource of FN2 since it is cheaper. However, a non-frugal ME may force service A to buy resource of FN1. In this case, service A has to spend more money to obtain the same utility, which may make it  unhappy.

Here we give a toy example of a non-frugal ME. The market consists of two services and two single-resource FNs with the same setting as in Appendix \ref{toy2}. Beside the market equilibria discussed in Appendix \ref{toy2}, we consider the following price vector and allocation: p = ($p_1,~p_2) = (1,2), x_1 = (x_{1,1},x_{1,2}) = (0,1), x_2 = (x_{2,1},x_{2,2}) = (1,0)$, and $u = (u_1, u_2) = (1,5)$. We can observe that service 1 obtains its optimal utility which is equal to its utility limit of 1. Also, service 2 spends full budget ($x_{2,1} p_1 + x_{2,2} p_2 = 1 = B_2$) to buy resource of the cheapest FN. Note that FN1 is the cheapest FN for service 2 since $\frac{5}{p_1} > \frac{2}{p_2}$ (i.e., utility gain over 1 unit of money spent). Therefore, the price vector and allocation above form an ME. 

However, this ME is not frugal since at the equilibrium, service 1 receives full resource of FN2 while the cheapest FN for service 1 is FN1 because  $\frac{8}{p_1} > \frac{2}{p_2}$. At this equilibrium, service 1 spends $x_{2,1} p_1 + x_{2,2} p_2 = 2$. However, at prices $p = (1,2)$, it is natural for service 1 to just buy $\frac{1}{8}$ unit of resource of FN1 to obtain its utility limit of 1. In this case, service 1 needs to spend only $\frac{1}{8} \times p_1 = 0.125$, which is much smaller than 2. Therefore, service 1 may not be happy with this non-frugal ME.

\subsection{Discussion of the Privacy Property in Market Design}

In an FC market, or in any market in general, it is  desirable for  market participants to keep their information private. For instance, if the  private data of an agent  is revealed,   other agents can exploit it to gain  benefits.
To illustrate this issue, consider an example with two services and two single-resource FNs. The services have the same budget of one. 
The utility functions of the services are $u_1(x_1) = 4 \times x_{1,1}+ 1 \times x_{1,2}$, and $u_2(x_2) = 4 \times x_{2,1}+ 4 \times x_{2,2}$, respectively. The GEG scheme produces the following ME: p = ($p_1,~p_2) = (1,1), x_1 = (x_{1,1},x_{1,2}) = (1,0), x_2 = (x_{2,1},x_{2,2}) = (0,1)$, and $u = (u_1, u_2) = (4,4)$.

Now assume that service 2 knows the utility of service 1 and knows that service 1 prefers FN1 to FN2. Then, service 2 can pretend that its utility is $u_2(x_2) = 4 \times x_{2,1}+ 2 \times x_{2,2}$. Consequently, the new ME computed by the GEG scheme is:  p = ($p_1,~p_2) = (4/3, 2/3), x_1 = (x_{1,1},x_{1,2}) = (0.75,0), x_2 = (x_{2,1},x_{2,2}) = (0.25,1)$, and $u = (u_1, u_2) = (3,5)$. We can see that $u_2$ increases while $u_1$ decreases. Service 1 receives less resources because FN1 becomes more expensive.
Indeed, service 2 has many options to manipulate its utility to gain benefit. For example, it can state its utility is $u_2(x_2) = 12 \times x_{2,1}+ 4 \times x_{2,2}$ (i.e., pretend to be more interested in FN1). The new ME becomes: p = ($p_1,~p_2) = (1.5, 0.5), x_1 = (x_{1,1},x_{1,2}) = (0.6666,0), x_2 = (x_{2,1},x_{2,2}) = (0.3333,1)$, and $u = (u_1, u_2) = (2.6666,5.3333)$. This new ME significantly increases the utility of service 2 while drastically reducing the utility of service 1.

Therefore, preserving the privacy of the agents is an important and desirable feature of any market mechanism. It is worth noting that the benefit that an agent can gain by playing strategically in a Fisher market when he knows the private information of the other agents has been studied in \cite{nche11, nche12,nche16}.

\subsection{Statistical Simulation Results over Multiple Runs}

\begin{figure}[ht]
	\centering
		\subfigure[Mean]{
		  \includegraphics[width=0.245\textwidth,height=0.10\textheight]{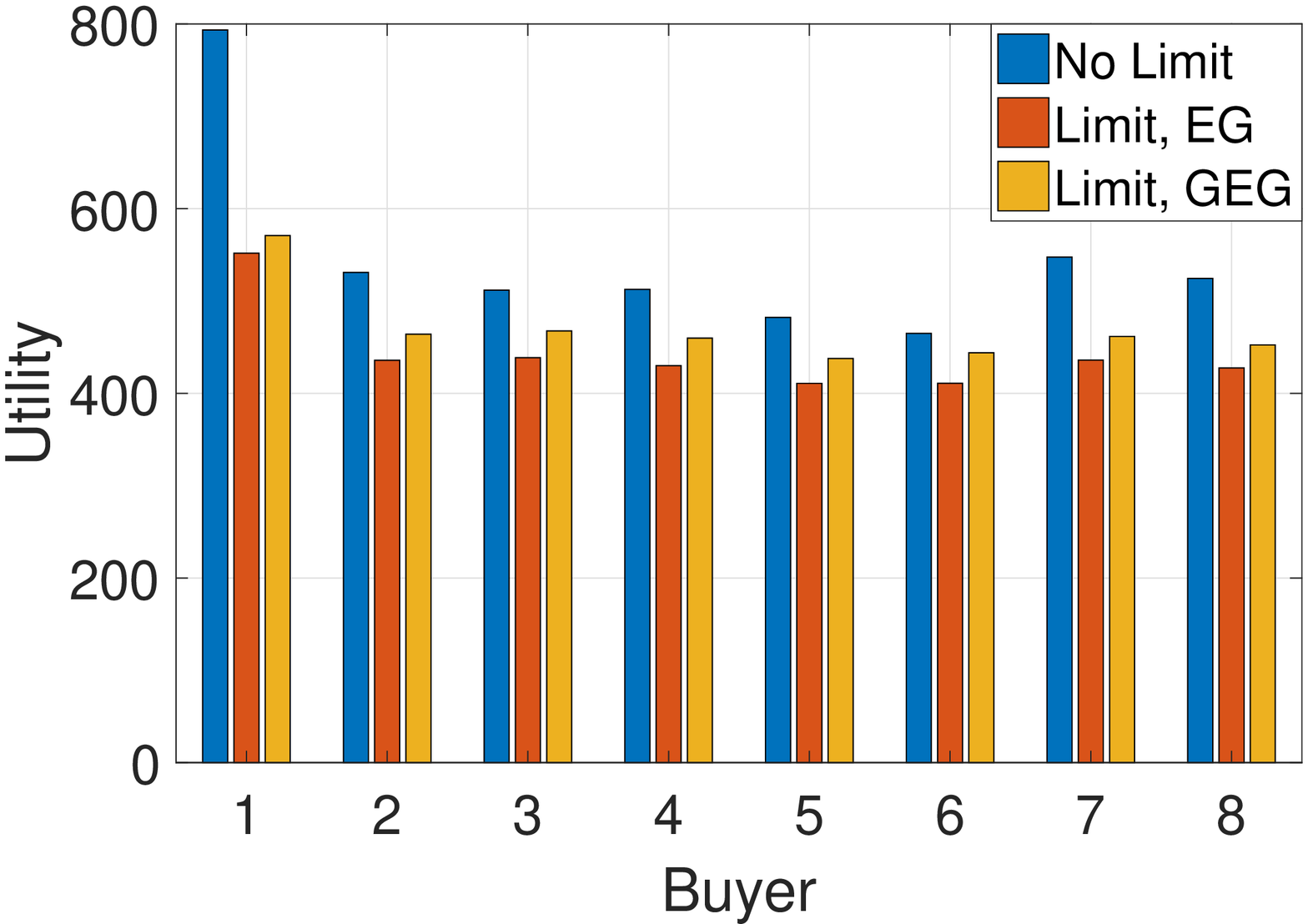}
	    \label{fig:EGGEG_a}
	} \hspace*{-2.1em}  
		 \subfigure[Boxplot]{
	     \includegraphics[width=0.245\textwidth,height=0.10\textheight]{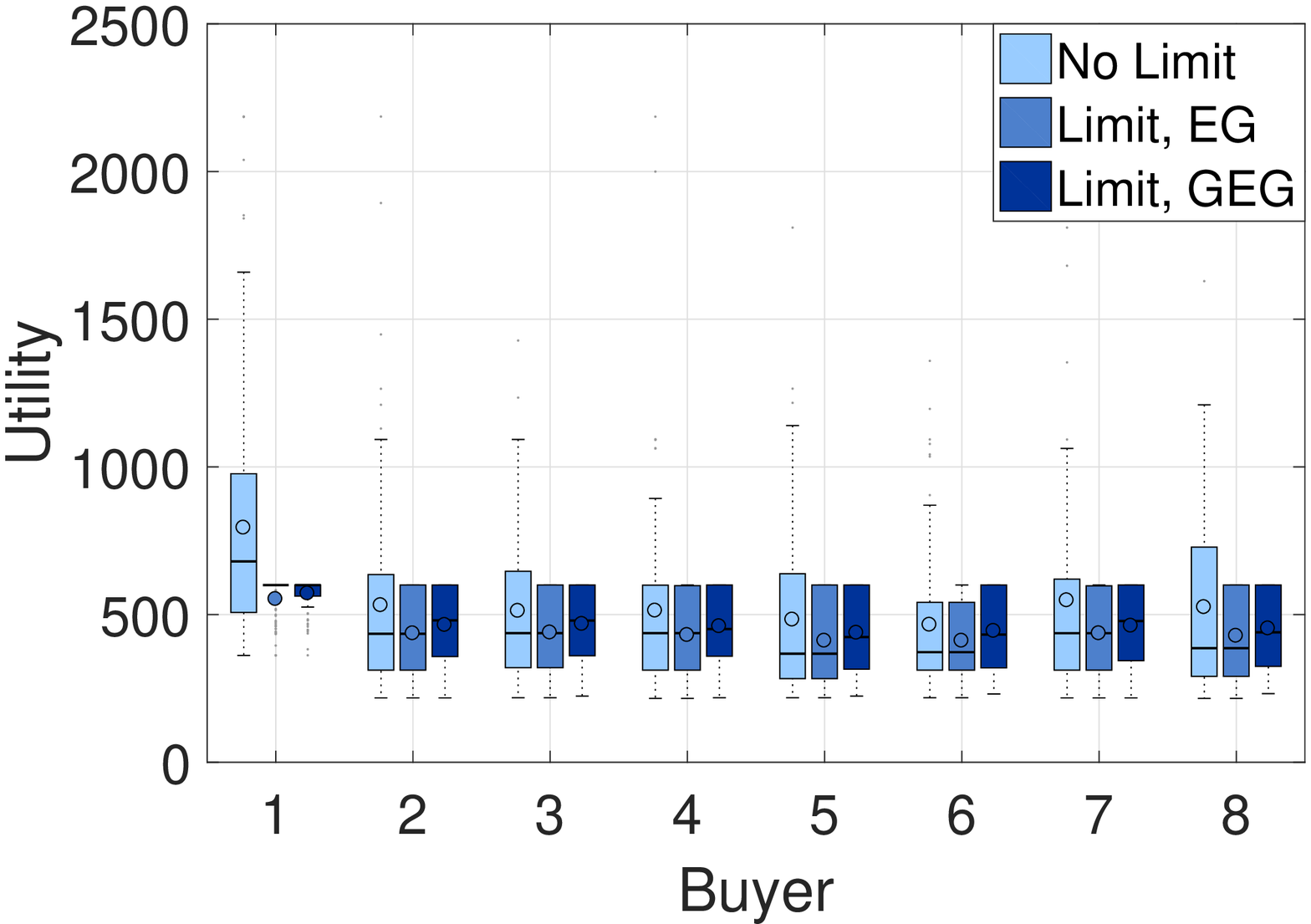}
	     \label{fig:EGGEG_a_b}
	}  
	
	\caption{Comparison between EG and GEG schemes}
\end{figure}

For the ease of explanation, in the main paper, we randomly generated the simulation data  for  one problem instance only.  
To further demonstrate the efficacy of the proposed resource allocation algorithm, for every experiment, we redo the simulation over 50 runs (i.e., 50 problem instances) to obtain the statistical results including the mean values and/or boxplot statistics.




\begin{figure}[ht]
	\centering
		\subfigure[$u^{\sf max} = \infty$]{
		  \includegraphics[width=0.45\textwidth,height=0.10\textheight]{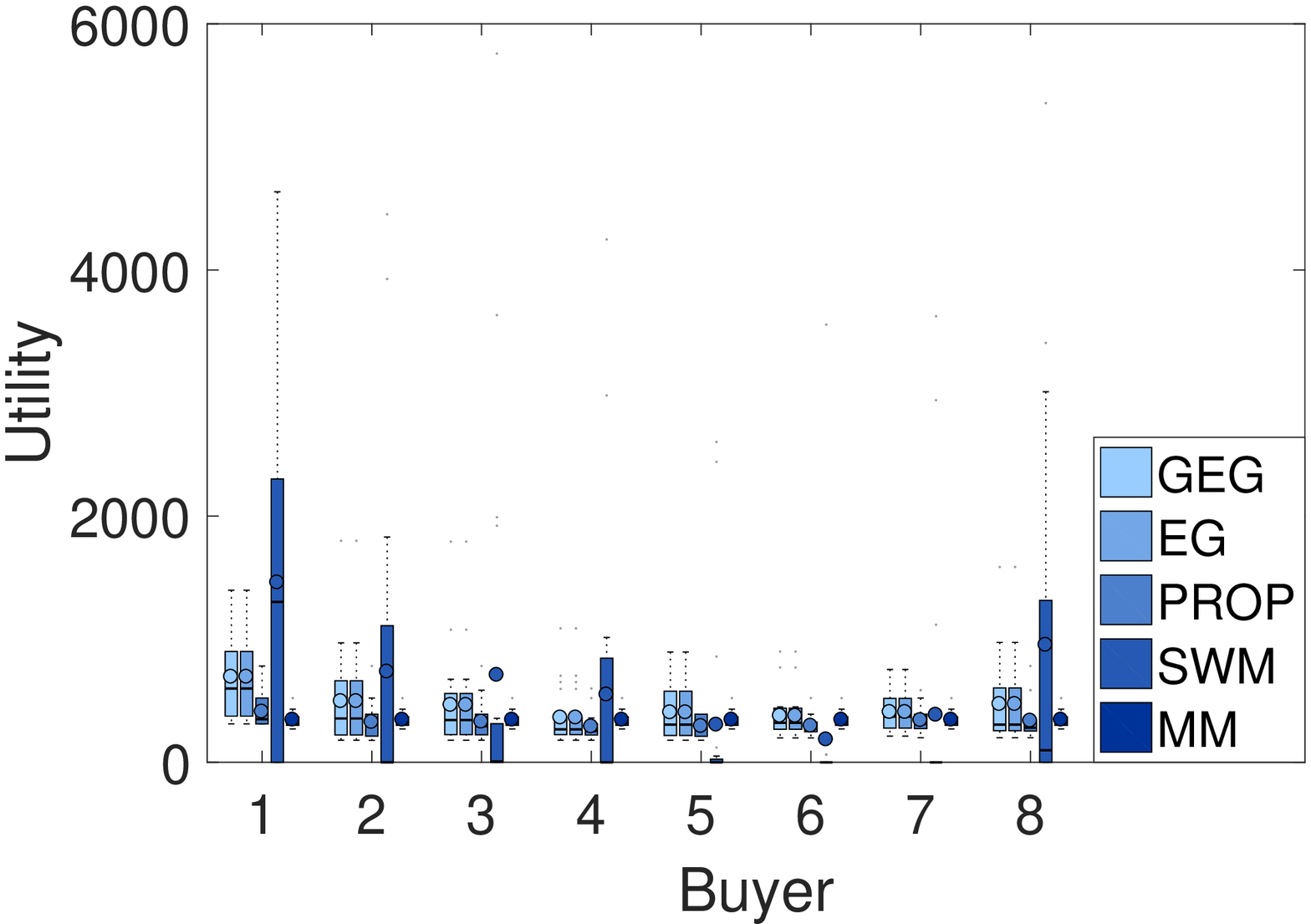}
	    \label{fig:5nocap_a_b}
	}  \\  \hspace*{0.6em} 
		 \subfigure[$u^{\sf max} = 600$]{
	     \includegraphics[width=0.45\textwidth,height=0.10\textheight]{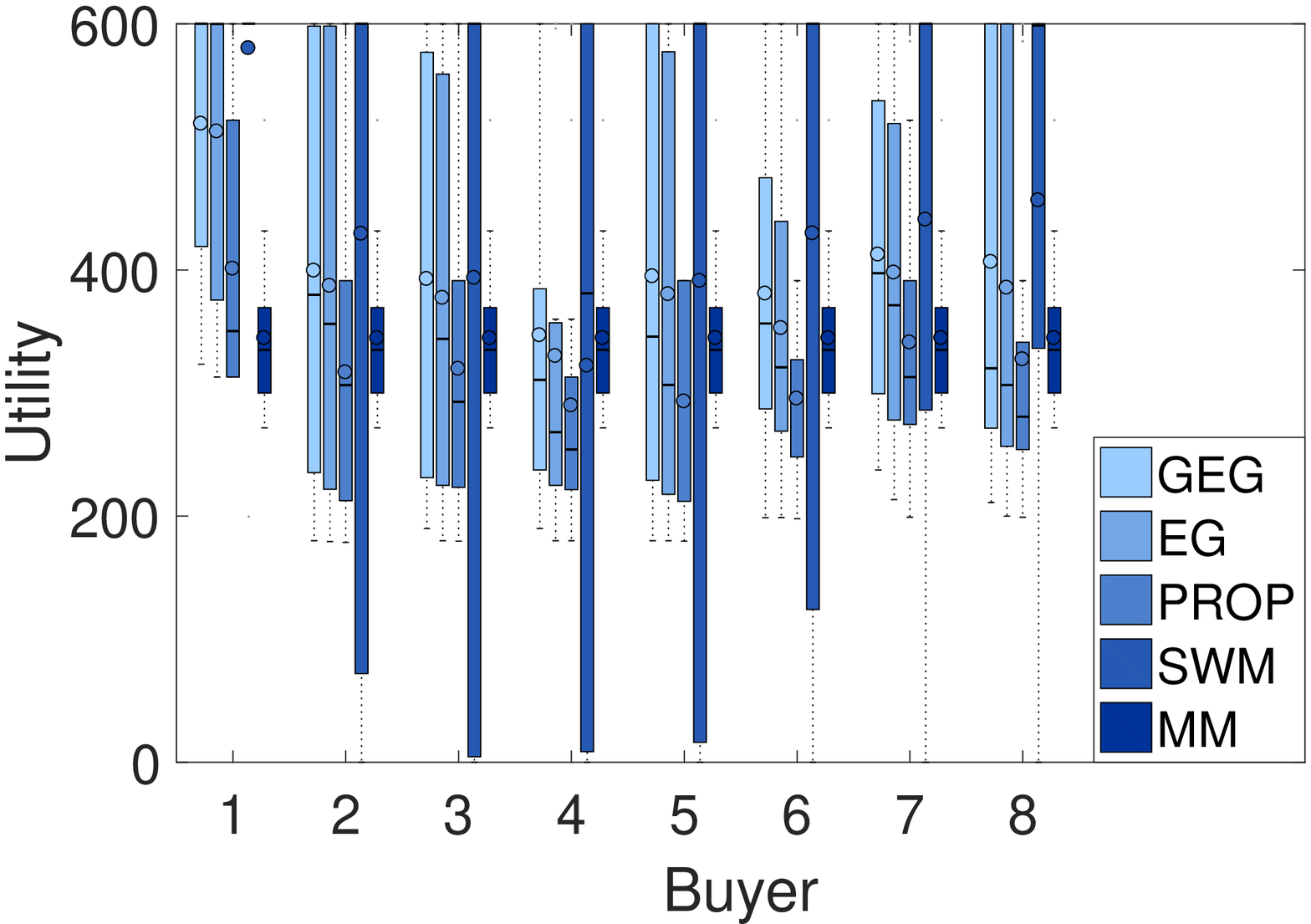}
	     \label{fig:5cap_a_b}
	}  \vspace{-0.2cm}
	
	\caption{Individual utility comparison (boxplot)}
\end{figure}

\begin{figure}[ht]
		\subfigure[$u^{\sf max} = \infty$]{
		  \includegraphics[width=0.24\textwidth,height=0.10\textheight]{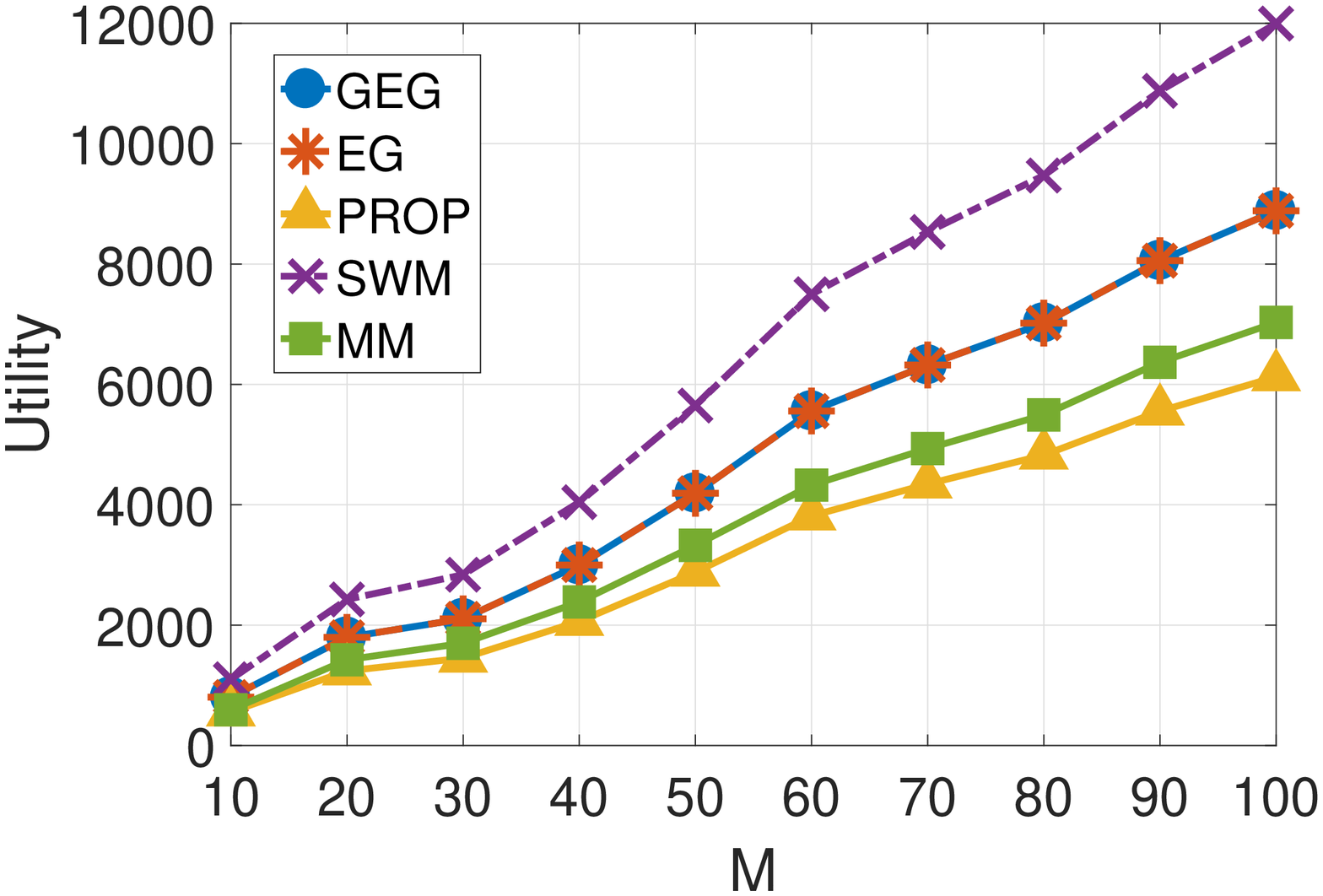}
	    \label{fig:nocaptu_a}
	}   \hspace*{-2.1em} 
		 \subfigure[$u^{\sf max} = 600$]{
	     \includegraphics[width=0.24\textwidth,height=0.10\textheight]{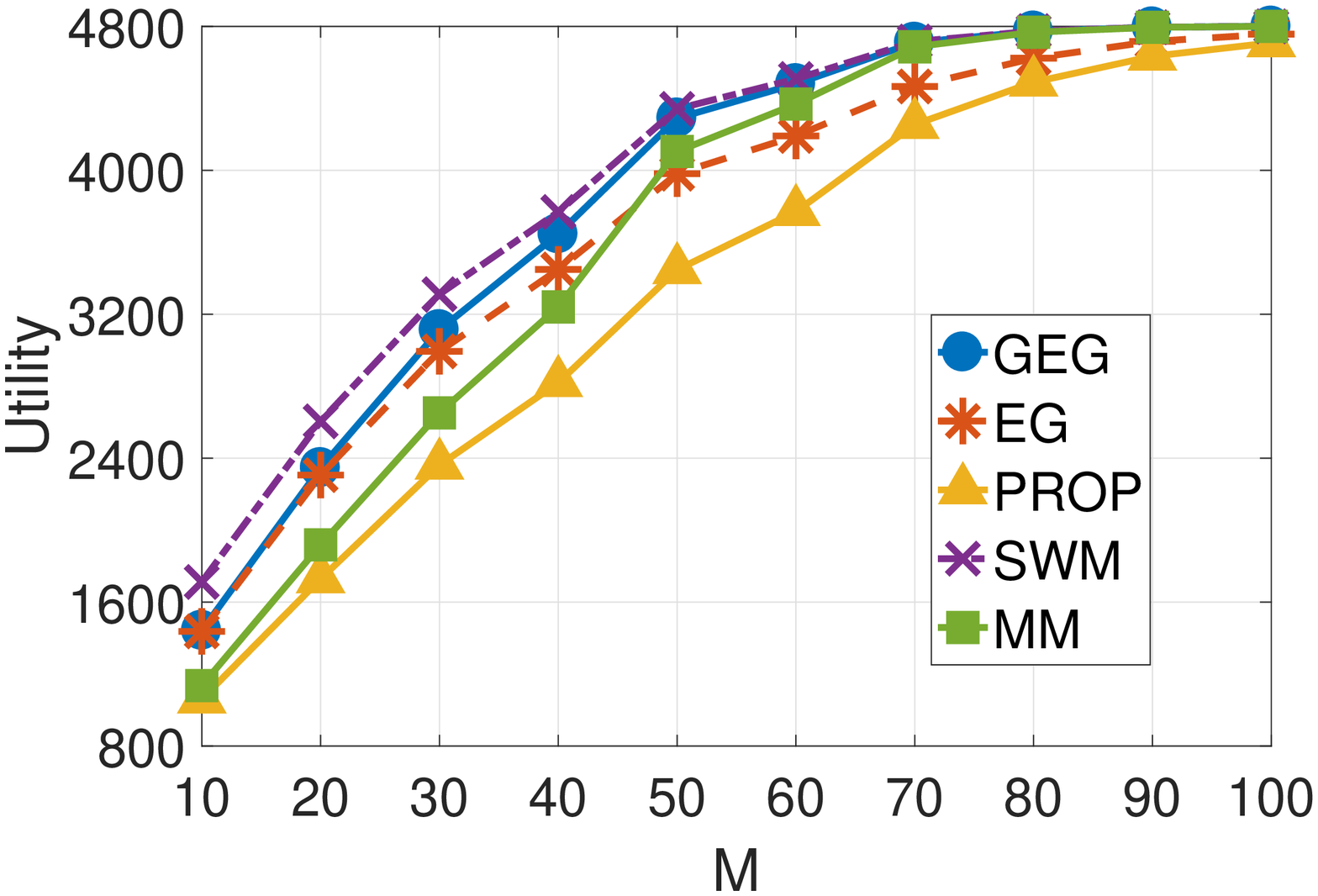}
	     \label{fig:captu_a}
	}  \vspace{-0.2cm}
	\caption{Comparison of the total utility (N = 8)}
\end{figure}

We can observe that the shapes of the figures and the  observations in the Performance Evaluation section remain unchanged. For example, as can be seen from Figs.~\ref{fig:EGGEG_a}--\ref{fig:EGGEG_a_b} (i.e., Fig.~\ref{fig:EGGEG} in the main paper), the GEG scheme outperforms the EG scheme. The boxplots in Figs.~\ref{fig:5nocap_a_b}--\ref{fig:5cap_a_b} (i.e., Figs.~\ref{fig:5nocap}--\ref{fig:5cap} in the main paper) show that the GEG and EG schemes are the same when we do not consider the utility limit and the GEG scheme dominates the EG scheme when the utility limit is considered. Furthermore, the utility of every buyer in the PROP scheme is  smaller or equal to that in the GEG scheme, which implies the sharing-incentive property of the proposed GEG scheme.

\begin{figure}[ht]
		\subfigure[ $u^{\sf max} = \infty$]{
		  \includegraphics[width=0.24\textwidth,height=0.10\textheight]{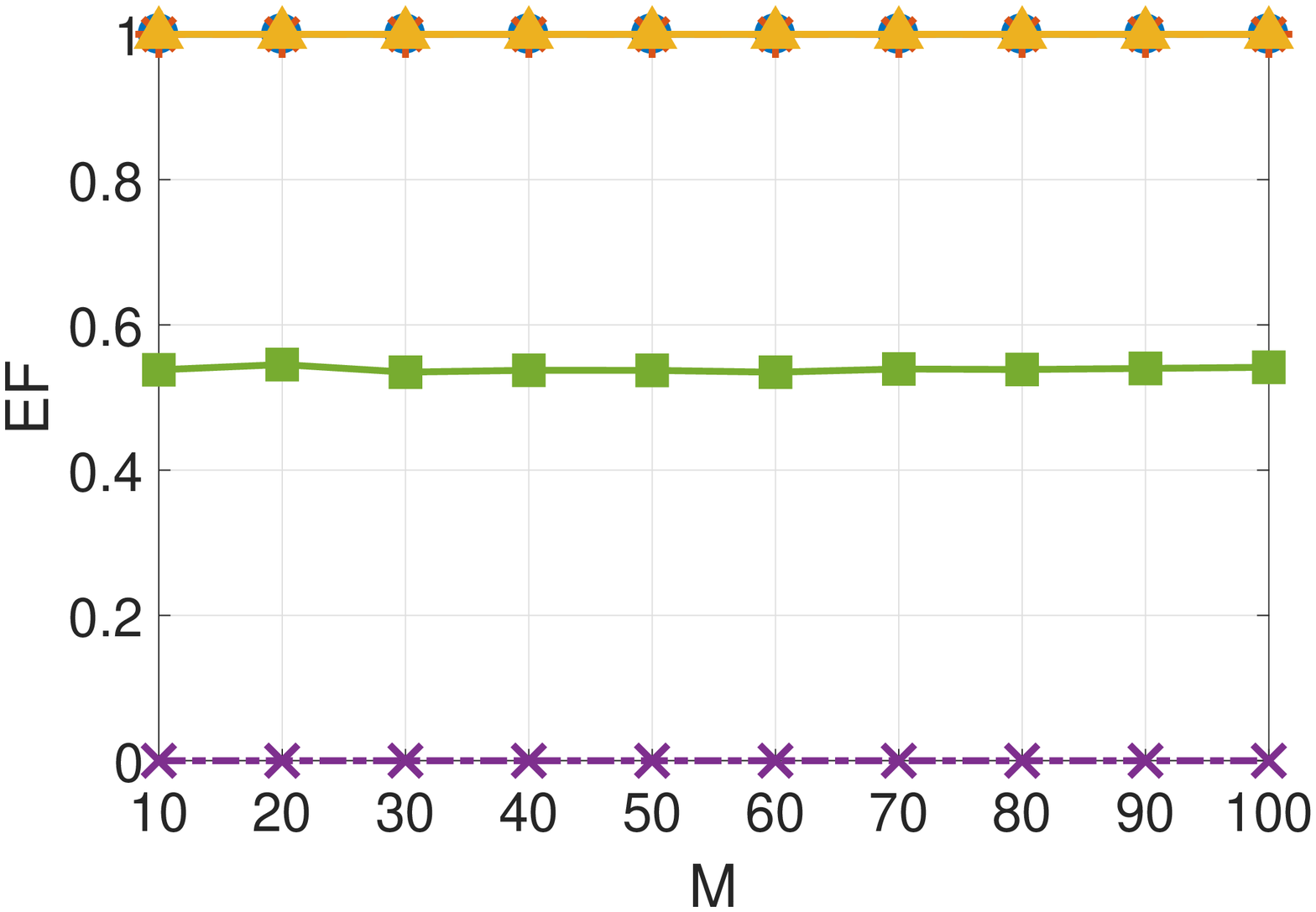}	    
	    \label{fig:uncapEF_a}
	}   \hspace*{-2.1em}  
		 \subfigure[ $u^{\sf max} = 600$]{
	     \includegraphics[width=0.24\textwidth,height=0.10\textheight]{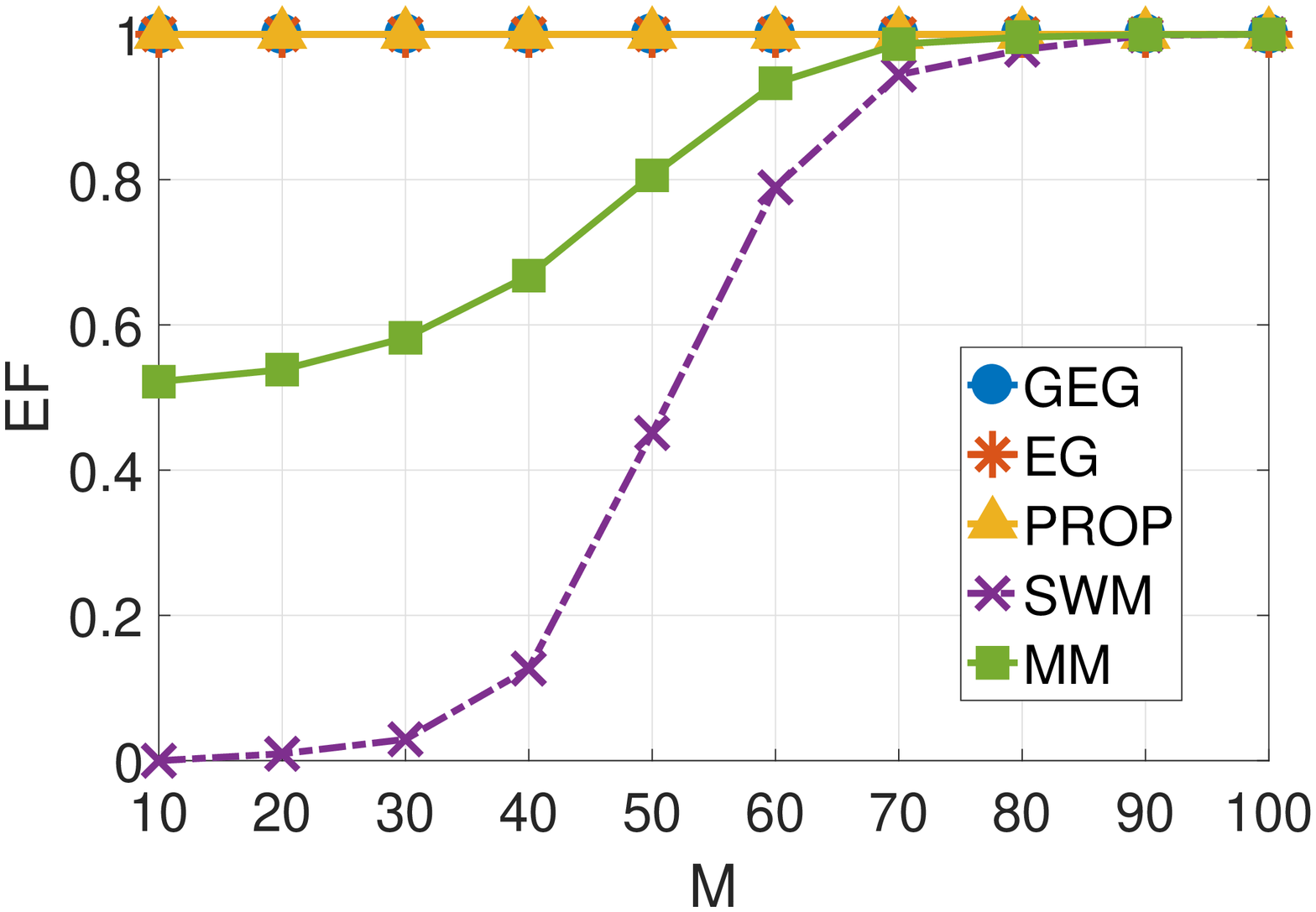}
	     \label{fig:capEF_a}
	} \vspace{-0.2cm}
	
	\caption{Envy-freeness comparison (N = 8, mean)}
\end{figure}

\begin{figure}[ht]
	\centering
		\subfigure[ MM scheme, $u^{\sf max} = \infty$]{
		  \includegraphics[width=0.24\textwidth,height=0.10\textheight]{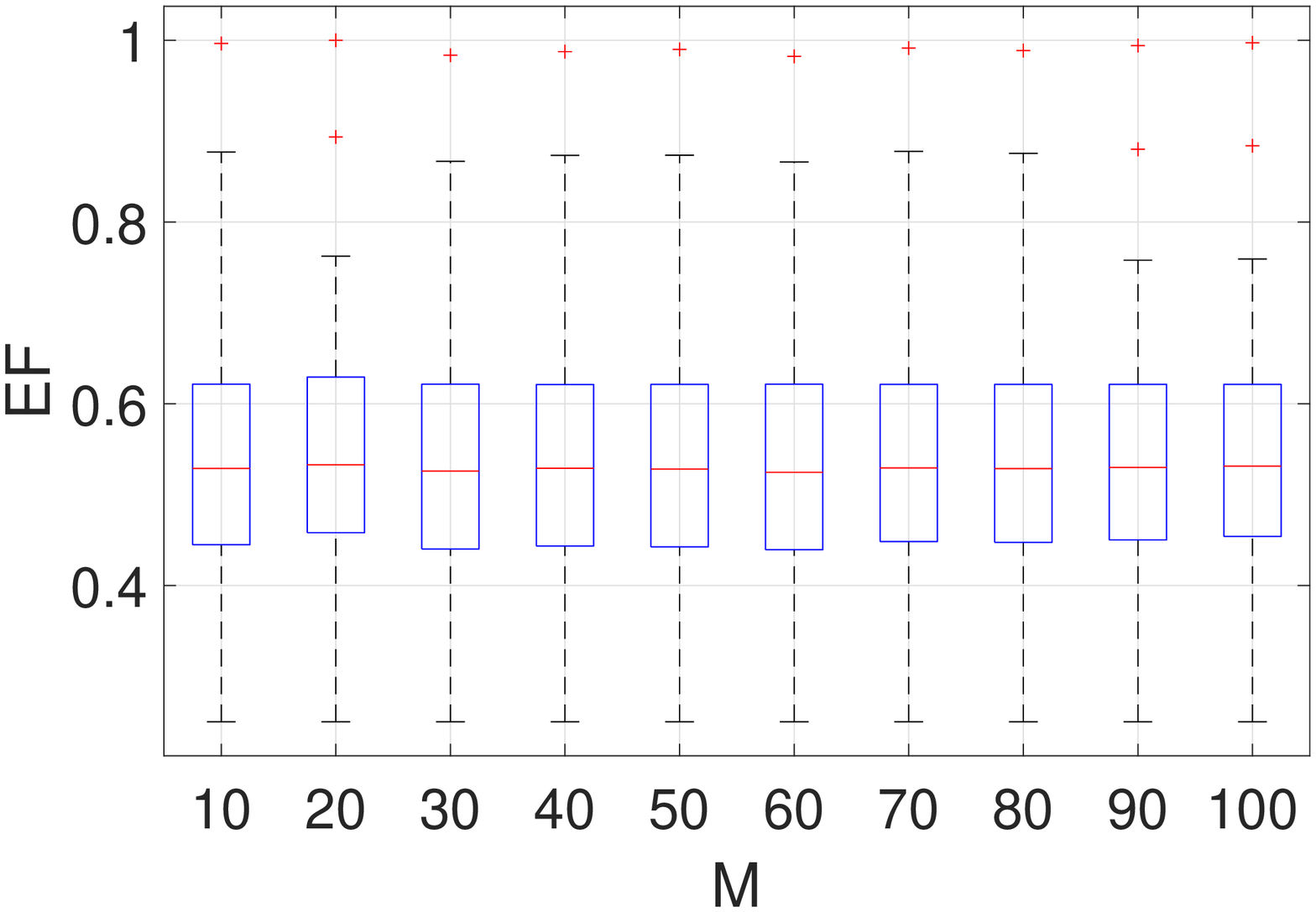}	    
	    \label{fig:uncapEF_a_b}
	}  \\ 
		 \subfigure[SW scheme, $u^{\sf max} = 600$]{
	     \includegraphics[width=0.24\textwidth,height=0.10\textheight]{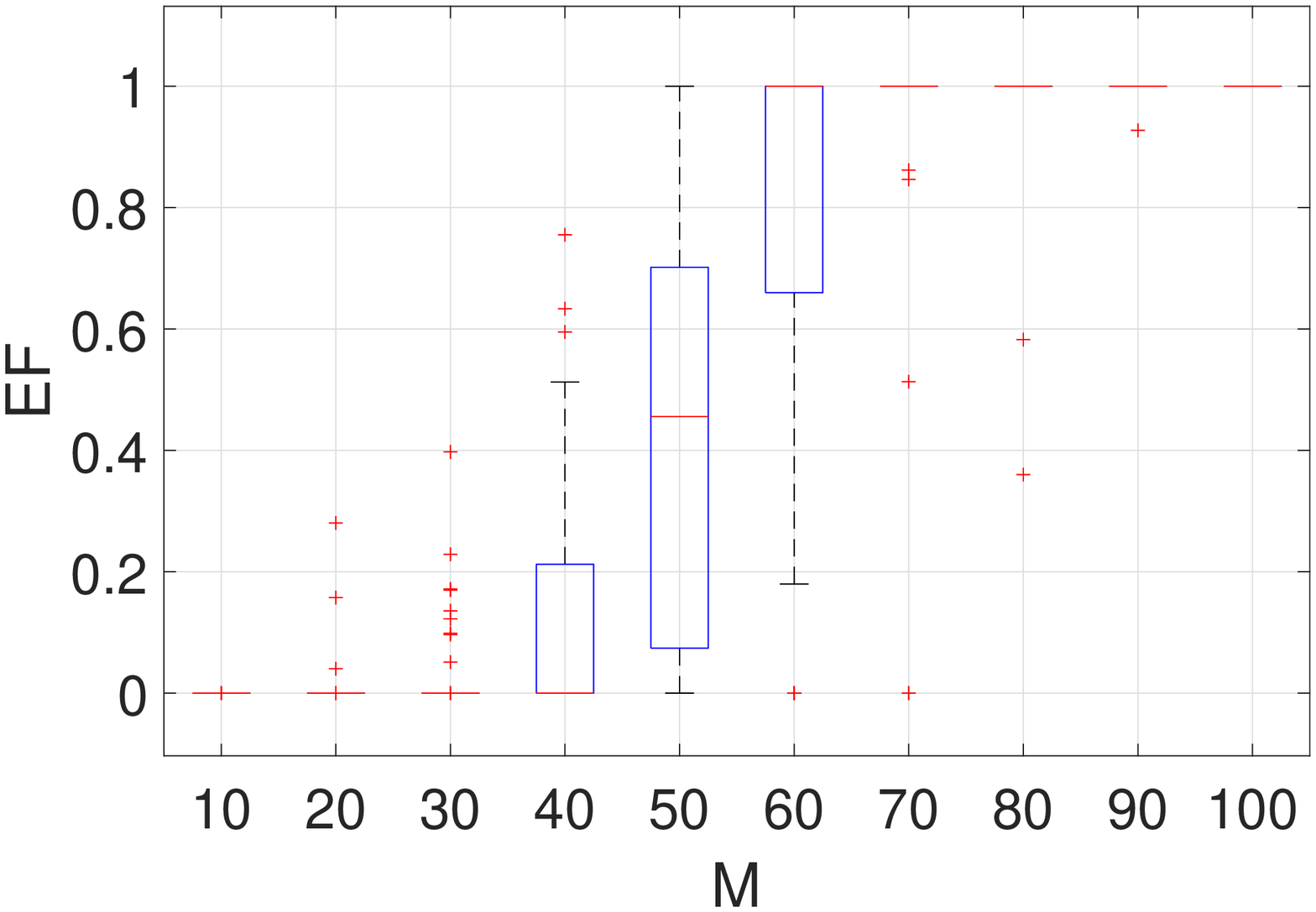}
	     \label{fig:capEF_a_bsw}
	}   \hspace*{-2.1em}  
		 \subfigure[MM scheme, $u^{\sf max} = 600$]{
	     \includegraphics[width=0.24\textwidth,height=0.10\textheight]{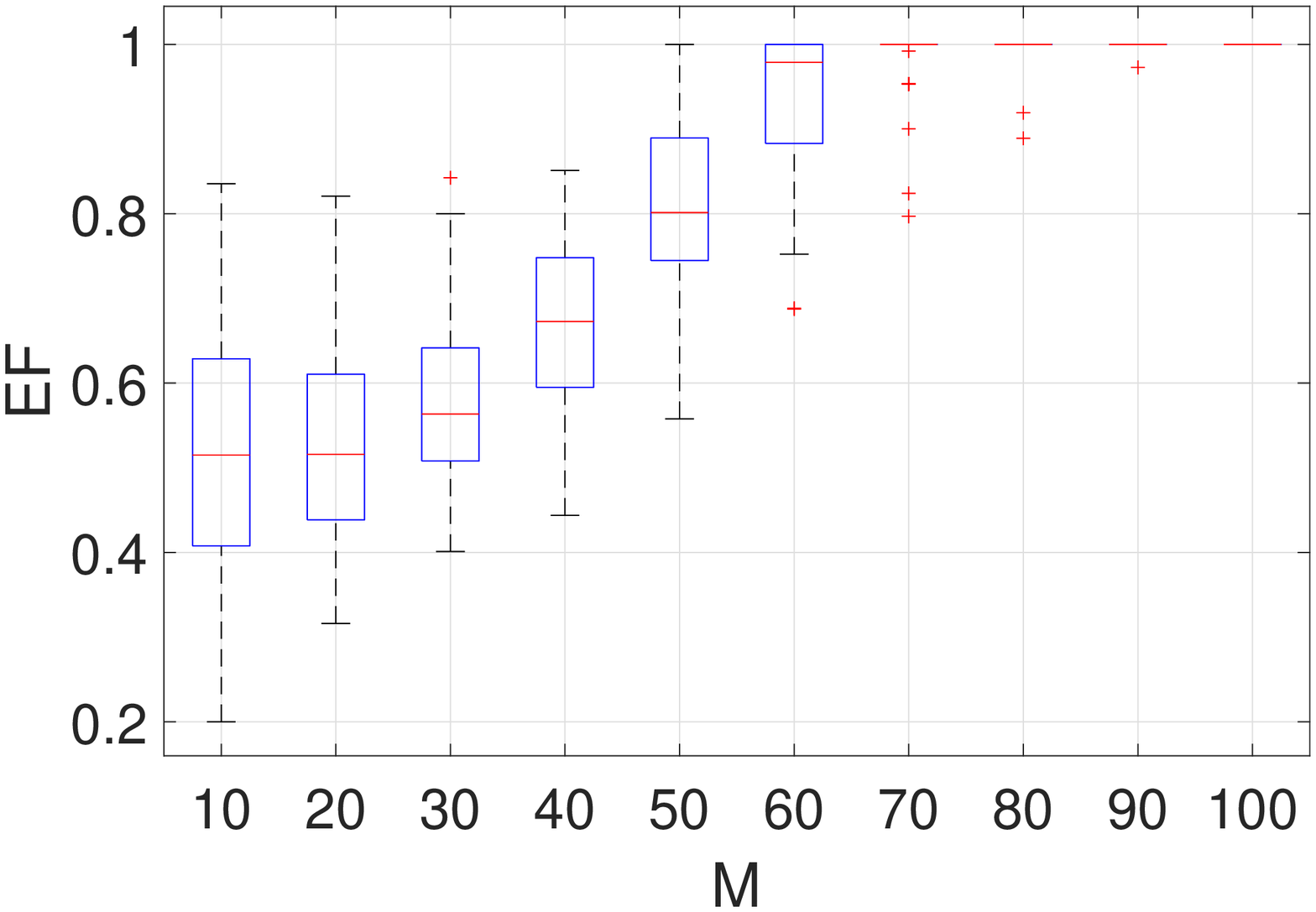}
	     \label{fig:capEF_a_bmm}
	}\vspace{-0.2cm}
	
	\caption{Envy-freeness comparison (N = 8, boxplot)}
\end{figure}

\begin{figure}[ht]
		\subfigure[Buyer 1, $u^{\sf max} = \infty$]{
		  \includegraphics[width=0.24\textwidth,height=0.10\textheight]{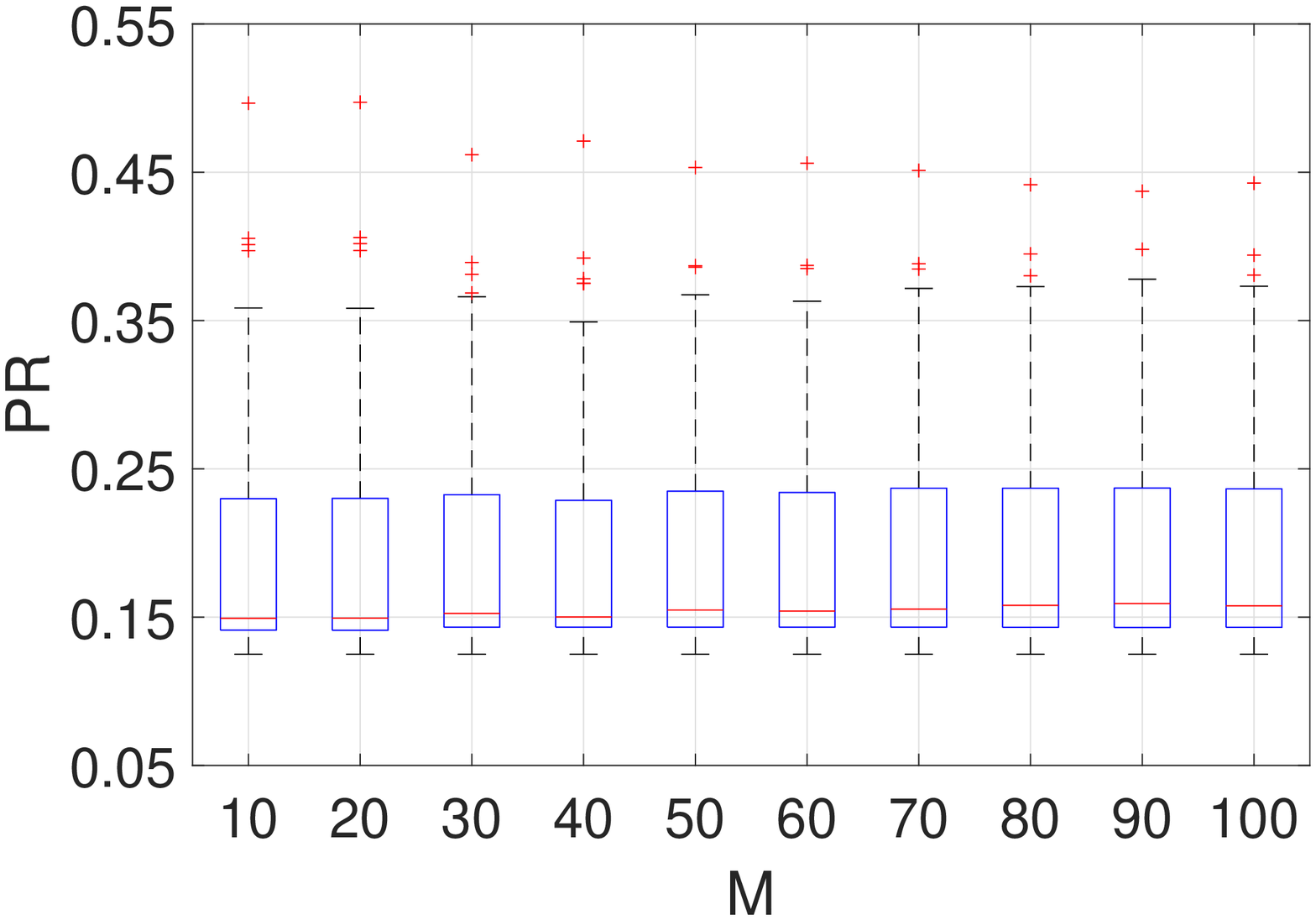}
	    \label{fig:uncapProp_a_b1}
	}   \hspace*{-1.8em} 
		 \subfigure[Buyer 1, $u^{\sf max} = 600$]{
	     \includegraphics[width=0.24\textwidth,height=0.10\textheight]{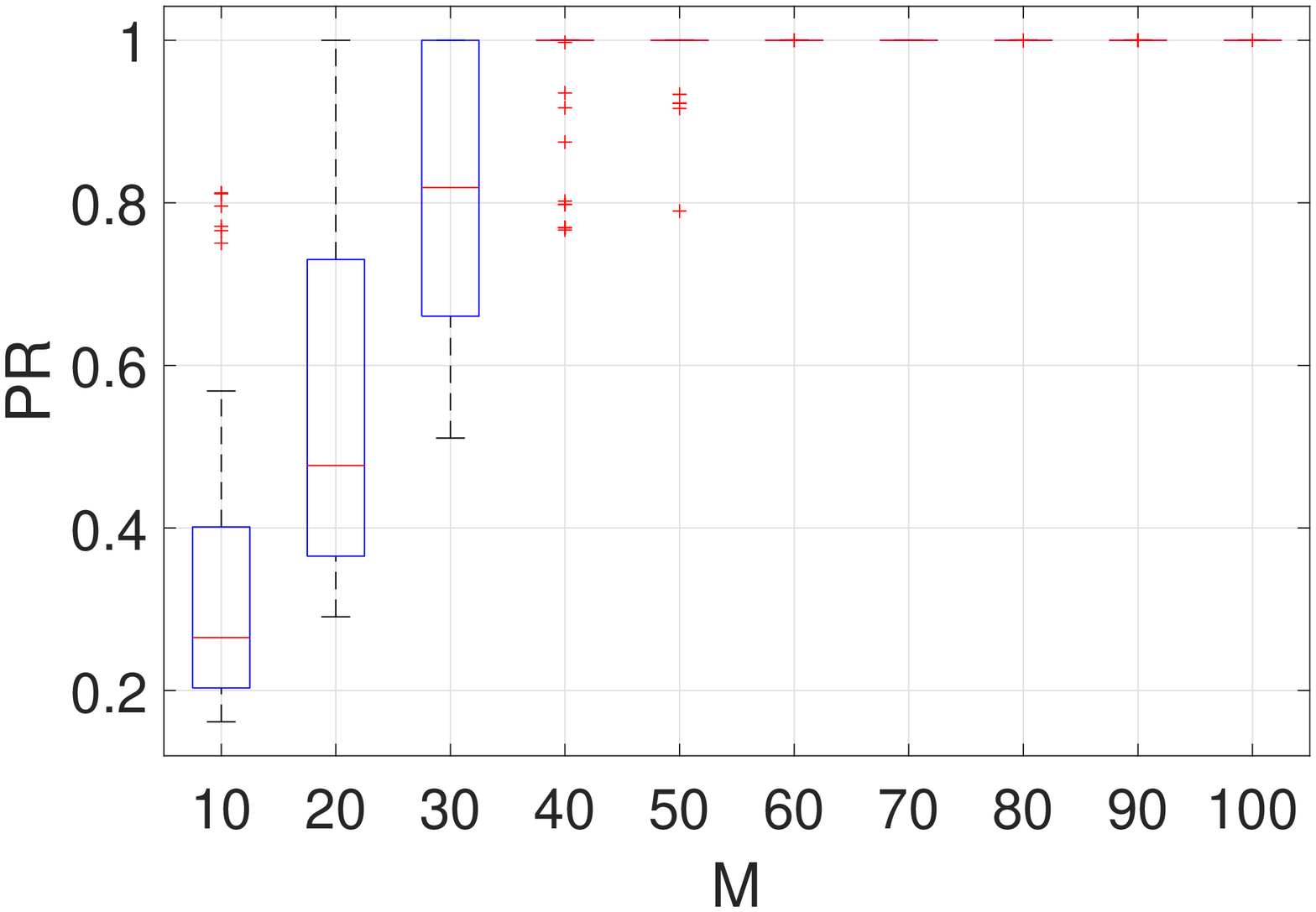}
	     \label{fig:capProp_a_b1}
	}  \vspace{-0.2cm}
			\subfigure[Buyer 2, $u^{\sf max} = \infty$]{
		  \includegraphics[width=0.24\textwidth,height=0.10\textheight]{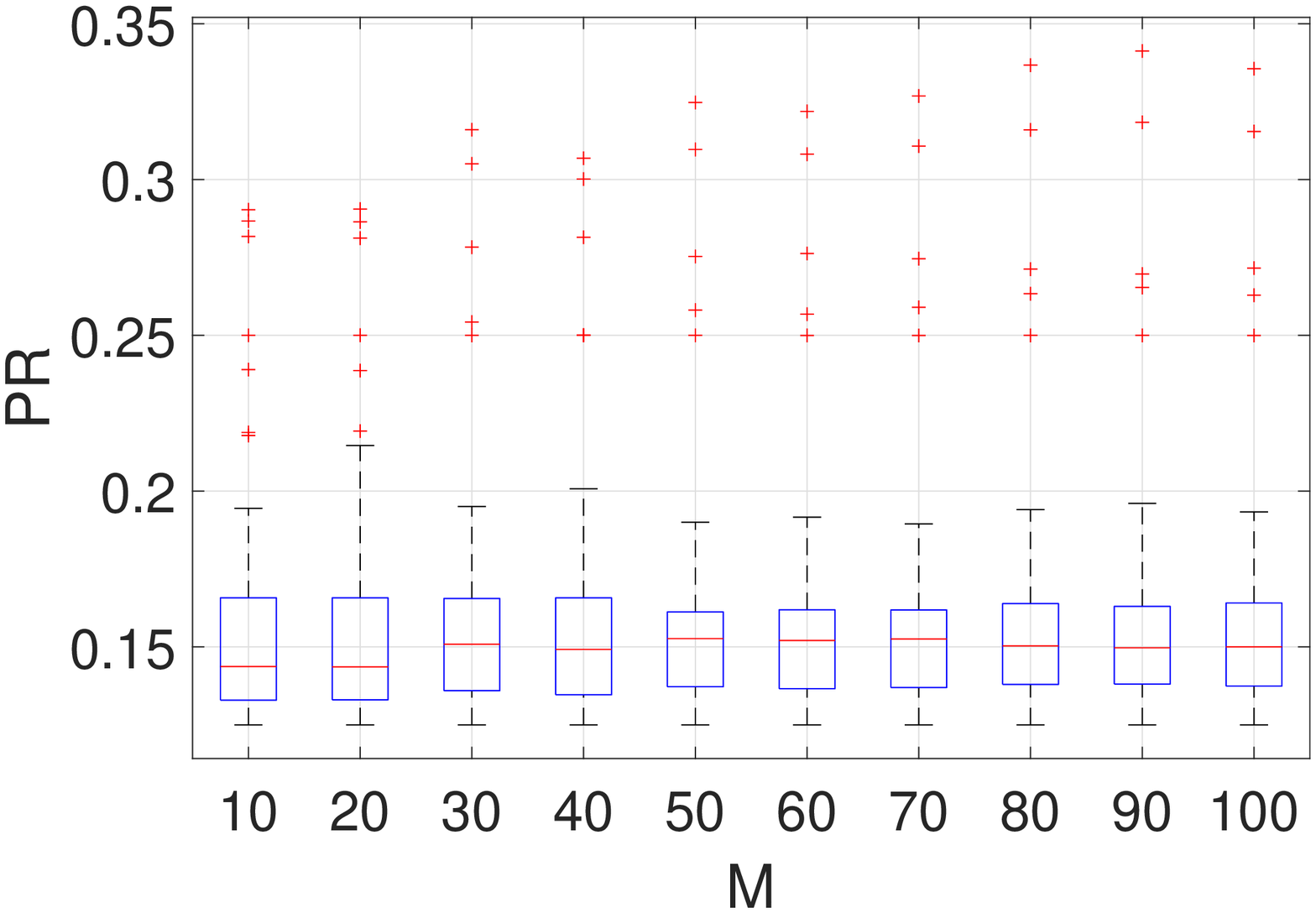}
	    \label{fig:uncapProp_a_b2}
	}   \hspace*{-1.8em} 
		 \subfigure[Buyer 2, $u^{\sf max} = 600$]{
	     \includegraphics[width=0.24\textwidth,height=0.10\textheight]{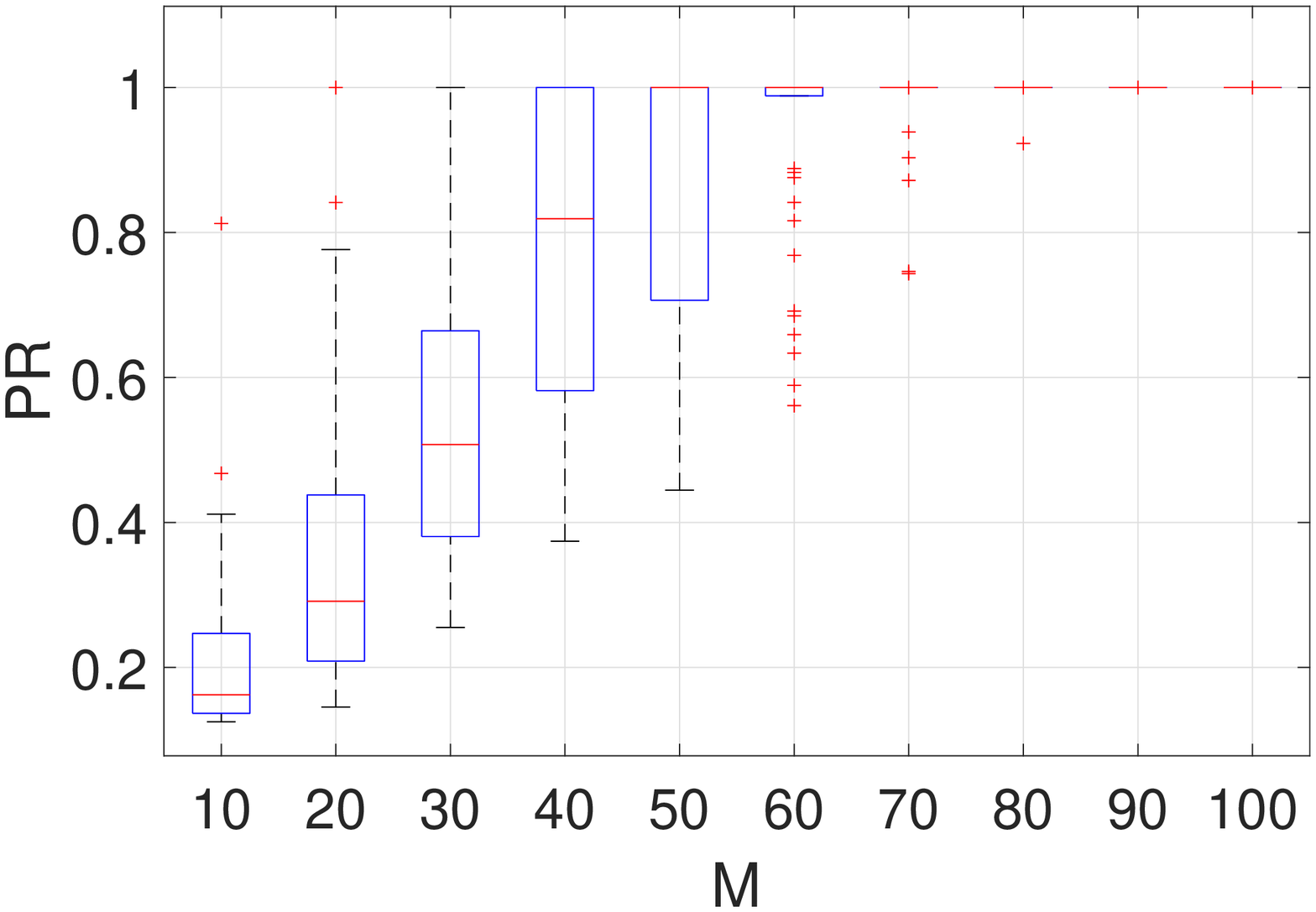}
	     \label{fig:capProp_a_b2}
	}  \vspace{-0.2cm}
	\caption{Proportionality property (N = 8, boxplot)}
\end{figure}

\begin{figure}[ht]
		\subfigure[$u^{\sf max} = \infty$]{
		  \includegraphics[width=0.24\textwidth,height=0.10\textheight]{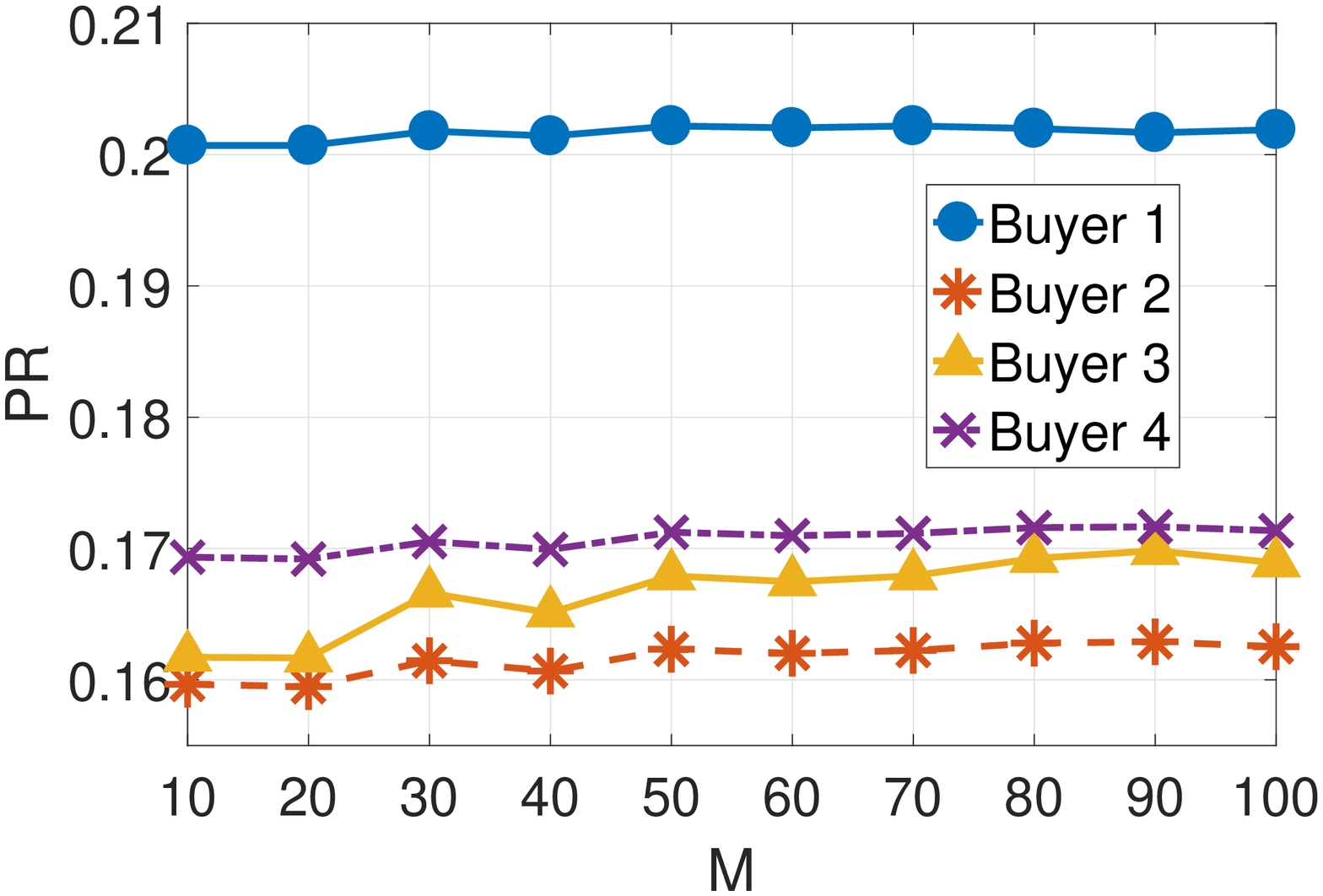}
	    \label{fig:uncapProp_a}
	}   \hspace*{-1.8em} 
		 \subfigure[$u^{\sf max} = 600$]{
	     \includegraphics[width=0.24\textwidth,height=0.10\textheight]{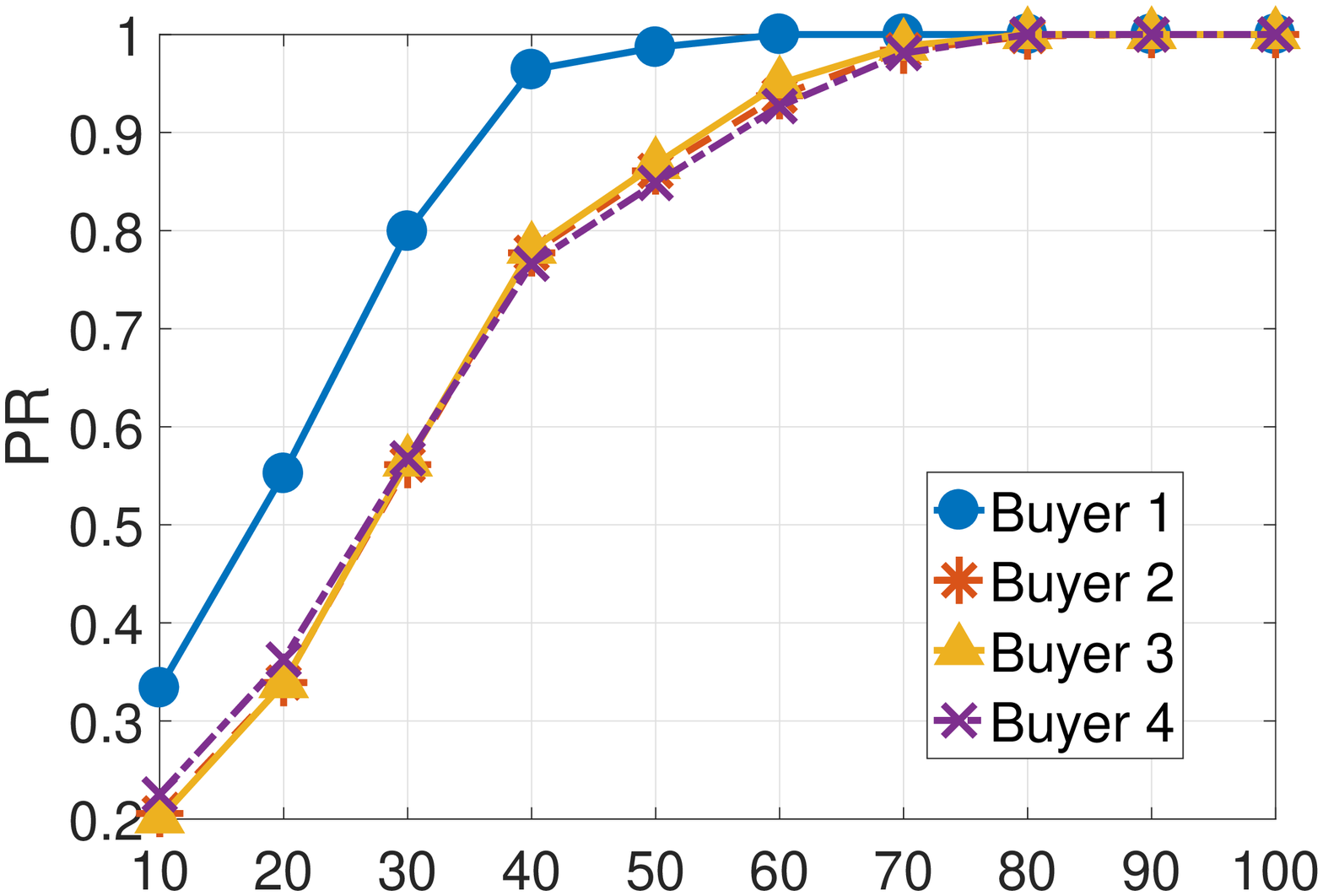}
	     \label{fig:capProp_a}
	}  \vspace{-0.2cm}
	\caption{Proportionality property (N = 8, mean)}
\end{figure}

The MM scheme produces a fair allocation in terms of utility at the cost of low system efficiency (i.e., total utility). Figs.~\ref{fig:nocaptu_a}--\ref{fig:captu_a} (i.e., Figs.~\ref{fig:nocaptu}--\ref{fig:captu}) present total utility comparison among the five schemes. Finally, we can observe that the SWM scheme has the highest total utility and gives high utilities for the buyers. However, it performs poorly in terms of fairness. Figs.~\ref{fig:5nocap_a_b}--\ref{fig:5cap_a_b} show that the utility of each buyer varies significantly over 50 runs. Indeed, for each problem instance, there always exist some buyers with very low (even zero) utilities. Furthermore, \ref{fig:uncapEF_a}--\ref{fig:capEF_a_bmm} (i.e., Figs.~\ref{fig:uncapEF}--\ref{fig:capEF}) show that SWM has bad envy-free index. Similar to Figs.~\ref{fig:uncapProp}--\ref{fig:capProp} in the main paper, Figs.~\ref{fig:uncapProp_a_b1}--\ref{fig:capProp_a} indicate the proportionality of the proposed resource allocation scheme.

\begin{figure}[ht]
	\centering
		\subfigure[ Mean]{
		  \includegraphics[width=0.42\textwidth,height=0.10\textheight]{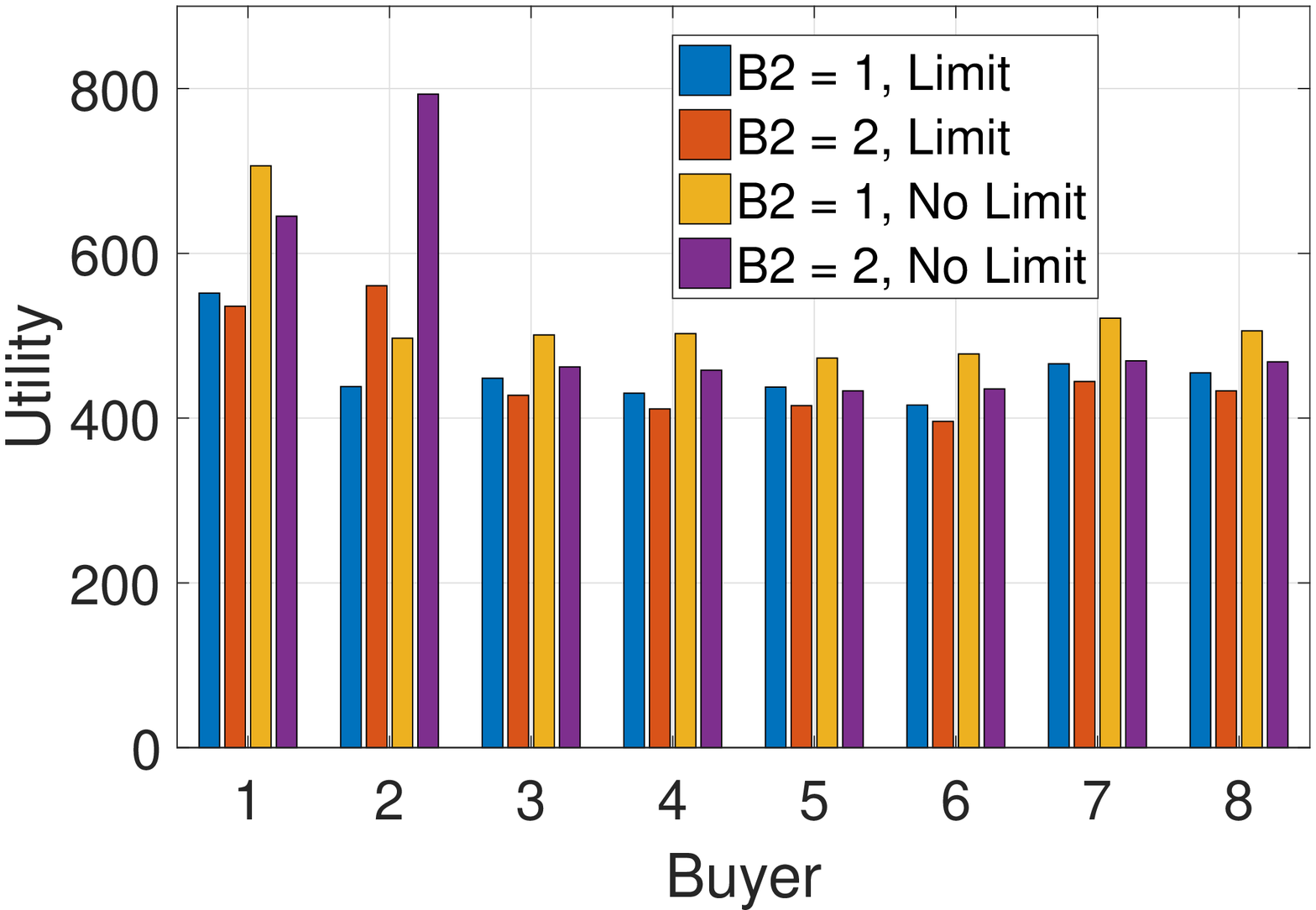}	    
	    \label{fig:budget_a}
	} 
		 \subfigure[Boxplot]{
	     \includegraphics[width=0.42\textwidth,height=0.10\textheight]{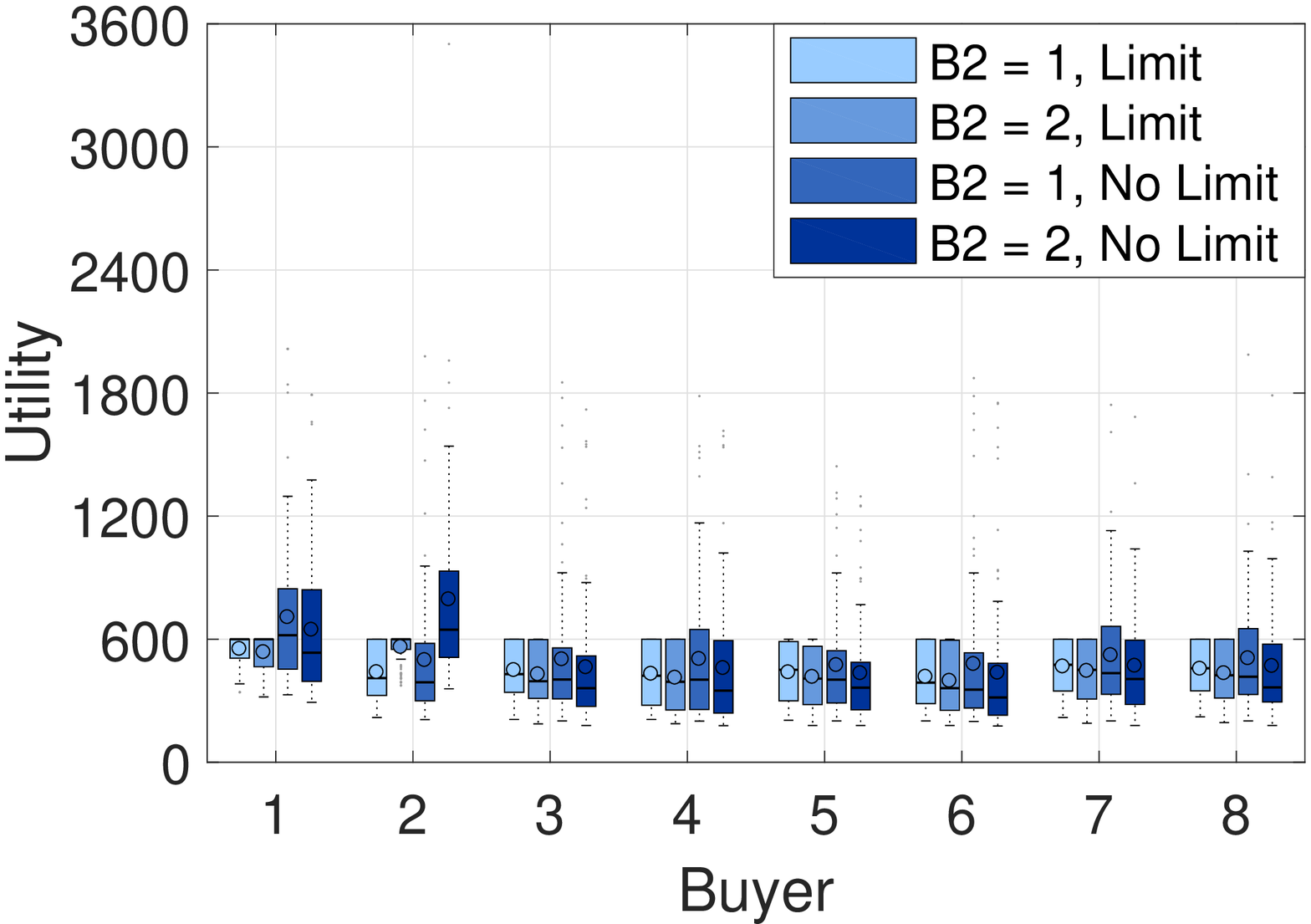}
	     \label{fig:budget_a_b}
	} \vspace{-0.2cm}
	\caption{Impact of budget on the equilibrium utilities}
\end{figure}

\begin{figure}[ht]
		\subfigure[N = 8, mean]{
		  \includegraphics[width=0.245\textwidth,height=0.10\textheight]{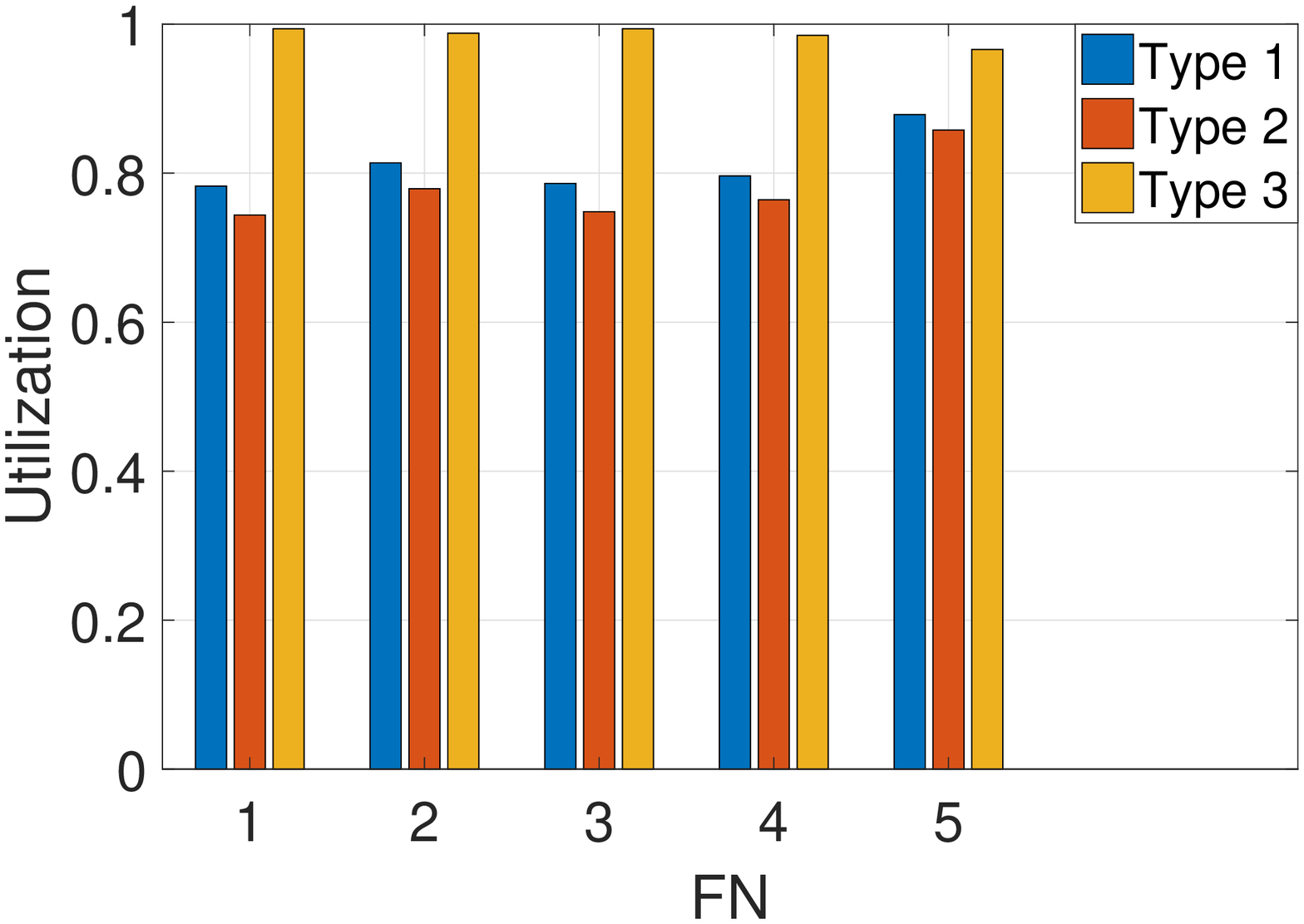}
	    \label{fig:ru1_a}
	}   \hspace*{-2.1em} 
		 \subfigure[N = 40, mean]{
	     \includegraphics[width=0.245\textwidth,height=0.10\textheight]{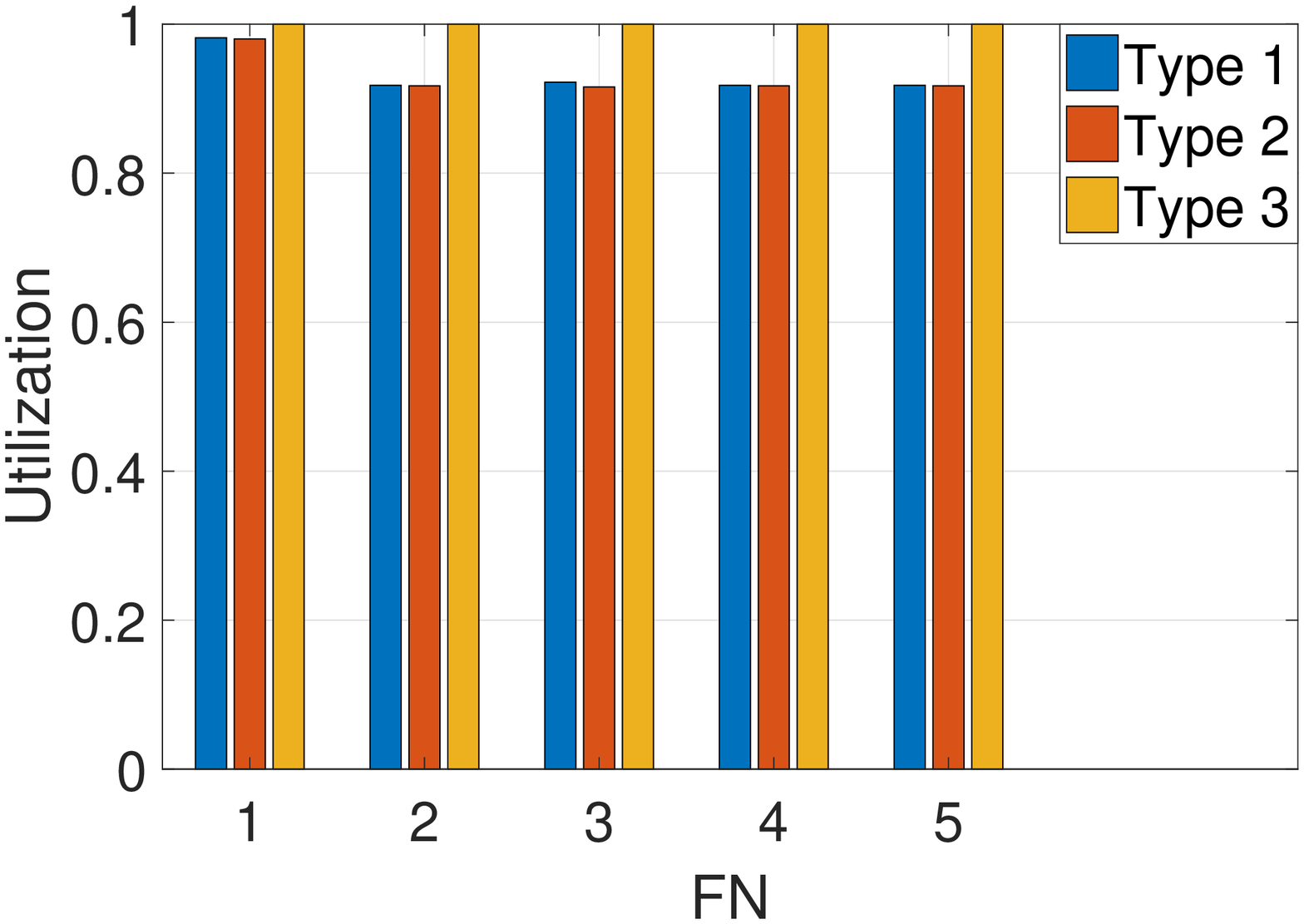}
	     \label{fig:ru2_a}
	}  \vspace{-0.2cm}
			\subfigure[N = 8, boxplot]{
		  \includegraphics[width=0.245\textwidth,height=0.10\textheight]{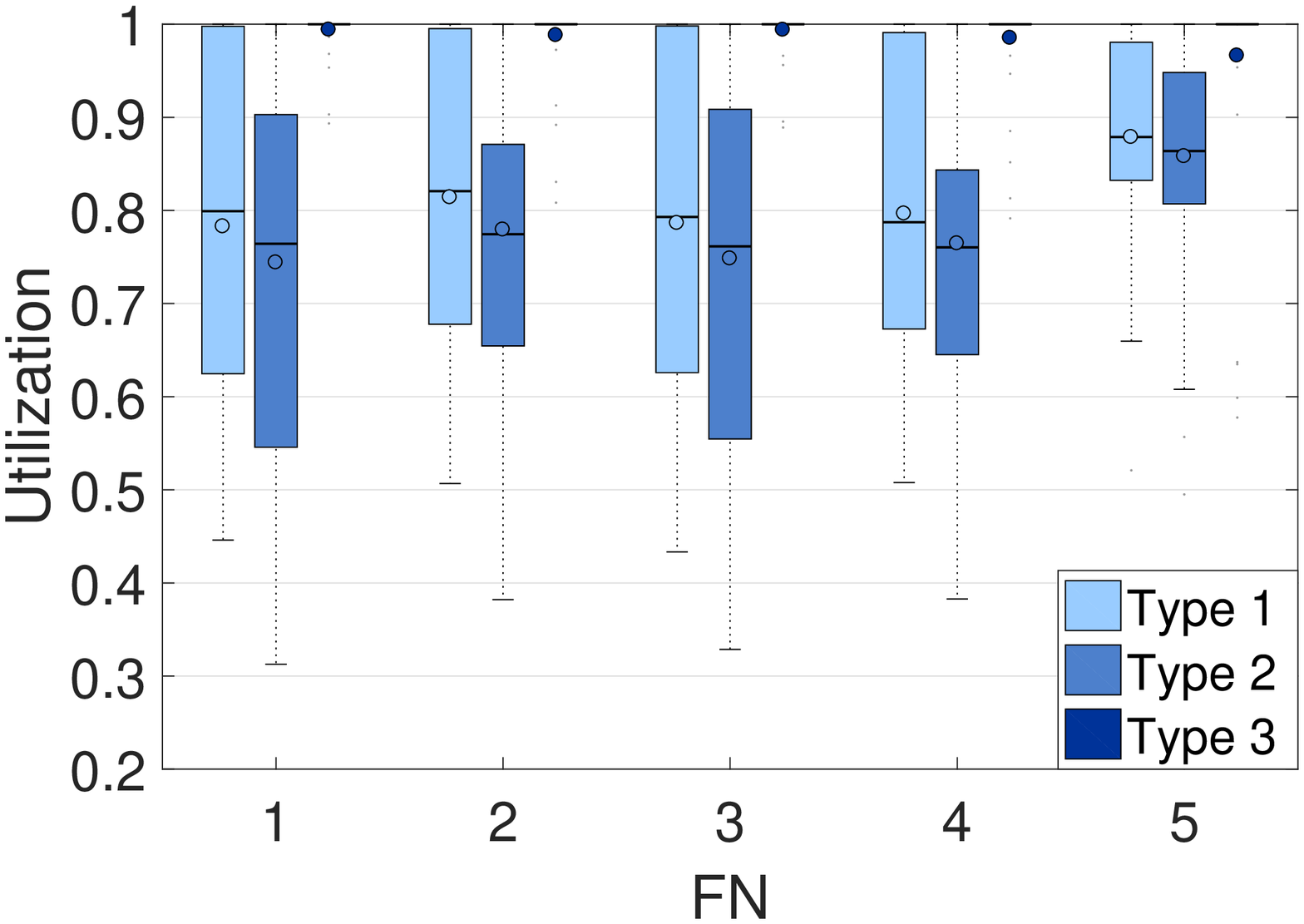}
	    \label{fig:ru1_a_b}
	}   \hspace*{-2.1em} 
		 \subfigure[N = 40, boxplot]{
	     \includegraphics[width=0.245\textwidth,height=0.10\textheight]{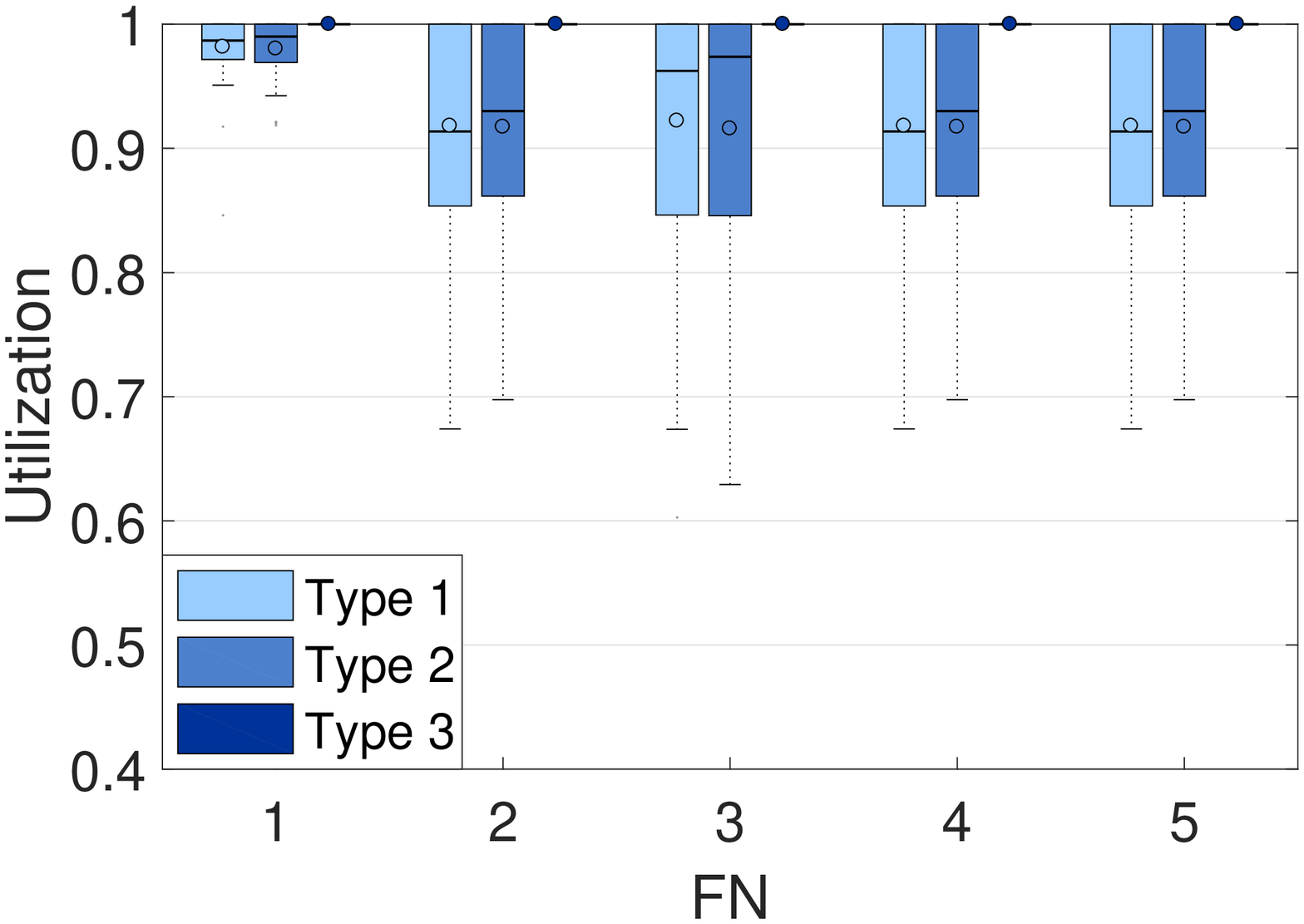}
	     \label{fig:ru2_a_b}
	}  \vspace{-0.2cm}
	\caption{Resource utilization (M = 40)}
\end{figure}

The impact of the  budget on the equilibrium utilities is presented in Figs. \ref{fig:budget_a}--\ref{fig:budget_a_b} (i.e., Fig.~\ref{fig:budget}). We can observe that when the budget of buyer 2 increases, his utility increases significantly while the utilities of the other buyers tend to decrease. Hence, the proposed scheme can effectively capture the service priority in the allocation decision. Finally, Figs.~\ref{fig:ru1_a}--\ref{fig:ru2_a_b} (i.e., Figs~~\ref{fig:ru1}--\ref{fig:ru2}) reveal that the proposed scheme produces an ME with high resource utilization and the resource utilization tends to increase as the number of buyers N  increases. It is worth mentioning that for each problem instance, each FN has one resource type that is fully utilized. However, the fully-utilized resource type of an FN is not necessarily the same for every run, which explains why the mean values of resource utilization of all resource types of some FNs (e.g., FN5 in \ref{fig:ru1_a}) are less than 1.

\newpage

\end{document}